\documentclass[preprint,5p,sort&compress]{elsarticle}

\usepackage{amsmath,amsthm,amssymb,amsfonts,mathtools}

\usepackage{parskip}
\usepackage{stfloats}
\usepackage{microtype} 
\usepackage{booktabs}
\usepackage[colorlinks]{hyperref}

\theoremstyle{definition}

\newtheorem{remark}{Remark}

\newcommand{\pVars}[3]{#1_{\text{#2}}^{#3}}

\begin{document}

\title{{\sc Predictive control of wastewater treatment plants as energy-autonomous water resource recovery facilities}}

\author[1]{Otac\'ilio B. L. Neto\corref{cor1}}
\author[2]{Michela Mulas}
\author[1]{Iiro Harjunkoski}
\author[1]{Francesco Corona}

\address[1]{School of Chemical Engineering, Aalto University, Finland}
\address[2]{Department of Teleinformatics Engineering, Federal University of Cear\'a, Brazil}

\cortext[cor1]{Corresponding author}

\begin{abstract}
    This work proposes an automatic control solution for the operation of conventional wastewater treatment plants (WWTPs) as energy-autonomous water resource recovery facilities.
    We first conceptualize a classification of the quality of treated water for three resource recovery applications (environmental, industrial, and agricultural water reuse). 
    We then present an output-feedback model predictive controller (Output MPC) that operates a plant to produce water of specific quality class, while also producing sufficient biogas to ensure nonpositive energy costs.
    The controller is demonstrated in the long-term operation of a full-scale WWTP subjected to typical influent loads and periodically changing quality targets.
    Our results provide a proof-of-concept on the energy-autonomous operation of existing wastewater treatment infrastructure with control strategies that are general enough to accommodate a wide range of resource recovery objectives.
\end{abstract}


\maketitle

\section{Introduction} \label{sec: Intro}

A paradigm shift in the management of water resources is taking place: 
Wastewater, traditionally seen as a pollutant, is now recognized as a sustainable source of clean water, energy, and nutrients \cite{Guest2009, Verstraete2011, UNESCO2017, Ricart2019, Kundu2022}.
The potential to commodify these resources, together with increasing water demands and environmental concerns, is driving the wastewater sector to rethink wastewater treatment plants (WWTPs) as \textit{water resource recovery facilities} (WRRFs) \citep{Sarpong2020, Kehrein2020}. 
Recovering nutrients directly to effluent streams, thus producing reuse water of tailored quality, has become a particularly attractive process goal:
Aside from promoting a circular economy and zero-waste principles, such a practice alleviates water stresses by conserving freshwater resources for the provision of potable water.
Moreover, a balanced water-energy nexus can be achieved by harvesting the excess sludge from this process for heat and electrical energy.
A technology able to produce reuse water of tailored quality while still having net-zero (or net-negative) energy balance costs thus contributes an essential step in the transition from wastewater treatment into water resource recovery.

Current research efforts focus on redesigning and upgrading treatment plants with new dedicated process units. 
Novel nutrient recovery processes include crystallizers for struvite precipitation \cite{Peng2018}, membrane-based processes for concentrating nitrogen via forward osmosis and ultrafiltration \cite{Xie2016, Wu2019}, and bioreactors promoting micro-algae cultivation \cite{ValverdePerez2015, Li2019}.
Energy recovery is commonly proposed by integrating combined heat and power (CHP) systems to treatment plants and rearranging their units to maximize the production of biofuels from waste \cite{Wan2016, Solon2019, Sarpong2020}.
In the aforementioned processes, source-separating systems for splitting domestic wastewater into urine, greywater, and blackwater is often a relevant pre-treatment step.
While promising, the costs of building and operating such technologies \cite{Chen2009, Farago2021} have undeniably slowed their implementation. 
In contrast, little has been investigated on the potential for repurposing the already in-place infrastructure by reprogramming their operation for water resource recovery goals.
In this direction, biological treatment of wastewater \cite{Henze2008} offer the most promising platform:
Such a treatment infrastructure \textit{i}) is common to most urban centers, \textit{ii}) has a recognized potential for energy recovery through the production of biogas, and \textit{iii}) is centred on controlling the nitrogen and phosphorus content within wastewater--two major nutrients lost by conventional treatment. 
Finally, these bioprocesses are increasingly instrumented and digitalized, thus facilitating the design and deployment of new operation policies.

Biological WWTPs are large-scale dynamical systems of notorious complexity.
Their real-time operation thus requires automation solutions capable of optimally driving actuators using data from \textit{in-situ} sensors.
Traditionally, the technology of choice has been low-level regulators that control key process variables to match operational targets set during the process design \cite{Olsson2013, Amand2013}.
Recently, advanced methods such as model predictive control (MPC, \cite{Rawlings2020}) have found noteworthy success:
The performance of biological treatment plants was shown to improve considerably when operated through both nonlinear \cite{Shen2008, Han2014, Vega2014, Sadeghassadi2018} and linear \cite{Holenda2008, Mulas2015, Santin2016} MPC strategies. 
Optimal plant-wide operation was also demonstrated with an MPC equipped only with a black-box model of the process dynamics \cite{Ekman2008, Foscoliano2016}. 
Model predictive controllers were employed in designing a cost-effective operation of WWTPs \cite{Zeng2015, Zhang2019b, Han2024} and in supporting the optimal selection of the controlled variables \cite{Francisco2015}. 
The technological scope has also been extended to incorporate predictive dynamical models in real-time state estimation for treatment plants \cite{Busch2013, Yin2018b, Yin2019}. 
However, while these contributions demonstrate the capabilities of model-based feedback control, they are limited to the classical goal of nitrogen removal. 
In our previous work, we built upon these solutions to propose a more flexible approach to the predictive control of WWTPs \cite{NMC_BSM1_Control}.
Our design was shown to be capable of operating an activated sludge process (the main section of a biological WWTP) given dynamic objectives and operational constraints.
Provided this control framework, now it remains to characterize and validate the task of energy-autonomous nutrient recovery for full-scale biological WWTPs.

In this work, we formalize the task of energy-autonomous production of reuse water as a control objective, then design a predictive controller for its solution.
Specifically, we
\begin{enumerate}[(i)]
    \item classify the quality of treated water, in terms of its biochemical profile, based on three target applications (environmental, industrial, and agricultural reuse);
    \item design an output-feedback MPC to operate a conventional WWTP to produce water of different quality classes, while enforcing nonpositive energy costs;
    \item demonstrate this controller on the long-term operation of a WWTP subjected to typical influent loads;
\end{enumerate}
The predictive controller is a digital solution whose functioning depends on a dynamic model of the process and its operation relies on the exchange of data with plants' actuators (the control actions) and sensors (the measurements).
We design its main components such that it achieves energy-autonomous resource recovery by deploying control actions that lead to the plant producing the desired water quality grade, while still ensuring that such an operation has nonpositive energy costs.
The controller is configured under the assumption that the targeted water quality changes periodically to supply alternating clients with reclaimed water.
While our setup is specific, this control strategy is general enough to accommodate a wider range of resource recovery objectives and treatment processes.
Our experiments are carried out at a simulation level and the results contribute a proof-of-concept on the flexibility of conventional wastewater treatment infrastructures. 

The rest of the paper is organized as follows: 
Section \ref{sec: ASP} overviews biological WWTPs, provides a classification of reuse water quality, and then presents the benchmark model used to represent and simulate real-world facilities.
Section \ref{sec: OutputMPC} overviews the output-feedback predictive controller, its individual components, and their specific configuration for the energy-autonomous operation of WWTPs as WRRFs.
Section \ref{sec: Results} presents and discusses the simulation results of operating a conventional WWTP using our predictive controller.
Finally, Section \ref{sec: Conclusion} provides concluding remarks.
For the model equations and additional implementation details, we refer to the Supplementary Material.

\section{The water resource recovery facility: Process layout, water quality classes, and simulation setup} \label{sec: ASP}

\begin{figure}[b!] \centering 
    \includegraphics[width=0.975\columnwidth]{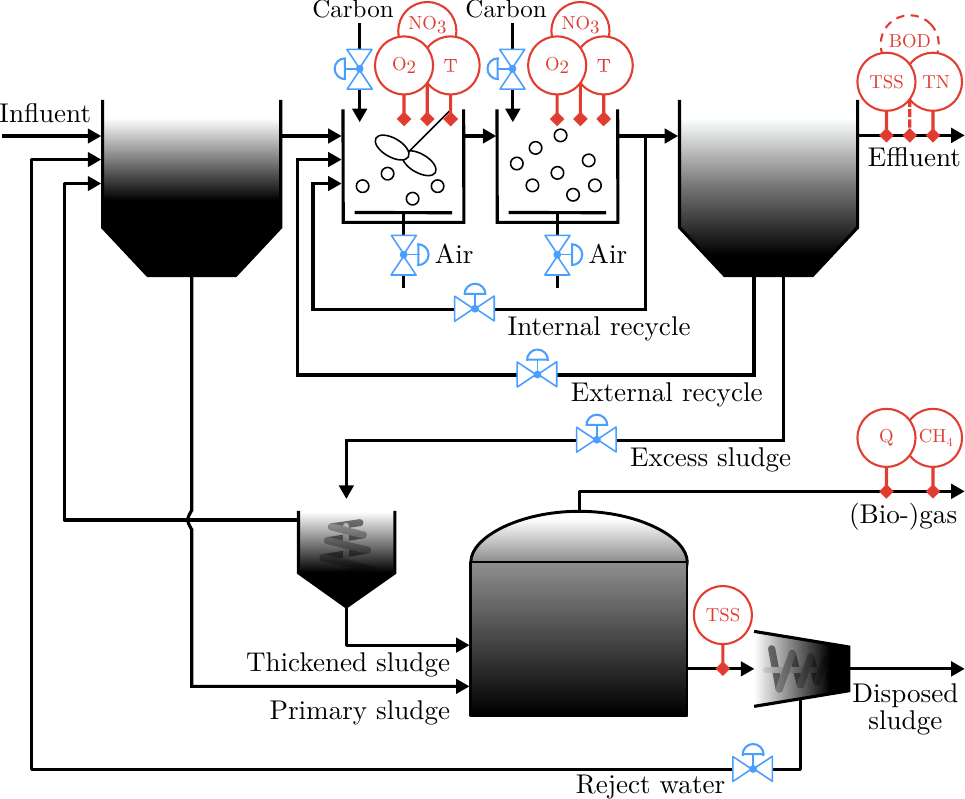}

    \caption{Process layout of a WWTP. The solid balloons refer to quantities measurable through online sensors, while the dashed balloons refer to quantities measurable only through laboratory analysis.}
    \label{fig: WWTP_Layout}
\end{figure}

We consider a conventional biological wastewater treatment plant (WWTP, \cite{Henze2008}) comprised of two main processes:
\begin{itemize}
    \item [--] The activated sludge process, in the \textit{water line}, based on nitrification-denitrification processes, in which microorganisms reduce ammonia present in the wastewater into nitrate, which is then further reduced into nitrogen gas and released into the atmosphere;
    \item [--] The anaerobic digestion process, in the \textit{sludge line}, in which organic material present in the excess sludge produced by the plant is converted into biogas, which is used as fuel for heat and energy generation.
\end{itemize}
In a typical process (Figure \ref{fig: WWTP_Layout}), the water line comprises two reacting sections and a settling tank. 
Each reacting section may consist of several zones for the oxidation of organic matter. 
The settler clarifies the treated water from suspended solids and flocculated particles before it is further processed or disposed. 
The sludge line typically comprises a thickener unit, an anaerobic digester, and a dewatering unit. 
The thickener and dewatering units are solid-separation processes that reduce the volume of the sludge fed to and disposed from the digester, respectively. 

In general, a WWTP can be understood as a process that converts influent wastewater into three main products: treated (or reclaimed) water, biogas, and disposed sludge.
When no further processes exist downstream, any quality requirements must be applied directly to these effluents.
The quality of the treated water is characterized by its concentrations of total suspended solids (TSS), organic matter (quantified as biochemical oxygen demand, BOD), and total nitrogen (TN, the sum of organic nitrogen, ammonia and nitrates/nitrites). 
The quality of the biogas is characterized by its concentrations of methane (CH$_4$). 
The quality of the disposed sludge is indirectly characterized by the concentrations of total suspended solids (TSS) in the outflow of the digester.
During operation, these quantities are routinely monitored through instrumentation consisting of both analyzers and online sensors.
Additional sensors (e.g., flow-rate, temperature, volume) are often deployed to further observe the operating conditions of the process.

\begin{remark}
    The quality of the treated water can include additional metrics such as its concentrations of phosphorus and pathogens.
    As these are treated by external or tertiary processes, they are not considered in our study.
    Regardless, our control framework (Section \ref{sec: OutputMPC}) can still be configured to incorporate such processes in the plant's layout-- provided that a model of their dynamics is available.
\end{remark}

A modern treatment plant might employ different solutions from control engineering to automate its operation \citep{Olsson2013}. 
This is required to ensure effluents of consistent quality despite the variability of the influent. 
Many control strategies are built upon two conventional feedback loops:
\begin{itemize}
    \item[--] Nitrate nitrogen (NO$_3$-N) in the anoxic section is controlled by manipulating the internal recirculation;
    \item[--] Dissolved oxygen (O$_2$) in the aerated section is controlled by manipulating the air flow-rate.
\end{itemize}
Using low-level regulators \cite{Seborg2016}, those measurable quantities are driven towards set-points provided manually by the plant's operators to achieve a specified effluent quality.
In more advanced strategies, these set-points are instead determined by higher-level control systems to optimally achieve a given objective under operational constraints.
To operate a WWTP as an energy-autonomous WRRF (that is, to produce reuse water of tailored quality with net-zero energy costs) such advanced solutions become necessary.

In the following, we define control objectives for this class of treatment plants when operated for resource recovery.
For the task of producing reuse water with net-zero energy costs, we i) characterize three water quality classes based on environmental, industrial and agricultural demands, and ii) present a quantitative metric of the energy being consumed (or generated) by the plant's operation.
We conclude the section by presenting the model and data used to simulate a WWTP for the validation of our proposed control system (Section \ref{sec: OutputMPC}) in face of these newly established targets.

\subsection{Water quality classes and control objectives} \label{subsec: Water_Class}

The primary control objective of an energy-autonomous WRRF consists of targeting different qualities for its effluent water.
Considering three major applications of reclaimed water (environmental, industrial, and agricultural reuse), the plant should be able to switch between producing:
\begin{enumerate}[{Class} A)]
    \item Water suitable for sustaining and/or augmenting natural water bodies;
    \item Water suitable for urban (e.g., toilet flushing) and industrial (e.g., cooling systems) activities;
    \item Water suitable for irrigation of both food and non-food crops.
\end{enumerate}
The classification is defined via upper limits on the effluent quality metrics (that is, concentrations of suspended solids, organic matter, and nitrogen), such that, would the treated water be used for the aforementioned purposes, no harmful side-effects are expected.
We consider the limits reported in Table \ref{tab: QualityClasses}, based on existing regulations and well-established standards \cite{EPA2012, EUreuse2017, EU2020, Shoushtarian2020}.
In practice, solids and organic matter must always be kept at low levels to avoid bacterial growth and accumulation of inert matter.
Conversely, being a nutrient, nitrogen has less stringent limits depending on the application: 
While not critical for industry, it is a valuable resource for agriculture. 
The limit on the nitrogen content of fertigation water (Class C) is then only meant to avoid issues related to the oversupply of nutrients \cite{Albornoz2016}.

The real-time operation of the WWTP to produce effluents of either quality requires an external (time-varying) demand for that specific water.
In the absence of explicit clients, we instead assume the control objective to change periodically (see Table \ref{tab: QualityClasses}). 
Specifically, we setup a scenario in which the plant alternates between targeting Class A and Class B on a monthly-basis.
In this case, the plant is to compensate an energy-intensive task (targeting Class A) with a less expensive operation (targeting Class B).
Exceptionally, Class C is the targeted quality during cold seasons:
This is to exploit the performance of the plant during winter, considering that \textit{i}) nitrogen removal is especially difficult at low temperatures and \textit{ii}) greenhouse farming would still create demands for irrigation water during this period.

\begin{remark}
    The scenario used for this study is specific but not restrictive:
    Our control strategy is also valid when the quality targets are given in real-time by external clients.
\end{remark}

\begin{table}[hb!] \centering
    \caption{Water quality: Demand period and classification in terms of the effluent concentrations of total suspended solids (TSS), biochemical oxygen demand (BOD), and total nitrogen (TN).}
    \label{tab: QualityClasses}
    \smallskip
    \begin{tabular}{@{}cccc@{}} \toprule
        \textbf{Class}   
                &      \textbf{A}   &     \textbf{B}    &    \textbf{C}     \\\midrule[1pt]
        Period  &     Odd months    &     Even months   &    Cold season    \\\midrule
        TSS     & $\leq 30$ g/m$^3$ & $\leq 30$ g/m$^3$ & $\leq 30$ g/m$^3$ \\
        BOD     & $\leq 10$ g/m$^3$ & $\leq 15$ g/m$^3$ & $\leq 20$ g/m$^3$ \\
        TN      & $\leq 15$ g/m$^3$ & $\leq 30$ g/m$^3$ & $\leq 45$ g/m$^3$ \\\bottomrule
    \end{tabular}
\end{table}

The secondary control objective of a energy-autonomous WRRF consists of adjusting its biogas production to recover the energetic costs incurred by its operation.
For this class of WWTPs, we quantify the net demand of electricity and heating using the energy cost index (ECI, in kWh/d),
\begin{equation}
    \text{ECI} = \text{AE} + \text{PE} + \text{ME} - \eta_{E} \text{MP} + \max(0, \text{HE} - \eta_{H}\text{MP})
\end{equation}
given the aeration (AE, kWh/d), pumping (PE, kWh/d), mixing (ME, kWh/d), and heating (HE, kWh/d) energy required for operation.
Energy generated from effluent biogas is expressed as methane production (MP, in kg\,CH$_4$/d) converted to electricity (with efficiency $\eta_{E} > 0$) and heat (with efficiency $\eta_H > 0$).
At any given time, these metrics are determined from the value of the process state and of the control actions requested to its actuators.
A WWTP is then characterized as energy-autonomous when its ECI, averaged over a specified period, is nonpositive.

\subsection{Model and simulation data} \label{subsec: BSM2}

The Benchmark Simulation Model no. 2 (BSM2, \cite{Gernaey2014}) is used in this work to represent a large-scale treatment plant. 
The BSM is the {\it de facto} platform for control design and simulation of biological treatment plants subject to typical municipal influents \citep{Olsson2013, Shen2008, Han2014, Vega2014, Sadeghassadi2018, Santin2016, Foscoliano2016, Zeng2015, Zhang2019b, Han2024, Francisco2015, Busch2013, Yin2018b, Yin2019, NMC_BSM1_Control}. 
It consists of a description of a plant's layout, a model of its dynamics and measurement process, and influent data for its simulation.

\begin{figure*}[b!] \centering
	\includegraphics[width=\textwidth]{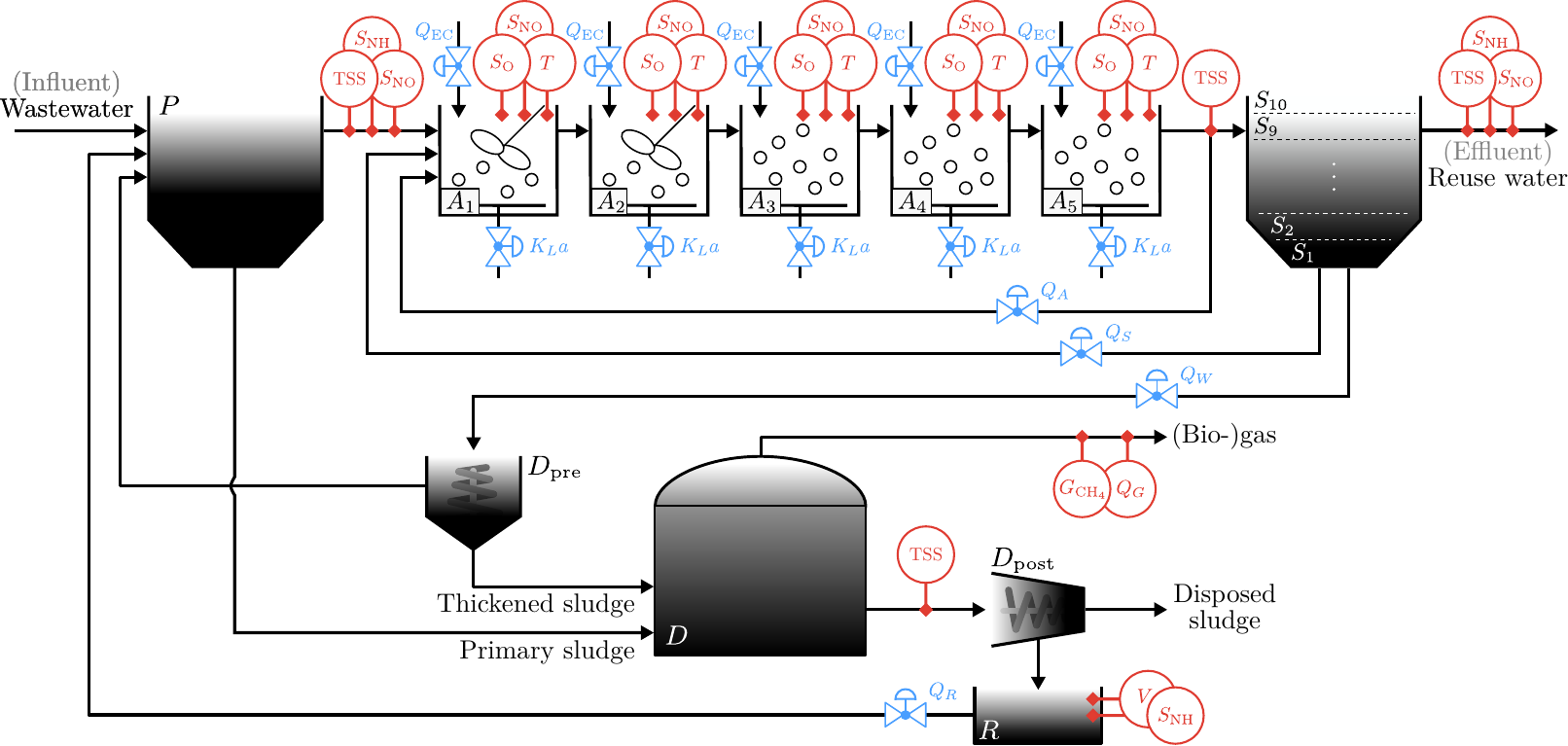}

    \caption{Process actuation and instrumentation layout for the Benchmark Simulation Model no. 2 (BSM2).} 
	\label{fig: BSM2}
\end{figure*}

The plant described by the benchmark (Figure \ref{fig: BSM2}) considers a biological WWTP whose water line is comprised of 
\begin{itemize}
    \item [--] 
    a series of $5$ bioreactors (denoted $A_1, \ldots, A_5$), with dynamics by the Activated Sludge Model no. 1 \citep{Henze2000}.
    For each $r$-th reactor, we control the inflow of air ($K_La^{A_r}$) and carbon ($Q_{\text{EC}}^{A_r}$), and measure the temperature ($T^{A_r}$) and concentrations of dissolved oxygen ($\pVars{S}{O}{A_r}$) and nitrate-plus-nitrite nitrogen ($\pVars{S}{NO}{A_r}$). 
    Additionally, the outflow concentrations of total suspended solids from the water line ($\pVars{\text{TSS}}{}{A_5}$) is also monitored;
    \item [--]
    a non-reactive settler represented by 10 vertical layers ($S_1, \ldots, S_{10}$) with dynamics by \citep{Takacs1991}.
    From its topmost layer ($S_{10}$), the concentrations of suspended solids ($\pVars{\text{TSS}}{}{S_{10}}$), soluble ammonia ($\pVars{S}{NH}{S_{10}}$), and nitrate-plus-nitrite nitrogen ($\pVars{S}{NO}{S_{10}}$), are monitored.
    The water in $S_{10}$ characterizes the plant's water effluent;
\end{itemize}
and the sludge line is comprised of
\begin{itemize}
    \item [--] 
    a thickener ($D_{pre}$) and a dewatering ($D_{post}$) unit, both described as ideal solid separation processes;
    \item [--] 
    an anaerobic digester ($D$), with dynamics by the Anaerobic Digestion Model no. 1 \citep{Batstone2002}.
    From this unit, we measure the flow-rate of effluent gas ($Q_{\text{gas}}^{D}$) and its methane composition ($\pVars{G}{CH$_4$}{D}$), and the outflow concentrations of total suspended solids ($\pVars{\text{TSS}}{}{D}$);
    \item [--] 
    a non-reactive storage tank for reject water ($R$), with basic mass flow dynamics.
    Its liquid volume ($V^{R}$) and concentrations of ammonia ($\pVars{S}{NH}{R}$) are measured.
\end{itemize}
The primary clarifier ($P$) has dynamics described by \cite{Otterpohl1995}.
The quality of its overflow ($P_{\text{eff}}$, fed to the water line) is monitored through its concentrations of total suspended solids ($\pVars{\text{TSS}}{}{P_{\text{eff}}}$), ammonia ($\pVars{S}{NH}{P_{\text{eff}}}$) and nitrate-plus-nitrite ($\pVars{S}{NO}{P_{\text{eff}}}$) nitrogen.
Finally, water and sludge can be recirculated in the plant by manipulating the internal ($Q_A$), external ($Q_S$), excess sludge ($Q_W$) and reject water ($Q_R$) recirculation.
We refer to Table \ref{tab: BSM2_Variables} for the summary of the output and actionable input variables for this process.

\begin{table}[tb!] \centering
    \caption{Actionable inputs, measurable outputs, and KPIs, for the Benchmark Simulation Model no. 2 (BSM2).}
    \label{tab: BSM2_Variables}
    \smallskip
    {\footnotesize\begin{tabular}{@{}p{3em}ll@{}}\toprule
                                & Description                           [Units]         & Location \\\midrule
    \textbf{Input}              &                                                       & \\\cmidrule{1-1}
    $Q_A$                       & Internal recycle flow-rate            [m$^3$/d]       & -- \\
    $Q_S$                       & External recycle flow-rate            [m$^3$/d]       & -- \\
    $Q_W$                       & Excess sludge flow-rate               [m$^3$/d]       & -- \\
    $Q_R$                       & Reject water flow-rate                [m$^3$/d]       & -- \\
    $K_La$                      & Oxygen transfer coefficient           [1/d]           & $A_{1 \leadsto 5}$ \\
    $Q_{\text{EC}}$             & External carbon flow-rate             [m$^3$/d]       & $A_{1 \leadsto 5}$ \\[1ex]
    \textbf{Output}             &                                                       & \\\cmidrule{1-1}
    $\pVars{\text{TSS}}{}{}$    & Total suspended solids                [g/m$^{3}$]     & $P_{\text{eff}}$, $A_5$, $S_{10}$, $D$ \\
    $\pVars{S}{NH}{}$           & NH$_4^+$+NH$_3$ nitrogen              [g/m$^{3}$]     & $P_{\text{eff}}$, $S_{10}$, $R$ \\
    $\pVars{S}{NO}{}$           & NO$_3^-$+NO$_2^-$ nitrogen            [g/m$^{3}$]     & $P_{\text{eff}}$, $A_{1 \leadsto 5}$, $S_{10}$ \\
    $\pVars{S}{O}{}$            & Dissolved oxygen                      [g/m$^{3}$]     & $A_{1 \leadsto 5}$ \\
    $\pVars{T}{}{}$             & Temperature                           [$^{\circ}$C]   & $A_{1 \leadsto 5}$ \\
    $\pVars{G}{CH$_4$}{}$       & Methane                               [kg/m$^3$]      & $D$ \\
    $Q_G$                       & Gas flow-rate                         [m$^3$/d]       & $D$ \\
    $\pVars{V}{}{}$             & Volume                                [m$^3$]         & $R$ \\[1ex]
    \textbf{KPI}                &                                                       & \\\cmidrule{1-1}
    TSS$^{\text{eff}}$          & Effluent total suspended solids       [g/m$^{3}$]     & -- \\
    BOD$_5^{\text{eff}}$        & Effluent 5-days measured BOD          [g/m$^{3}$]     & -- \\
    TN$^{\text{eff}}$           & Effluent total nitrogen               [g/m$^{3}$]     & -- \\
    ECI                         & Energy cost index                     [kWh/d]         & -- \\
    \bottomrule
    \end{tabular}}
\end{table}

\begin{figure*}[b!] \centering
	\includegraphics[width=\linewidth]{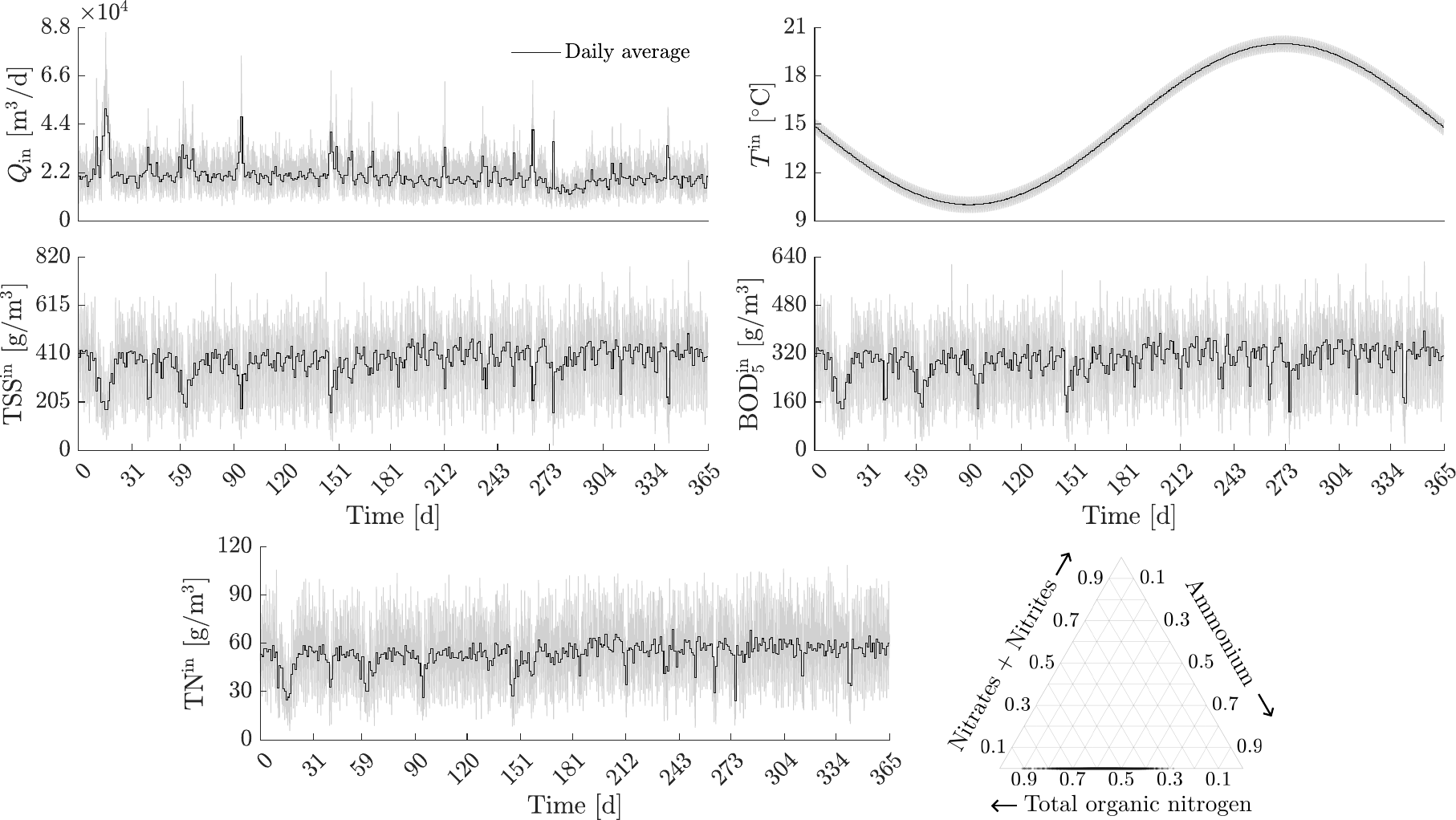}

  \caption{Wastewater, 1-year period: Influent flow-rate ($Q_{\text{in}}$) and temperature ($T^{\text{in}}$), and its quality in terms of total suspended solids (TSS$^{\text{in}}$), biochemical oxygen demand (BOD$_5^{\text{in}}$) and total nitrogen (TN$^{\text{in}}$). The ternary plot shows the composition of TN$^{\text{in}}$ at each time-step in terms of inorganic (nitrate, nitrite, and ammonium) and total organic nitrogen.}
  \label{fig: Influent_Summary}
\end{figure*}

The benchmark WWTP consists of a process with a total of $N_u = 14$ actionable inputs and $N_y = 27$ measurable outputs. 
Under this layout, the BSM2 provides a unified dynamical model describing the plant in terms of $N_x = 225$ state variables (the chemical concentrations and temperature in all units), how they evolve in response to external inputs (the influent wastewater and control actions), and how they are emitted in the measurements.
Moreover, it also describes how the state is used to determine the instantaneous value of $N_z = 4$ key performance indicators (KPIs): The effluent total suspended solids (TSS$^{\text{eff}}$), 5-days measured biochemical oxygen demand (BOD$_5^{\text{eff}}$), total nitrogen (TN$^{\text{eff}}$), and the plant's energy cost index (ECI).
Importantly, the actuation and instrumentation layout in this plant extends that of Figure \ref{fig: WWTP_Layout}: 
It assumes that air and extra carbon can be supplied to each bioreactor, and that solids ($\pVars{\text{TSS}}{}{}$) and inorganic nitrogen ($\pVars{S}{NO}{}$ and $\pVars{S}{NH}{}$) are monitored at both the primary outflow and effluent streams. 
Such an extension provides more degrees-of-freedom for designing control strategies, while still being representative of commercially available sensors and actuators \cite{Vanrolleghem2003}.

\subsubsection*{Characterization of the influent wastewater} \label{subsubsec: Influent}

The BSM2 is accompanied by wastewater data (Figure \ref{fig: Influent_Summary}) describing the influent flow-rate, its temperature, and its biochemical profile in terms of $13$ chemical concentrations. 
The data consists of $N_w = 1+1+13 = 15$ time series sampled every $15$ minutes over a period of $609$ days.
The influent is always of the municipal kind and non-actionable: Despite being the main driver of the plant, it is treated as a disturbance.
Exceptionally, the influent flow-rate and its temperature are assumed to be measurable.

We focus on controlling the WWTP when subjected to the first $365$ days of this influent scenario.
In Figure \ref{fig: Influent_Summary}, we show the influent wastewater during this period in terms of its flow-rate ($Q_{\text{in}}$), temperature ($T^{\text{in}}$), concentrations of total suspended solids (TSS$^{\text{in}}$), biochemical oxygen demand (BOD$_5^{\text{in}}$) and total nitrogen (TN$^{\text{in}}$).
For the latter, we also show that most influent nitrogen occurs in the form of either organic or ammonium nitrogen.
In addition to the seasonality (as per the temperature changes), we remark on the storm and rain events occurring over this scenario: They are characterized by sudden increases in the influent flow-rate and simultaneous dilution of its solids and nitrogen content.
Since deviating noticeably from the average conditions, these are considered extreme events from the perspective of designing control strategies.

\section{The predictive controller: Overview and specific setup for resource recovery tasks}\label{sec: OutputMPC} 

\begin{figure}[b!] \centering   
    \includegraphics[width=\columnwidth]{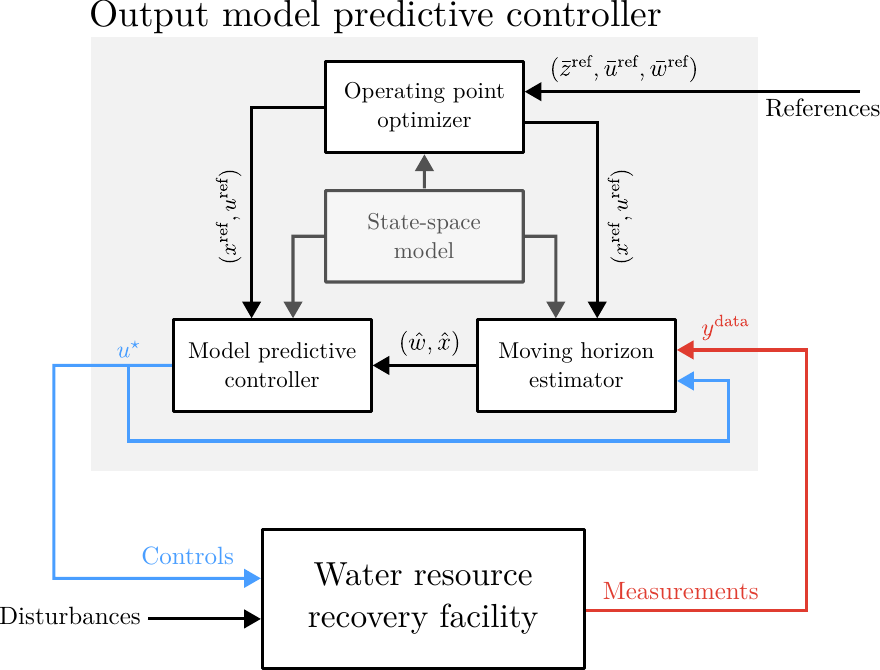}  

    \caption{Main components of the Output MPC. To follow a reference, the controller uses a state-space model and measurement data to determine the control actions to deploy to the plant.} 
    \label{fig: OutputMPC_Scheme} 
\end{figure}

Output model predictive control (Output MPC, \cite{NMC_BSM1_Control, Rawlings2020}) is a systematic approach to design and execute controllers to operate physical systems according to practical objectives while subject to operational and technological restrictions. 
For operating WWTPs as energy-autonomous WRRFs, we design an Output MPC (Figure \ref{fig: OutputMPC_Scheme}) consisting of three main components: 
\textit{i)} an operating-point optimizer (OPO); 
\textit{ii)} a model predictive controller (MPC); and 
\textit{ii)} a moving horizon state estimator (MHE). 
Control actions are computed as solutions to optimization problems based on a predictive model of the system's response to external inputs. 

The output-feedback controller is a device which computes, in real-time, the actions that evolve the state of the process such that its KPIs follow references provided by the plant's operators. 
This is achieved by an interoperability of its main components:
The OPO determines operating conditions for the process that, if enforced, lead to the desired KPIs.
The MHE determines the current state (and incoming disturbances) of the process based on the measurement data from its instrumentation.
Finally, the MPC determines the corrective control actions that drive the current state towards these optimal operating conditions. 

The Output MPC operates in cycles of duration $\Delta t > 0$, the \textit{control period}, each starting at time $t_k = k\Delta t$ ($k = 1, 2, \ldots$).
The computation of the control actions must be amenable to a digital implementation: 
This can be achieved by adopting a \textit{discretize-then-optimize} strategy \cite{Betts2010}, according to which the continuous-time predictive model (Section \ref{subsec: SS}) and the MPC and MHE optimizations (Section \ref{subsec: MPC} and \ref{subsec: MHE}) are discretized in time and then solved numerically. 
The reader is referred to the Supplementary Material for the technical details of its implementation. 
In the following, we overview the main components of this controller and present their configuration for operating our benchmark plant (Section \ref{subsec: BSM2}) for energy-autonomous water resource recovery. 

\subsection{The predictive model} \label{subsec: SS}

The predictive controller is based on a representation in state-space form of the wastewater treatment process,
\begin{subequations}
\begin{align}
    \textstyle\frac{d}{dt}{x}(t) & = f(x(t), u(t), w(t) \mid \theta_x);   \label{eq: SS_Nonlinear_x}\\
          y(t) & = g(x(t), u(t) \mid \theta_y);         \label{eq: SS_Nonlinear_y}\\
          z(t) & = h(x(t), u(t) \mid \theta_z).         \label{eq: SS_Nonlinear_z}
\end{align} \label{eq: SS_Nonlinear}%
\end{subequations} 
The state equation \eqref{eq: SS_Nonlinear_x} models how the $N_x$ state variables evolve given their value $x(t) \in \mathbb{R}^{N_x}$, the value $u(t) \in \mathbb{R}^{N_u}$ of the $N_u$ actionable input (or control variables), and the value $w(t) \in \mathbb{R}^{N_w}$ of the $N_w$ disturbances, at time $t > 0$. 
The output equation \eqref{eq: SS_Nonlinear_y} models how the $N_y$ measured outputs $y(t) \in \mathbb{R}^{N_y}$ are formed as emissions of the state $x(t)$ and actions $u(t)$. 
The performance equation \eqref{eq: SS_Nonlinear_z} models how the $N_z$ key performance indicators $z(t) \in \mathbb{R}^{N_z}$ are computed from the value of the state $x(t)$ and actions $u(t)$. 
The vectors $\theta_x$, $\theta_y$, and $\theta_z$ collect the constant parameters in the functions $f$, $g$, and $h$, respectively.
Since the process has no input feedthrough, we hereafter write $y(t) = g(x(t)|\theta_y)$.

\begin{remark}
    During operation, only the measurement data need to be collected from the instruments in real-time. 
    While the KPIs could be monitored (e.g., to assess the efficacy of the controller), the Output MPC operates by computing these quantities using its predictive model.
\end{remark}

The predictive model is assumed to be that provided by the Benchmark Simulation Model no. 2 (Section \ref{subsec: BSM2})--as such, the controller has knowledge of the \textit{true dynamics} of the system being controlled.
While this is a strong assumption, it enables our proof-of-concept to highlight the capabilities of automatic control of WRRFs without the additional problem of plant-model mismatch.
Naturally, the same design can be applied when only a surrogate model of the dynamics is available.
Regardless, we remark that the controller has no knowledge of the incoming disturbances and thus its predictions are still imperfect and must rely on accurate estimations of the process' state from data.

\subsection{The operating-point optimizer (OPO)} \label{subsec: OPO}

The operating-point optimizer determines operating conditions that, if enforced by a controller, achieve the desired performance while satisfying the operational restrictions. 
Here, optimality is quantified by \textit{closeness between the plant's KPIs and some target reference profile} set by the plant's management to reflect their economic and environmental objectives. 
An optimal operating point thus consists of a state-, input-, and disturbance-vector at which the plant is in steady-state and its KPIs match the reference profile, according to the model in Eq. \eqref{eq: SS_Nonlinear}. 

For a cycle starting at $t_k$, the optimal operating conditions are the state $x_{k}^{\text{ref}} \in \mathbb{R}^{N_x}$ and controls $u_{k}^{\text{ref}} \in \mathbb{R}^{N_u}$ solving
\begin{subequations}
\begin{align} 
    \underset{x_{k}, u_{k}}{\text{minimize}}
        & \quad \| W_z^{\text{ref}}(z_{k} {-} \bar{z}^{\text{ref}}_{k}) \|^2_2 + \| W_u^{\text{ref}} (u_{k} {-} \bar{u}^{\text{ref}}_{k}) \|^2_2 \label{eq: OPO_a}\\
    \text{subject to} 
        & \quad 0 = f(x_{k}, u_{k}, \bar{w}^{\text{ref}}_{k} \mid \theta_x),	                       \label{eq: OPO_b}\\
        & \quad x_{k} \in \mathcal{X}^{\text{ref}}, ~~ u_{k} \in \mathcal{U}^{\text{ref}}, ~~ z_{k} \in \mathcal{Z}^{\text{ref}}, \label{eq: OPO_c}
\end{align} \label{eq: OPO}%
\end{subequations}
with $z_{k} = h(x_{k}, u_{k})$. 
The references $(\bar{z}^{\text{ref}}_{k}, \bar{u}^{\text{ref}}_{k}, \bar{w}^{\text{ref}}_{k})$ are, respectively, the desired KPIs, the desired control configuration, and the expected disturbance conditions for the $k$-th cycle. 
The objective function (Eq. \ref{eq: OPO_a}) is also parametrized by the weighting matrices $W_z^{\text{ref}}, W_u^{\text{ref}} \succeq 0$: 
As customary practice, we consider diagonal matrices (leading to $N_z + N_u$ tuning parameters) with $W_z^{\text{ref}} \succ W_u^{\text{ref}}$ to prioritize matching the desired performance rather than the reference control configuration.
The requirement that $(x_{k}, u_{k}, \bar{w}^{\text{ref}}_{k})$ be a steady-state (Eq. \ref{eq: OPO_b}) enables the MPC to stabilize the system around the reference (consequently, achieving $\bar{z}^{\text{ref}}_{k}$) with zero-offset. 
Finally, the constraint sets $\mathcal{X}^{\text{ref}}$, $\mathcal{U}^{\text{ref}}$, and $\mathcal{Z}^{\text{ref}}$, (Eq. \ref{eq: OPO_c}) define which pairs $(x_{k}, u_{k})$ are feasible based on restrictions to be satisfied during the operation of the process.
Here, they are taken to describe bound constraints (that is, minimum and maximum values the corresponding variables can take).
Formally, 
\begin{subequations}
\begin{align}
    \mathcal{X}^{\text{ref}} &= \{ x \in \mathbb{R}^{N_x} : x^{\text{ref}}_{lb} \preceq x \preceq x^{\text{ref}}_{ub}\}; \label{eq: OPO_Constraint_Sets_Linear_A}\\
    \mathcal{U}^{\text{ref}} &= \{ u \in \mathbb{R}^{N_u} : u^{\text{ref}}_{lb} \preceq u \preceq u^{\text{ref}}_{ub}\}; \label{eq: OPO_Constraint_Sets_Linear_B}\\
    \mathcal{Z}^{\text{ref}} &= \{ z \in \mathbb{R}^{N_z} : z^{\text{ref}}_{lb} \preceq z \preceq z^{\text{ref}}_{ub}\}, \label{eq: OPO_Constraint_Sets_Linear_C}
\end{align} \label{eq: OPO_Constraint_Sets_Linear}%
\end{subequations}
given the collections of lower $\{x^{\text{ref}}_{lb}, u^{\text{ref}}_{lb}, z^{\text{ref}}_{lb}\}$ and upper $\{x^{\text{ref}}_{ub}, u^{\text{ref}}_{ub}, z^{\text{ref}}_{ub}\}$ bounds of appropriate dimensions. 
When a variable is unbounded below (respectively, above), the corresponding entry of $(\cdot)_{lb}$ ($(\cdot)_{ub}$) is set to $-\infty$ ($\infty$). 

We configure the OPO to implement energy-autonomous resource recovery by having i) the references $\bar{z}^{\text{ref}}$ enforce a desired quality class, and ii) the constraints $\mathcal{Z}^{\text{ref}}$ enforce nonpositive energy costs.
We define the references $(\bar{z}^{\text{ref}}, \bar{u}^{\text{ref}}, \bar{w}^{\text{ref}})$, bound constraints $(\mathcal{X}^{\text{ref}}, \mathcal{U}^{\text{ref}}, \mathcal{Z}^{\text{ref}})$, and weighting factors $(W_z^{\text{ref}}, W_u^{\text{ref}})$ as reported in Table \ref{tab: OPO_Tuning}.
The weighting for ECI ($z_4$) removes this variable from the objective function; it is not required to match a specific value, but only to be nonpositive.
This setup is based on extensive experimental campaigns, with the references for $(\bar{u}^{\text{ref}}, \bar{w}^{\text{ref}})$ taken directly from the benchmark: 
They are the conventional open-loop controls ($\bar{u}^{\text{ref}}$) when the process performs nitrogen removal under average influent conditions ($\bar{w}^{\text{ref}}$).

\begin{table}[tb!] \centering
    \caption{
        Output MPC, operating point optimizer (OPO): 
        Tuning for energy-autonomous resource recovery. 
        References for the KPIs $z_{1 \leadsto 4}$ depend on the targeted water quality class (A, B, or C). 
    }
    \label{tab: OPO_Tuning}
    \smallskip
    {\small\begin{tabular}{@{}rccccc@{}} \toprule
        \textbf{Variable}            & \multicolumn{3}{c}{\textbf{Reference}} & \textbf{Bounds} & \textbf{Weight} \\\midrule
                                     &  A  &   B  &   C                       &                 &   \\\cmidrule{2-4}
        TSS$^{\text{eff}}$   ($z_1$) &  10 &   10 &   10                      & $[0, \infty)$   & 1 \\
        BOD$_5^{\text{eff}}$ ($z_2$) &   4 &    4 &    4                      & $[0, \infty)$   & 1 \\
        TN$^{\text{eff}}$    ($z_3$) & 7.5 & 22.5 & 37.5                      & $[0, \infty)$   & 1 \\
        ECI                  ($z_4$) &   0 &    0 &    0                      & $(-\infty, 0]$  & 0 \\\midrule
        $Q_A$                ($u_1$) & \multicolumn{3}{c}{61944}              & $[0, 92230]$    & 6e-5 \\
        $Q_S$                ($u_2$) & \multicolumn{3}{c}{20648}              & $[0, 36892]$    & 6e-4 \\
        $Q_W$                ($u_3$) & \multicolumn{3}{c}{300}                & $[0,  1844]$    & 6e-4 \\
        $Q_R$                ($u_4$) & \multicolumn{3}{c}{100}                & $[0,   500]$    & 6e-3 \\
        $K_La^{A_1}$         ($u_5$) & \multicolumn{3}{c}{0}                  & $[0, 360]$      & 6e-3 \\
        $K_La^{A_2}$         ($u_6$) & \multicolumn{3}{c}{0}                  & $[0, 360]$      & 6e-3 \\
        $K_La^{A_3}$         ($u_7$) & \multicolumn{3}{c}{120}                & $[0, 360]$      & 6e-3 \\
        $K_La^{A_4}$         ($u_8$) & \multicolumn{3}{c}{120}                & $[0, 360]$      & 6e-3 \\
        $K_La^{A_5}$         ($u_9$) & \multicolumn{3}{c}{60}                 & $[0, 360]$      & 6e-3 \\
        $\pVars{Q}{EC}{A_{1\leadsto5}}$ ($u_{10\leadsto14}$) & \multicolumn{3}{c}{0} & $[0, 5]$ & 2e-1 \\\midrule
        ($w_{1\leadsto 15}$)        & \multicolumn{5}{c}{See the Supplementary Material}  \\\midrule
        ($x_{1\leadsto 225}$)       & \multicolumn{5}{c}{See the Supplementary Material}  \\\bottomrule
    \end{tabular}}
\end{table}

\begin{remark}
    Problem \eqref{eq: OPO} belongs to the class of nonconvex optimization problems and, as such, its solution is computationally demanding. 
    However, the OPO needs to be executed only when changes occur for either the references, the operational constraints, or the expected disturbance profile. 
    As such, its computational burden is not critical. 
\end{remark}

\subsection{The model predictive controller (MPC)} \label{subsec: MPC}

The model predictive controller plans a sequence of control actions which optimally evolve the system from its current state $x(t_k)$, at time $t_{k}$, towards the reference state $x^{\text{ref}}_{k}$, with minimal deviation from the reference input $u^{\text{ref}}_{k}$. 
The planning is done for a period of time of duration $H_c > 0$, the \textit{control-horizon}, under the assumption that \textit{i}) the system evolves as the model predicts and that \textit{ii}) the disturbances remain constant throughout the control horizon. 
Accounting for disturbance and model uncertainties, only the first optimal action $u(t_k)$ is deployed to the system then held constant until the next cycle starting at $t_{k+1} = t_k + \Delta t$.

For a cycle starting at $t_k$, the sequence of optimal actions is the function (of time) $u^{\star}: [t_{k},t_{k}{+}H_c] \to \mathbb{R}^{N_u}$ solving
\begin{subequations}
\begin{align}
    \underset{u(\cdot), x(\cdot)}{\text{minimize}}
        & \quad \int_{t_k}^{t_k + H_c}{\hskip-0.50em L^c(x(t), u(t))dt} + L^f (x(t_k{+}H_c))	\label{eq: MPC_cost}\\
    \underset{t \in [t_k,t_k+H_c]}{\text{subject to}}
        & \quad \textstyle\frac{d}{dt}{x}(t) = f(x(t),u(t),\widehat{w}(t_k)),			\label{eq: MPC_dynamics}\\[-1.5ex]
        & \quad x(t) \in \mathcal{X}^c, ~ u(t) \in \mathcal{U}^c,	\label{eq: MPC_sets}\\
        & \quad x(t_k) = \widehat{x}(t_k).						    \label{eq: MPC_init}
\end{align} \label{eq: MPC}%
\end{subequations}
The controller is designed to achieve optimal tracking by minimizing the (squared) distance of the states ($x(t)$) and inputs ($u(t)$) through the control horizon ($t \in [t_k, t_k{+}H_c]$) to the desired reference ($x^{\text{ref}}_{k}$, $u^{\text{ref}}_{k}$) obtained by the OPO. 
Formally, this is expressed in Eq. \eqref{eq: MPC_cost} by the stage and terminal cost functions
\begin{align*} 
    L^c(\cdot, \cdot)	
        & = \| W_x \big( x(t) - x^{\text{ref}}_{k} \big) \|^2_2 + \| W_u \big( u(t) - u^{\text{ref}}_{k} \big) \|^2_2;  \\ 
    L^f(\cdot) 
        & = \| W_f \big( x(t_k{+}H_c) - x^{\text{ref}}_{k} \big) \|^2_2.
\end{align*}
parametrized by the matrices $W_x, W_f \succeq 0$ and $W_u \succ 0$: 
We consider diagonal matrices (leading to $N_x + N_u$ tuning parameters) with $W_x \succ W_u$ to prioritize correcting deviations from the reference state. 
The terminal $W_{f}$ is of particular importance and a sensible design can improve the robustness of the controller \cite{Rawlings2020}.
We refer to the Supplementary Material for the explicit tuning of these parameters.

The performance of the predictive controller depends on a frequent and consistent deployment of control actions. 
In its current formulation, however, the problem \eqref{eq: MPC} is nonconvex and thus computationally demanding. 
We substantially reduce this computational load by converting the dynamic constraints (Eq. \ref{eq: MPC_dynamics}) into linear equalities: 
Specifically, at the $k$-th cycle, we consider the linearization of the plant's dynamics around the operating point $p_{k} = (x_{k}^{\text{ref}}, u_{k}^{\text{ref}}, \bar{w}_{k}^{\text{ref}})$,
\begin{align}
    \textstyle\frac{d}{dt}x(t) &= f(x(t), u(t), \widehat{w}(t_k)) \nonumber\\
        & \approx A_{k} (x(t) {-} x_{k}^{\text{ref}}) {+} B_{k} (u(t) {-} u_{k}^{\text{ref}}) {+} G_{k} (\widehat{w}(t_k) {-} \bar{w}_{k}^{\text{ref}}), \nonumber
\end{align} %
with $A_{k} \in \mathbb{R}^{N_x \times N_x}$, $B_{k} \in \mathbb{R}^{N_x \times N_u}$, and $G_{k} \in \mathbb{R}^{N_x \times N_w}$ being the first-order variations of $f(\cdot)$ evaluated at $p_{k}$,
$$
    A_{k} {=} (\partial f / \partial x)\vert_{p_{k}}, ~ B_{k} {=} (\partial f / \partial u)\vert_{p_{k}}, ~ G_{k} {=} (\partial f / \partial w)\vert_{p_{k}}.
$$
This is an accurate approximation of the nonlinear dynamics provided that the controller maintains the plant's state and inputs close to the fixed-point $p_{k}$; a valid assumption aside from extreme events. 
Importantly, the Jacobian matrices $(A_{k}, B_{k}, G_{k})$ can be computed efficiently and with high accuracy even for large-scale processes \cite{Rawlings2020}.

The design of our MPC is concluded by having the static constraints sets, $\mathcal{U}^c$ and $\mathcal{X}^c$ (Eq. \ref{eq: MPC_sets}), to describe bound constraints (that is, minimum and maximum values that control and state variables can take). 
We enforce the input signal to satisfy the actuator limits reported in Table \ref{tab: OPO_Tuning} (thus, $\mathcal{U}^c = \mathcal{U}^{\text{ref}}$), while the state signal is only constrained by the linearized dynamics (thus, $\mathcal{X}^c = \mathbb{R}^{N_x}$).

\subsection{The moving-horizon state estimator (MHE)}  \label{subsec: MHE}

The moving-horizon estimator is responsible for determining the state- $\widehat{x}(t_k)$ and disturbance-vector $\widehat{w}(t_k)$ used in the MPC (Problem \ref{eq: MPC}): 
They are the terminal values of the state and disturbance trajectories that solve an optimal state estimation problem, over a past \textit{estimation-horizon} of duration $H_e$. 
Again, the problem is solved under the assumption that the system evolves as the model predicts. 
In this case, optimality is defined in terms of \textit{closeness between the past model outputs and plant's measurements} under the controls deployed to the plant over the estimation-horizon. 

For a cycle starting at $t_k$, the state and disturbance estimates are the functions (of time) $\widehat{x} : [t_k{-}H_e,t_k] \to \mathbb{R}^{N_x}$ and $\widehat{w} : [t_k{-}H_e,t_k] \to \mathbb{R}^{N_w}$ solving the problem
\begin{subequations}
\begin{align}
    \underset{w(\cdot), x(\cdot)}{\text{minimize}} \quad
        & L^i (x(t_k{-}H_e)) + \int_{t_k{-}H_e}^{t_k}{\hskip-0.5em L^e(x(t),w(t))dt}	\label{eq: MHE_cost}\\ 	
    \underset{t \in [t_{k}{-}H_e,t_{k}]}{\text{subject to }} \quad
        & \textstyle\frac{d}{dt}{x}(t) = f(x(t),u^{\star}(t),w(t)),					\label{eq: MHE_dynamics}\\[-1.5ex]
        & x(t) \in \mathcal{X}^e, \ w(t) \in \mathcal{W}^e.       \label{eq: MHE_sets}
\end{align} \label{eq: MHE}%
\end{subequations}
The estimator is designed to determine the state and disturbance that best explain the plant's data by minimizing the (squared) distance of the model output ($y(t) = g(x(t))$) to the actual measurements ($y^{\text{data}}(t)$) obtained over the past estimation horizon ($t \in [t_k{-}H_e,t_k]$). 
This is expressed in Eq. \eqref{eq: MHE_cost} by the stage and initial cost functions
\begin{align*} 
    L^e(\cdot, \cdot)	
        & = \| W_y \big( y(t) - y^{\text{data}}(t) \big) \|^2_2 + \| W_w \big( w(t) - \bar{w}^{\text{ref}}_{k} \big) \|^2_2;\\
    L^i(\cdot)
        & = \| W_i \big( x(t_k{-}H_e) - \widehat{x}^{\text{ref}}_{k} \big) \|^2_2,
\end{align*}%
parametrized by matrices $W_y, W_w, W_i \succ 0$: 
We consider diagonal matrices (leading to $N_y + N_w$ tuning parameters) with $W_y \succ W_w$ to prioritize matching the measurement data $y^{\text{data}}(t)$ instead of the expected disturbances $\bar{w}^{\text{ref}}_{k}$. 
Again, we refer to the Supplementary Material for the explicit tuning of these parameters.
The initial state reference is defined recursively as $\widehat{x}_{k}^{\text{ref}} = \widehat{x}(t_k{-}H_e{+}\Delta t)$ with $\widehat{x}(\cdot)$ being the estimates of the previous MHE cycle. 

The computational load of the estimator is reduced by approximating the plant's model with the linearization,
\begin{align}
    \textstyle\frac{d}{dt}{x}(t) 
        &= f(x(t), u(t), w(t)) \nonumber\\
        & \approx A_{k} (x(t) {-} x_{k}^{\text{ref}}) {+} B_{k} (u(t) {-} u_{k}^{\text{ref}}) {+} G_{k} (w(t) {-} \bar{w}_{k}^{\text{ref}}), \nonumber\\
    y(t) 
        &= g(x(t)) \nonumber\\
        & \approx g(x_{k}^{\text{ref}}) + C_{k} (x(t) {-} x_{k}^{\text{ref}}), \nonumber
\end{align} %
with the same Jacobian matrices ($A_{k}$, $B_{k}$, $G_{k}$) as defined in Section \ref{subsec: MPC} and  $C_{k} = (\partial g / \partial x)\vert_{p_{k}} \in \mathbb{R}^{N_y \times N_x}$ being the first-order variations of $g(\cdot)$ evaluated at $p_k = (x_k^{\text{ref}}, u_k^{\text{ref}}, \bar{w}_k^{\text{ref}})$. 

Finally, the design of our MHE is concluded by having the static constraints sets, $\mathcal{W}^{e}$ and $\mathcal{X}^{e}$ (Eq. \ref{eq: MHE_sets}), to describe bound constraints (that is, minimum and maximum values that disturbance and state variables can take).
In this case, the disturbances are known to be positive quantities (thus, we set $\mathcal{W}^e = \mathbb{R}_{\geq 0}^{N_w}$), while the state signal is again only constrained by the linearized dynamics ($\mathcal{X}^e = \mathbb{R}^{N_x}$).

\section{Simulation results and discussion}\label{sec: Results} 

This section presents the results obtained by the benchmark WWTP (Section \ref{subsec: BSM2}) when operated by our Output MPC (Section \ref{sec: OutputMPC}) for energy-autonomous resource recovery (Section \ref{subsec: Water_Class}). 
For a realistic simulation, the process data ($y^{\text{data}}$) is corrupted with a random process representing real-occurring sensor noise (see the Supplementary Material).
We first discuss how the automatic operation of the plant achieves our primary objective (production of reuse water) then show how it simultaneously achieves our secondary objective (energy-autonomous operation).
We provide a qualitative assessment of the controller's performance and discuss its potential in transitioning conventional treatment plants towards water resource recovery facilities.

\subsection{Primary objective: Reuse water of tailored quality}

\begin{figure*}[htb!] \centering
    \includegraphics[width=1\textwidth]{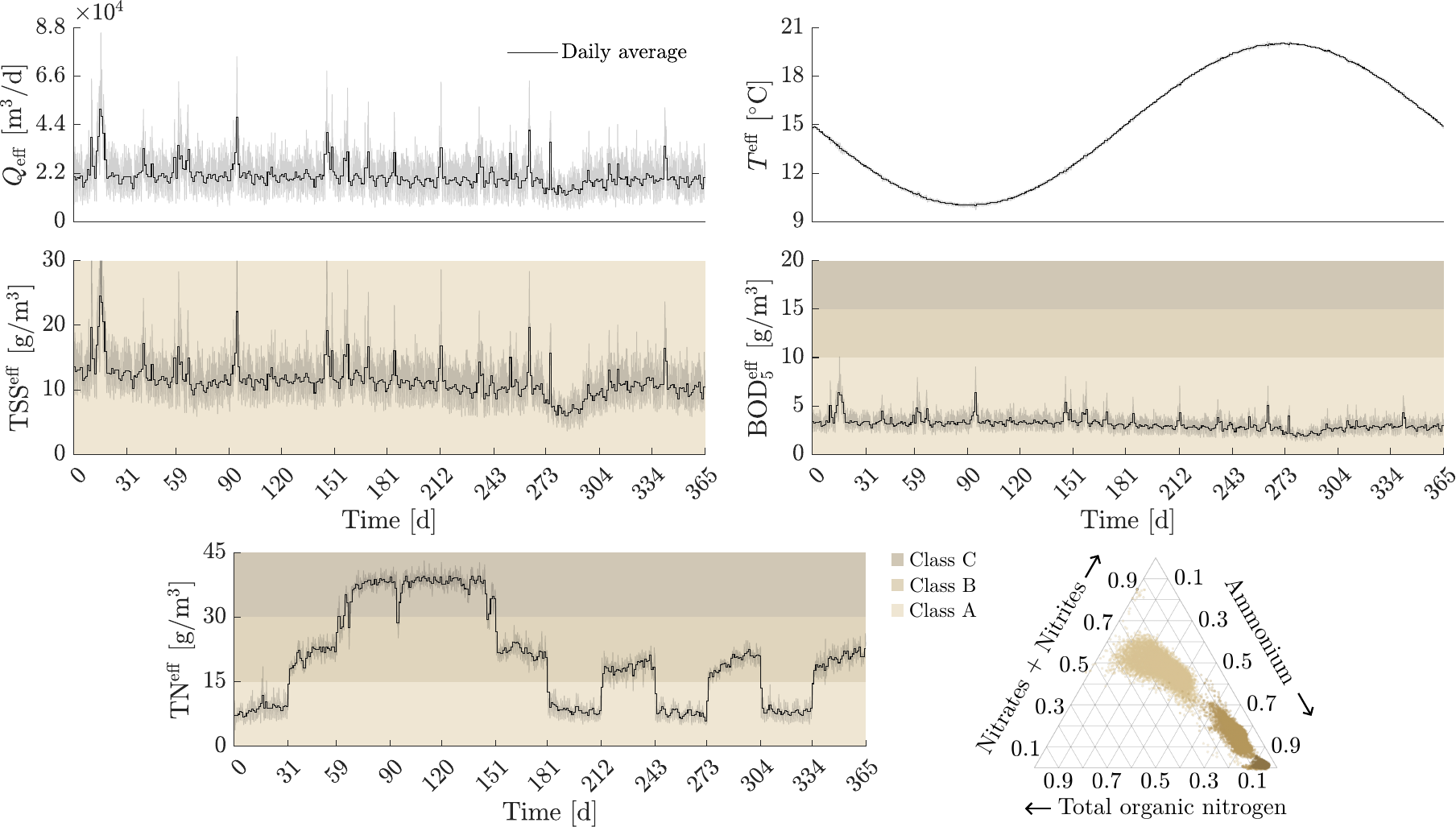}
    \vskip-1em
    \caption{%
        Simulation, 1-year period: 
        Effluent flow-rate ($Q_{\text{eff}}$) and temperature ($T^{\text{eff}}$), and its quality in terms of total suspended solids (TSS$^{\text{eff}}$), biochemical oxygen demand (BOD$_5^{\text{eff}}$) and total nitrogen (TN$^{\text{eff}}$). 
        The shaded regions indicate the different water quality classes (A, B, and C). 
        The ternary plot shows the composition of TN$^{\text{eff}}$ at each time-step in terms of inorganic and organic nitrogen forms.
    }
    \label{fig: Simulation_Effluent}
\end{figure*}

The results (Figure \ref{fig: Simulation_Effluent}) show that the Output MPC can operate the WWTP to provide the requested quality class, consistently, despite influent disturbances.
The effluent flow-rate $Q_{\text{eff}}$ and temperature $T^{\text{eff}}$ are shown to follow closely their influent counterpart: This is expected, as the water line has a constant volume and its reactions are modelled as isothermal.
The biochemical profile of the treated water (in terms of TSS$^{\text{eff}}$, BOD$_5^{\text{eff}}$, and TN$^{\text{eff}}$) shows that the controller can regulate its quality with satisfactory performance except for the sporadic rain and storm events.
For the solids and organic matter:
\begin{itemize}
    \item [--] 
    The effluent total suspended solids (TSS$^{\text{eff}}$) are kept slightly above the $10$ g/m$^3$ reference throughout this 1-year period.
    Besides the occasional storm, when TSS$^{\text{eff}}$ increases abruptly, we remark on the extreme events occurring at the $1^{\text{st}}$ month ($t \in [0,31)$ days) and $10^{\text{th}}$ month ($t \in [273, 304)$ days).
    During these events, TSS$^{\text{eff}}$ varies notably from the reference and stays uncorrected for a long period.
    However, the daily-average signal is always below the $30$ g/m$^3$ limit, and thus this KPI complies with the requirements for all quality classes (A, B, and C).

    \item [--]
    The effluent 5-days measured biochemical oxygen demand (BOD$_5^{\text{eff}}$) is kept around the $4$ g/m$^3$ reference throughout this 1-year period.
    As with TSS$^{\text{eff}}$, the performance is less satisfactory only during extreme events. 
    Because we set a constant BOD$_5^{\text{eff}}$ reference for all quality classes (see Table \ref{tab: OPO_Tuning}), the values of this KPI remain consistent throughout the year.
    Consequently, the daily-average signal is always below the $10$ g/m$^3$ limit, and thus complies with the requirements for all the quality classes (A, B, and C).
\end{itemize}

The effluent total nitrogen (TN$^{\text{eff}}$), being the nutrient we are interested to recover, is controlled to change periodically according to the targeted water quality class:
\begin{itemize}
    \item [--] 
    During odd months, the controller targets quality Class A by regulating TN$^{\text{eff}}$ around 7.5 g/m$^3$. 
    The performance is mostly consistent, with a noticeable but small decline in the $1^{\text{st}}$ month ($t \in [0,31)$): 
    In addition to the rain event, the task is more demanding at this period as the lower temperature conditions affect the nitrogen removal efficiency. 
    In contrast, the nitrogen removal efficiency is noticeably higher in the latter months, when temperatures increase. 
    The daily-average signal during odd months is always below the $15$ g/m$^3$ limit, and thus complies with the requirements for reuse water of quality Class A. 

    \item [--]
    During even months, the controller targets quality Class B by regulating TN$^{\text{eff}}$ around 22.5 g/m$^3$. 
    In this case, performance is less consistent but still satisfactory.
    The controller struggles to track the reference during the $8^{\text{th}}$ month ($t \in [212,243)$) and the $10^{\text{th}}$ month ($t \in [273,304)$). 
    Again, this is because higher temperature conditions amplify the nitrogen removal efficiency of this treatment process.
    Regardless, the daily-average signal during even months is within the $[15, 30)$ g/m$^3$ limits and thus complies with the requirements for reuse water of quality Class B. 

    \item [--]
    During the cold season (months 3--5), the controller targets quality Class C by regulating TN$^{\text{eff}}$ around 37.5 g/m$^3$. 
    The reference is tracked consistently, except for a few isolated events. 
    Specifically, the effluent concentrations of this nutrient decrease abruptly at days $t \approx 65$, $t \approx 95$, and $t \approx 145$. 
    These periods align with the occurrence of rain events, indicating that this operational regime is more sensitive to extreme influent disturbances. 
    Regardless, the daily-average signal during this period is mostly within the $[30, 45)$ g/m$^3$ limits and thus complies with the requirements for reuse water of quality Class C. 
\end{itemize}
In Figure \ref{fig: Simulation_Effluent}, we also visualize the composition of TN$^{\text{eff}}$ at each time-step based on the achieved quality class. 
The results show that the stricter the quality limits, the higher the focus on recovering nitrogen in the form of ammonium. 
This is not surprising, as the influent nitrogen (Figure \ref{fig: Influent_Summary}) is mostly comprised of organic and ammonium nitrogen: 
While the former must be removed to avoid violating the quality limits on organic matter (BOD$_5$), the latter can be recovered by relaxing the nitrogen removal efficiency of the process. 
The fact that inorganic nitrogen is preferred for crop fertigation, together with the known toxicity of nitrite, makes this chemical profile ideal for clients requesting water of quality Class C (agricultural reuse).
Interestingly, this behavior was not explicitly programmed into the controller, but resulted from its planning given the process model and the expected influent profile.

In summary, the results demonstrate the efficacy of this Output MPC strategy and highlight some facts on wastewater treatment and nitrogen recovery. 
Firstly, it reinforces the well-known fact that nitrogen removal via biological treatment is reduced and amplified, respectively, by low and high temperature conditions. 
The conjecture is that conventional WWTPs operated as WRRFs should be scheduled to recover nutrients during cold seasons, when producing nitrogen-rich reuse water is shown to be a less demanding task.
Similarly, the plant can be scheduled to focus on producing cleaner water during warm seasons.
Finally, the results show that conventional WWTPs can still remove solids and organic matter while simultaneously manipulating the chemical profile of its water effluents.

\begin{figure}[tb!] \centering
    \includegraphics[width=\columnwidth]{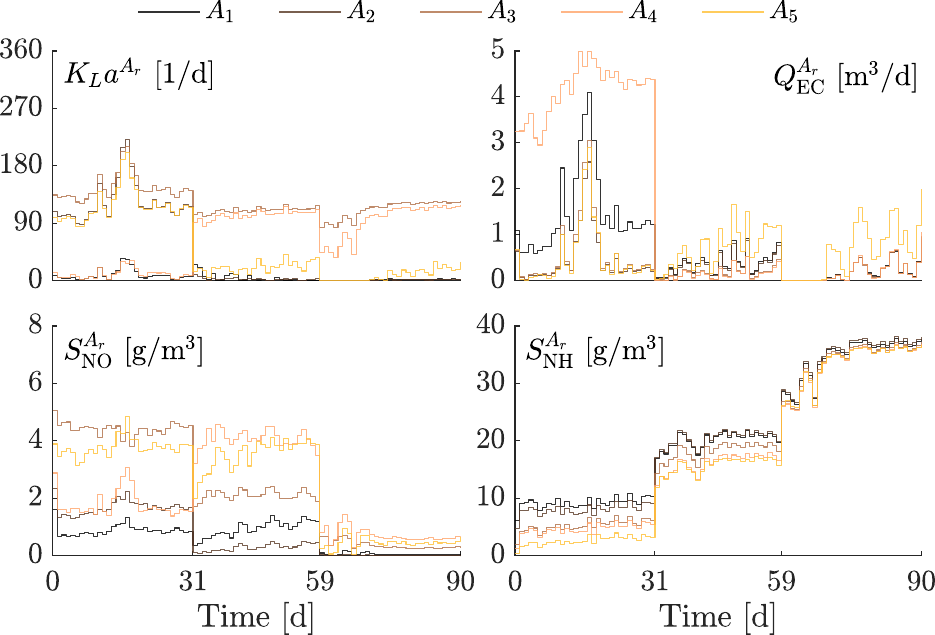}

    \caption{%
        Simulation, $t \in [0, 90]$: 
        Air ($K_La$) and external carbon ($Q_{\text{EC}}$) flow-rates, top panels, and nitrate-plus-nitrite ($\pVars{S}{NO}{}$) and ammonium ($\pVars{S}{NH}{}$) nitrogen, bottom panels, for each $A_r$ ($r = 1,\ldots,5$). 
        Only the daily-average values are displayed.
    }
    \label{fig: Simulation_Inputs}
\end{figure}

To understand how the nutrient recovery is achieved, we analyze a selection of the control actions deployed by the Output MPC (Figure \ref{fig: Simulation_Inputs}).
We focus on the first three months, each corresponding to a distinct quality target, and on the evolution of inorganic nitrogen within the water line.
We recall that the control actions determined by the controller are still constrained to enforce an operation with nonpositive energy costs.
The controller operates this section by manipulating the aeration ($K_La$) and addition of carbon ($Q_{\text{EC}}$) in all reactors $A_r$ ($r=1,\ldots,5$) to optimally adjust the nitrification-denitrification process:

\begin{itemize}
    \item [--]
    When targeting Class A (1$^{\text{st}}$ month), the controller increases and decreases, respectively, the aeration to reactors $A_2$ and $A_4$ with respect to the nominal values (Table \ref{tab: OPO_Tuning}). 
    Simultaneously, the controller requests extra carbon to all reactors, with $A_4$ being supplied the most. 
    Under this configuration, the water line implements a \textit{anoxic-aerated-anoxic-aerated} scheme with denitrification happening in reactors $\{ A_1, A_4 \}$ and nitrification happening in reactors $\{ A_2, A_3, A_5 \}$. 
    The state-responses show that nitrate-plus-nitrite nitrogen ($\pVars{S}{NO}{}$) increase while ammonium nitrogen ($\pVars{S}{NH}{}$) is greatly reduced in all reactors, leading to small levels of effluent total nitrogen TN$^{\text{eff}}$. 

    \item [--]
    When targeting Class B (2$^{\text{nd}}$ month), the controller decreases the aeration and slightly increases the carbon addition to all reactors $A_r$ ($r=1,\ldots,5$) with respect to the nominal values (Table \ref{tab: OPO_Tuning}). 
    This leads to the conventional \textit{anoxic-aerated} scheme for activated sludge plants, but with a more conservative aeration profile. 
    The state-responses show that nitrate-plus-nitrite nitrogen ($\pVars{S}{NO}{}$) increase while ammonium ($\pVars{S}{NH}{}$) is only moderately reduced in all reactors, leading to medium levels of effluent total nitrogen TN$^{\text{eff}}$. 

    \item [--]
    When targeting Class C (3$^{\text{rd}}$ month), the controller decreases the aeration to all reactors $A_r$ ($r=1,\ldots,5$) with respect to the nominal values (Table \ref{tab: OPO_Tuning}). 
    As a result, the reactor $A_5$ is operated under near-anoxic conditions. 
    Simultaneously, the controller requests a small supplement of carbon to all reactors, with emphasis on $A_5$. 
    This leads to a \textit{anoxic-aerated-anoxic} scheme with only $\{ A_3, A_4 \}$ being constantly aerated. 
    The state-responses show that nitrate-plus-nitrite nitrogen ($\pVars{S}{NO}{}$) is hardly produced while ammonium ($\pVars{S}{NH}{}$) is kept close to influent levels. 
    Thus, the plant produces high levels of effluent total nitrogen TN$^{\text{eff}}$.
\end{itemize}

Adjusting the anoxic/aerated volume of the water line is a well-established strategy for operating biological WWTPs. 
The Output MPC autonomously finds this strategy and utilizes it to switch between operational modes, while also exploiting the external carbon sources in anoxic zones to compensate for the organic matter consumed in the aerated zones.
The degrees-of-freedom enabled by the external carbon sources seemingly led the controller to discover an unconventional treatment scheme (\textit{anoxic-aerated-anoxic-aerated}) for a conventional task (targeting Class A).
Formally, the implication is that (according to the predictive model) this is an optimal configuration for efficient nitrogen removal under energy-autonomous restrictions.
Finally, we remark on how relaxing the quality limits leads the controller to discover ever-cheaper control strategies-- an expected benefit of transitioning WWTPs into WRRFs.

\subsection{Secondary objective: Nonpositive energy costs}

\begin{figure}[tb!] \centering
    \includegraphics[width=\columnwidth]{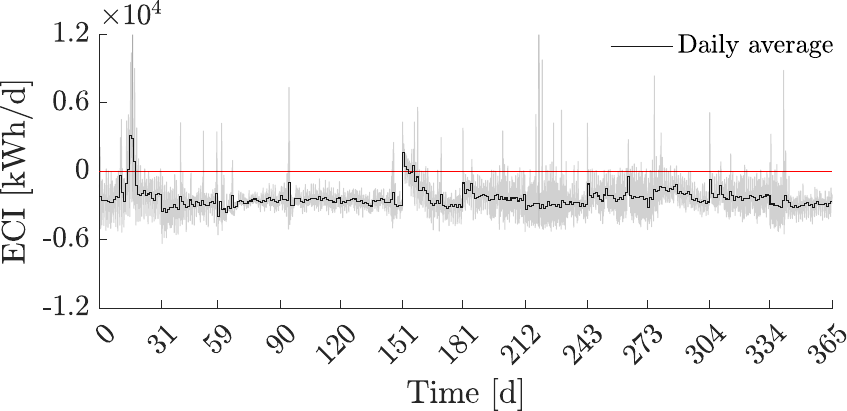}

    \caption{%
        Simulation, 1-year period: 
        Energy cost index (ECI) during operation. 
        Negative values imply that the process is recovering more heat and electrical energy than needed for driving the plant.
        }
    \label{fig: Simulation_Energy}
\end{figure}

The results (Figure \ref{fig: Simulation_Energy}) show that the operation implemented by the Output MPC is energy-autonomous: 
Except for two short periods (at days $t \approx 15$ and $t \approx 150$), the energy cost index (ECI) indicates that the plant can recover more heat and electrical energy daily than needed for its operation. 
The violations of the energy-autonomy constraints align with the occurrence of rain events, highlighting the effect of extreme influent disturbances. 
For this experiment, the treatment produced a total of $83$ GWh, with an average output of $2385$ kWh/d of surplus energy. 
We conjecture that the plant can eliminate the need for buying external electricity (and even benefit from selling its production) by incorporating energy-storage systems into its infrastructure. 
The achieved energy recovery is also shown to depend on the water quality being targeted:
\begin{itemize}
    \item [--]
    When targeting Class A, the output is $2022$ kWh/d, on average. 
    On a monthly-basis, this ranges from producing $1607$ kWh/d ($1^{\text{st}}$ month, $t \in [0,31)$ days) to $2178$ kWh/d ($7^{\text{th}}$ month, $t \in [181,212)$ days).

    \item [--]
    When targeting Class B, the output is $2534$ kWh/d, on average. 
    On a monthly-basis, this ranges from producing $1877$ kWh/d ($6^{\text{th}}$ month, $t \in [151,181)$ days) to $2933$ kWh/d ($2^{\text{nd}}$ month, $t \in [31,59)$ days).
    
    \item [--]
    When targeting Class C, the output is $2635$ kWh/d, on average. 
    This production is consistent throughout the cold season, with little variability.
\end{itemize}
The changes in the energy recovery performance concur with the conclusions from the nutrient recovery task: 
It is more advantageous to produce water of Class A during warm seasons, whereas producing water of Class B or C benefits from low-temperature conditions. 
In addition, the results show that set-point changes have a noticeable effect on the ECI. 
Specifically, switching the reference from Class B to Class A causes a short burst in the energy costs, whereas switching from Class A to Class B causes a sudden decrease. 
Such an effect is expected considering how quickly the controller changes the aeration regimes when switching the targeted water quality (see Figure \ref{fig: Simulation_Inputs}).

\begin{figure}[tb!] \centering
    \includegraphics[width=\columnwidth]{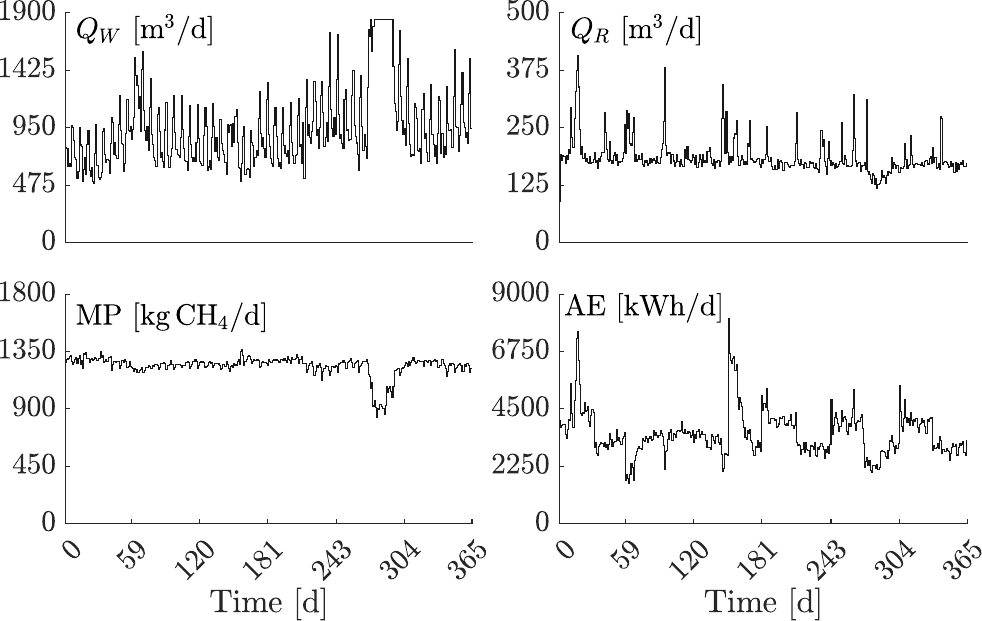}

    \caption{%
        Simulation, 1-year period: 
        Excess sludge ($Q_W$) and reject water ($Q_R$) recycle flow-rates, top panels, and the aeration energy (AE) costs and methane production (MP), bottom panels. 
        Only the daily-average values are displayed.
    }
    \label{fig: Simulation_Energy_Summary}
\end{figure}

To understand how the energy recovery is achieved, we again analyze a selection of the control actions deployed by the Output MPC (Figure \ref{fig: Simulation_Energy_Summary}). 
We focus on the excess sludge ($Q_W$) and reject water ($Q_R$) recycles--the control handles more directly related to the sludge line--, and on the methane production (MP) from the anaerobic digester. 
Since it comprises most of the costs, we also show the aeration energy (AE) demanded by the water line. 
\begin{itemize}
    \item [--]
    The results show that the methane production is consistent throughout most of the year, with a noticeable decrease only during the $11^{\text{th}}$ month.
    The controller adjusts this production by increasing the wastage flow-rate $Q_W$ when more organic matter need to be supplied for the anaerobic digester. 
    We remark on the $10^{\text{th}}$ month ($t \in [273, 304)$ days), when this control handle is saturated while trying to compensate for the decreased influx of organic matter in this period.
    
    \item [--]
    The aeration energy costs are lower the more relaxed the quality limits, as already discussed in Figure \ref{fig: Simulation_Inputs}. 
    The reject water storage tank, rich in organic matter, is used as a buffer: 
    When energy costs increase (e.g., during set-point changes), the recycle flow-rate $Q_R$ is used to send reject water into the primary clarifier, whose underflow is then sent back to the digester. 
\end{itemize}

In conclusion, the controller enforces an energy-autonomous operation not by boosting biogas production but instead by implementing a regular methane production alongside cheaper aeration schemes. 
Such a strategy reflects the fact that the controller was explicitly programmed not to minimize energy costs but only to ensure they are nonpositive. 
Under an alternative tuning, the controller would be able to prioritize energy production, likely at the expense of worsened nutrient recovery.
Since our configuration achieves both the primary (resource recovery) and secondary (energy recovery) goals, it contributes a solution to the operation of conventional WWTPs as energy-autonomous WRRFs.

\section{Concluding remarks} \label{sec: Conclusion} 

This work presented an automatic control solution for the operation of conventional WWTPs as energy-autonomous WRRFs.
We first formalize the task as the production of reclaimed water of different quality classes (for either environmental, industrial, or agricultural reuse), while having such operation result in net-zero energy costs. 
The effluent water quality is described in terms of its concentrations of total suspended solids (TSS$^{\text{eff}}$), biochemical oxygen demand (BOD$_5^{\text{eff}}$), and total nitrogen (TN$^{\text{eff}}$), and quality classes are defined by upper limits on these metrics.
Once the task was formalized, we proposed and configured an output-feedback model-predictive controller (Output MPC) comprised of three main computational units: 
$i$) The OPO, which determines an operating point that achieves the desired performance; 
$ii$) The MHE, which estimates the current state of the process based on measurement data; and 
$iii$) The MPC, which computes the control actions that steer the process state to the optimal operating point. 

Experimental results demonstrated that such a control strategy is capable to operate a full-scale conventional WWTP as an energy-autonomous WRRF. 
Specifically, we showed that the controller manipulates the actionable inputs (mainly, the influx of air and supplementary carbon to the water line) to adjust the effluent water quality according to the desired class. 
The control actions determined by the Output MPC enforce an energy-autonomous operation by maintaining a regular production of biogas while decreasing aeration energy costs. 
The results provide a proof-of-concept on the transition of conventional wastewater treatment infrastructure into WRRFs by adapting their operation using automation solutions. 

In this study, the well-established Benchmark Simulation Model no. 2 (BSM2) is taken as representative of a real-world large-scale treatment plant. 
This is a pragmatic, but not restrictive, choice. 
We remark that, under the availability of a predictive model for the process of interest, our control strategy is general enough to accommodate different resource recovery tasks (e.g., phosphorus recovery), influent conditions, and treatment technologies (e.g., membrane bioreactors)--A relevant direction for future work. 

\bibliographystyle{elsarticle-num}          

\begin{thebibliography}{10}
    \expandafter\ifx\csname url\endcsname\relax
    \def\url#1{\texttt{#1}}\fi
    \expandafter\ifx\csname urlprefix\endcsname\relax\def\urlprefix{URL }\fi
    \expandafter\ifx\csname href\endcsname\relax
    \def\href#1#2{#2} \def\path#1{#1}\fi

    \bibitem{Guest2009}
    J.~S. Guest, S.~J. Skerlos, J.~L. Barnard, M.~B. Beck, G.~T. Daigger, H.~Hilger, S.~J. Jackson, K.~Karvazy, L.~Kelly, L.~Macpherson, J.~R. Mihelcic, A.~Pramanik, L.~Raskin, M.~C.~M. Van~Loosdrecht, D.~Yeh, N.~G. Love, A {New} {Planning} and {Design} {Paradigm} to {Achieve} {Sustainable} {Resource} {Recovery} from {Wastewater}, Environ. Sci. Technol. 43~(16) (2009) 6126--6130.

    \bibitem{Verstraete2011}
    W.~Verstraete, , S.~E. Vlaeminck, {ZeroWasteWater}: short-cycling of wastewater resources for sustainable cities of the future, Int. J. of Sustain. Dev. \& World Ecol. 18~(3) (2011) 253--264.

    \bibitem{UNESCO2017}
    {United Nations World Water Assessment Programme}, Wastewater: {The} untapped resource, The {United} {Nations} {World} {Water} {Development} {Report} 2017, UNESCO, 2017.

    \bibitem{Ricart2019}
    S.~Ricart, A.~M. Rico, Assessing technical and social driving factors of water reuse in agriculture: A review on risks, regulation and the yuck factor, Agric. Water Manag. 217 (2019) 426--439.

    \bibitem{Kundu2022}
    S.~Kundu, B.~K. Pramanik, P.~Halder, S.~Patel, M.~Ramezani, M.~Khairul, M.~H. Marzbali, J.~Paz-Ferreiro, S.~Crosher, G.~Short, A.~Surapaneni, D.~Bergmann, K.~Shah, Source and central level recovery of nutrients from urine and wastewater: A state-of-art on nutrients mapping and potential technological solutions, J. of Environ. Chem. Eng. 10~(2) (2022) 107146.

    \bibitem{Sarpong2020}
    G.~Sarpong, V.~G. Gude, Near future energy self-sufficient wastewater treatment schemes, Int. J. of Environ. Res. 14 (2020) 479--488.

    \bibitem{Kehrein2020}
    P.~Kehrein, M.~van Loosdrecht, P.~Osseweijer, M.~Garfí, J.~Dewulf, J.~Posada, A critical review of resource recovery from municipal wastewater treatment plants - market supply potentials{,} technologies and bottlenecks, Environ. Sci.: Water Res. \& Technol. 6 (2020) 877--910.

    \bibitem{Peng2018}
    L.~Peng, H.~Dai, Y.~Wu, Y.~Peng, X.~Lu, A comprehensive review of phosphorus recovery from wastewater by crystallization processes, Chemosphere 197 (2018) 768--781.

    \bibitem{Xie2016}
    M.~Xie, H.~K. Shon, S.~R. Gray, M.~Elimelech, Membrane-based processes for wastewater nutrient recovery: {Technology}, challenges, and future direction, Water Res. 89 (2016) 210--221.

    \bibitem{Wu2019}
    B.~Wu, Membrane-based technology in greywater reclamation: {A} review, Sci. of the Total Env. 656 (2019) 184--200.

    \bibitem{ValverdePerez2015}
    B.~Valverde-Pérez, E.~Ramin, B.~F. Smets, B.~G. Plósz, {EBP2R} - an innovative enhanced biological nutrient recovery activated sludge system to produce growth medium for green microalgae cultivation, Water Res. 68 (2015) 821--830.

    \bibitem{Li2019}
    K.~Li, Q.~Liu, F.~Fang, R.~Luo, Q.~Lu, W.~Zhou, S.~Huo, P.~Cheng, J.~Liu, M.~Addy, P.~Chen, D.~Chen, R.~Ruan, Microalgae-based wastewater treatment for nutrients recovery: {A} review, Bioresour. Tech. 291 (2019) 121934.

    \bibitem{Wan2016}
    J.~Wan, J.~Gu, Q.~Zhao, Y.~Liu, {COD} capture: a feasible option towards energy self-sufficient domestic wastewater treatment, Sci. Rep. 6~(1) (2016) 25054.

    \bibitem{Solon2019}
    K.~Solon, M.~Jia, E.~I.~P. Volcke, Process schemes for future energy-positive water resource recovery facilities, Water Sci. and Tech. 79~(9) (2019) 1808--1820.

    \bibitem{Chen2009}
    R.~Chen, X.~C. Wang, Cost-benefit evaluation of a decentralized water system for wastewater reuse and environmental protection, Water Sci. and Tech. 59~(8) (2009) 1515--1522.

    \bibitem{Farago2021}
    M.~Faragò, A.~Damgaard, J.~A. Madsen, J.~K. Andersen, D.~Thornberg, M.~H. Andersen, M.~Rygaard, From wastewater treatment to water resource recovery: {Environmental} and economic impacts of full-scale implementation, Water Res. 204 (2021) 117554.

    \bibitem{Henze2008}
    M.~Henze, M.~C. van Loosdrecht, G.~A. Ekama, D.~Brdjanovic, Biological wastewater treatment, IWA publishing, 2008.

    \bibitem{Olsson2013}
    G.~Olsson, B.~Carlsson, J.~Comas, J.~Copp, K.~Gernaey, P.~Ingildsen, U.~Jeppsson, C.~Kim, L.~Rieger, I.~Rodr\`{i}guez-Roda, J.~Steyer, I.~Tak{\'{a}}cs, P.~Vanrolleghem, A.~Vargas, Z.~Yuan, L.~{\AA}mand, Instrumentation, control and automation in wastewater - from {L}ondon 1973 to {N}arbonne 2013, Water Sci. \& Technol. 69~(7) (2014) 1372--1385.

    \bibitem{Amand2013}
    L.~{\AA}mand, G.~Olsson, B.~Carlsson, Aeration control - {A} review, Water Sci. and Technol. 67~(1) (2013) 2374--2398.

    \bibitem{Rawlings2020}
    J.~B. Rawlings, D.~Q. Mayne, M.~M. Diehl, Model Predictive Control: Theory, Computation and Design, 2nd Edition, Nob Hill Publ., LLC., 2020.

    \bibitem{Shen2008}
    W.~Shen, X.~Chen, J.~Corriou, Application of model predictive control to the {BSM1} benchmark of wastewater treatment process, Comput. and Chem. Eng. 22~(12) (2008) 2849--2856.

    \bibitem{Han2014}
    H.-G. Han, H.-H. Qian, J.-F. Qiao, Nonlinear multiobjective model-predictive control scheme for wastewater treatment process, J. of Process Control 24~(3) (2014) 47--59.

    \bibitem{Vega2014}
    P.~Vega, S.~Revollar, M.~Francisco, J.~M. Martín, Integration of set point optimization techniques into nonlinear {MPC} for improving the operation of {WWTPs}, Comput. and Chem. Eng. 68 (2014) 78--95.

    \bibitem{Sadeghassadi2018}
    M.~Sadeghassadi, C.~J.~B. Macnab, B.~Gopaluni, D.~Westwick, Application of neural networks for optimal-setpoint design and {MPC} control in biological wastewater treatment, Comput. and Chem. Eng. 115 (2018) 150--160.

    \bibitem{Holenda2008}
    B.~Holenda, E.~Dokomosa, A.~R\`{e}dey, J.~Fazakas, Dissolved oxygen control of the activated sludge wastewater treatment process using model predictive control, Comput. \& Chem. Eng. 33~(6) (2008) 1270--1278.

    \bibitem{Mulas2015}
    M.~Mulas, S.~Tronci, F.~Corona, H.~Haimi, P.~Lindell, M.~Heinonen, R.~Vahala, R.~Baratti, Predictive control of an activated sludge process: An application to the {Viikinmäki} wastewater treatment plant, J. of Process Control 35 (2015) 89--100.

    \bibitem{Santin2016}
    I.~Santín, C.~Pedret, R.~Vilanova, M.~Meneses, Advanced decision control system for effluent violations removal in wastewater treatment plants, Control Eng. Pract. 49 (2016) 60--75.

    \bibitem{Ekman2008}
    M.~Ekman, Bilinear black-box identification and {MPC} of the activated sludge process, J. of Process Control 18~(7--8) (2008) 643--653.

    \bibitem{Foscoliano2016}
    C.~Foscoliano, S.~{Del Vigo}, M.~Mulas, S.~Tronci, Predictive control of an activated sludge process for long term operation, Chem. Eng. J. 304 (2016) 1031--1044.

    \bibitem{Zeng2015}
    J.~Zeng, J.~Liu, Economic model predictive control of wastewater treatment processes, Ind. and Eng. Chem. Res. 54~(21) (2015) 5710--5721.

    \bibitem{Zhang2019b}
    A.~Zhang, X.~Yin, S.~Liu, J.~Zeng, J.~Liu, Distributed economic model predictive control of wastewater treatment plants, Chem. Eng. Res. and Des. 141 (2019) 144--155.

    \bibitem{Han2024}
    M.~Han, J.~Yao, A.~W.-K. Law, X.~Yin, Efficient economic model predictive control of water treatment process with learning-based {Koopman} operator, Control Eng. Pract. 149 (2024) 105975.

    \bibitem{Francisco2015}
    M.~Francisco, S.~Skogestad, P.~Vega, Model predictive control for the self-optimized operation in wastewater treatment plants: {A}nalysis of dynamic issues, Comput. and Chem. Eng. 82 (2015) 259--272.

    \bibitem{Busch2013}
    J.~Busch, D.~Elixmann, P.~K\"uhl, C.~Gerkens, J.~P. Schl\"oder, H.~G. Bock, W.~Marquardt, State estimation for large-scale wastewater treatment plants, Water Res. 47~(13) (2013) 4774--4787.

    \bibitem{Yin2018b}
    X.~Yin, B.~Decardi-Nelson, J.~Liu, Subsystem decomposition and distributed moving horizon estimation of wastewater treatment plants, Chem. Eng. Res. and Des. 134 (2018) 405 -- 419.

    \bibitem{Yin2019}
    X.~Yin, J.~Liu, Subsystem decomposition of process networks for simultaneous distributed state estimation and control, AIChE J. 65~(3) (2019) 904--914.

    \bibitem{NMC_BSM1_Control}
    O.~B.~L. {Neto}, M.~{Mulas}, F.~{Corona}, A model-based framework for controlling activated sludge plants, Chem. Eng. J. 488 (2024) 150750.

    \bibitem{Seborg2016}
    D.~E. Seborg, T.~F. Edgar, D.~A. Mellichamp, F.~J. Doyle~III, Process Dynamics and Control, 4th Edition, John Wiley \& Sons, 2016.

    \bibitem{EPA2012}
    {U.S. EPA}, Guidelines for water reuse, 2012.

    \bibitem{EUreuse2017}
    B.~M.~G. L.~Alcalde-Sanz, Minimum quality requirements for water reuse in agricultural irrigation and aquifer recharge - towards a water reuse regulatory instrument at {EU} level, Tech. rep., {EUR} 28962 {EN}, Publ. Office of the {EU} (2017).

    \bibitem{EU2020}
    {Council of the European Union}, {Regulation ({EU}) 2020/741 of the European Parliament and of the Council}, \newline\url{https://eur-lex.europa.eu/legal-content/EN/TXT/PDF/?uri=CELEX:32020R0741&from=EN} (2020).

    \bibitem{Shoushtarian2020}
    F.~Shoushtarian, M.~Negahban-Azar, Worldwide regulations and guidelines for agricultural water reuse: a critical review, Water 12~(4) (2020) 971.

    \bibitem{Albornoz2016}
    F.~Albornoz, Crop responses to nitrogen overfertilization: {A} review, Scientia Horticulturae 205 (2016) 79--83.

    \bibitem{Gernaey2014}
    K.~Gernaey, U.~Jeppsson, P.~Vanrolleghem, J.~Copp, Benchmarking of Control Strategies for Wastewater Treatment Plants, Scientific and Technical Report Series No. 23, IWA Publ., 2014.

    \bibitem{Henze2000}
    M.~Henze, W.~Gujer, T.~Mino, M.~C.~M. van Loosdrecht, Activated Sludge Models {ASM1}, {ASM2}, {ASM2d} and {ASM3}, Scientific and Technical Report Series No. 9, IWA Publ., 2000.

    \bibitem{Takacs1991}
    I.~{Takács}, G.~{Patry}, D.~{Nolasco}, A dynamic model of the clarification-thickening process, Water Res. 25~(10) (1991) 1263--1271.

    \bibitem{Batstone2002}
    D.~Batstone, J.~Keller, I.~Angelidaki, S.~Kalyuzhnyi, S.~Pavlostathis, A.~Rozzi, W.~Sanders, H.~Siegrist, V.~Vavilin, {The {IWA} Anaerobic Digestion Model No 1 ({ADM1})}, Water Sci. and Technol. 45~(10) (2002) 65--73.

    \bibitem{Otterpohl1995}
    R.~Otterpohl, Technical university of aachen, Ph.D. thesis, Dynamische Simulation zur Unterstützung der Planung und des Betriebes von kommunalen Kläranlagen, Germany (1995).

    \bibitem{Vanrolleghem2003}
    P.~Vanrolleghem, D.~Lee, On-line monitoring equipment for wastewater treatment processes: state of the art, Water Sci. and Technol. 47~(2) (2003) 1--34.

    \bibitem{Betts2010}
    J.~T. Betts, Practical methods for optimal control and estimation using nonlinear programming, Advances in Design and Control, SIAM, 2010.

\end{thebibliography}

\end{document}


\sloppy \allowdisplaybreaks

\title{{\sc (Supplementary Material)}\\{\sc Predictive control of wastewater treatment plants as energy-autonomous water resource recovery facilities}}

\author[1]{Otac\'ilio B. L. Neto\corref{cor1}}
\author[2]{Michela Mulas}
\author[1]{Iiro Harjunkoski}
\author[1]{Francesco Corona}

\address[1]{School of Chemical Engineering, Aalto University, Finland}
\address[2]{Department of Teleinformatics Engineering, Federal University of Cear\'a, Brazil}

\cortext[cor1]{Corresponding author}
\begin{abstract}%
    This document provides supplementary material for the article ``\textit{Predictive control of wastewater treatment plants as energy-autonomous water resource recovery facilities}''. 
    In Section S1, provide the full details on the benchmark simulation model used in this study. 
    Section S2 provides further technical and implementation details for the output-feedback model predictive controller (Output MPC) proposed in the main text.
    Finally, Section S3 reports the complete data obtained from the simulation experiments that were performed.
\end{abstract}

\maketitle
\setlength{\parskip}{0.25cm}

\section{The Benchmark Model No. 2: State-space, parameters and equilibrium point} \label{sec: BSM2_SS}

The section provides the main component of the BSM2: Its state-space model.
Firstly, we define the set of variables that describe the state of the treatment plant at any given time.
Then, we present the first-order differential equations describing how these state-variables change in time in response to inputs (controls and disturbances) given their current values.
Simultaneously, we also present a set of algebraic equations relating these quantities to the set of measurement (or output) variables.
Finally, we conclude the section presenting the benchmark values of any constant model parameters and the default equilibrium point used in our study.

A list of the symbols and abbreviations used in this section is provided in Table \ref{tab: WRRF_Notation}. 
Symbols specific to each process unit (i.e., constant parameters and auxiliary variables) are defined in the appropriate sections.

\begin{table}[htb!] \centering
    \caption{Benchmark Simulation Model no. 2: List of symbols and abbreviations.}
    \label{tab: WRRF_Notation}
    \smallskip
    {\footnotesize\begin{tabular}{@{}p{2.8em}p{22em} p{2.8em}p{22em}@{}}
        \toprule
        COD	                        & Chemical oxygen demand	                  [g\,COD/m$^{3}$]          & $\pVars{S}{IN}{}    $         & Inorganic nitrogen                          [kg\,COD/m$^{3}$]         \\
        TSS		                    & Total suspended solids			          [g\,SS/m$^{3}$]           & $\pVars{X}{C}{}     $         & Composites    					          [kg\,COD/m$^{3}$]         \\
        BOD$_5$	                    & Biochemical oxygen demand	                  [g\,O$_2$/m$^{3}$]        & $\pVars{X}{CH}{}   $          & Carbohydrates                               [kg\,COD/m$^{3}$]         \\
        TN		                    & Total nitrogen					          [g\,COD/m$^{3}$]          & $\pVars{X}{PR}{}   $          & Proteins                                    [kg\,COD/m$^{3}$]         \\
        ECI		                    & Energy cost index						      [kWh/d]                   & $\pVars{X}{LI}{}   $          & Lipids                                      [kg\,COD/m$^{3}$]         \\
        $Q$		                    & Flow-rate			                          [m$^{3}$/d  ]             & $\pVars{X}{SU}{}   $          & Sugar degraders                             [kg\,COD/m$^{3}$]         \\ 
        $\pVars{T}{}{}	$           & Temperature						          [$^{\circ}$C]             & $\pVars{X}{AA}{}   $          & Amino acid degraders                        [kg\,COD/m$^{3}$]         \\ 
        $\pVars{V}{}{}$             & Volume                                      [m$^3$]                   & $\pVars{X}{FA}{}   $          & Long chain fatty acids degraders            [kg\,COD/m$^{3}$]         \\ 
        $\pVars{S}{I}{}	$           & Soluble inert organic matter				  [g\,COD/m$^{3}$]          & $\pVars{X}{C$_4$}{}$          & Valerate and butyrate degraders             [kg\,COD/m$^{3}$]         \\ 
        $\pVars{S}{S}{}	$           & Readily biodegradable substrate			  [g\,COD/m$^{3}$]          & $\pVars{X}{PRO}{}  $          & Propionate degraders                        [kg\,COD/m$^{3}$]         \\ 
        $\pVars{X}{I}{}	$           & Particulate inert organic matter			  [g\,COD/m$^{3}$]          & $\pVars{X}{AC}{}   $          & Acetate degraders                           [kg\,COD/m$^{3}$]         \\ 
        $\pVars{X}{S}{}	$           & Slowly biodegradable substrate			  [g\,COD/m$^{3}$]          & $\pVars{X}{H$_2$}{}$          & Hydrogen degraders                          [kg\,COD/m$^{3}$]         \\ 
        $\pVars{X}{BH}{} $          & Active heterotrophic biomass				  [g\,COD/m$^{3}$]          & $\pVars{S}{CAT$^+$}{}  $      & Cations                                     [kmol/m$^{3}$]            \\ 
        $\pVars{X}{BA}{} $          & Active autotrophic biomass				  [g\,COD/m$^{3}$]          & $\pVars{S}{AN$^-$}{}   $      & Anions                                      [kmol/m$^{3}$]            \\ 
        $\pVars{X}{P}{}  $          & Organic matter from biomass decay	          [g\,COD/m$^{3}$]          & $\pVars{S}{VA$^-$}{}   $      & Valerate                                    [kg\,COD/m$^{3}$]         \\ 
        $\pVars{S}{O}{}  $          & Dissolved oxygen							  [g\,O$_2$/m$^{3}$]        & $\pVars{S}{BU$^-$}{}   $      & Butyrate                                    [kg\,COD/m$^{3}$]         \\ 
        $\pVars{S}{NO}{} $          & Nitrate and nitrite nitrogen				  [g\,N/m$^{3}$]            & $\pVars{S}{PRO$^-$}{}  $      & Propionate                                  [kg\,COD/m$^{3}$]         \\ 
        $\pVars{S}{NH}{} $          & Ammonium plus ammonia nitrogen			  [g\,N/m$^{3}$]            & $\pVars{S}{AC$^-$}{}   $      & Acetate                                     [kg\,COD/m$^{3}$]         \\ 
        $\pVars{S}{ND}{} $          & Soluble biodegradable organic nitrogen	  [g\,N/m$^{3}$]	        & $\pVars{S}{HCO$_3^-$}{}$      & Bicarbonate                                 [kmol/m$^{3}$]            \\ 
        $\pVars{X}{ND}{} $          & Particulate biodegradable organic nitrogen  [g\,N/m$^{3}$]	        & $\pVars{S}{NH$_3$}{}   $      & Ammonia nitrogen                            [kmol/m$^{3}$]            \\ 
        $\pVars{S}{ALK}{}$          & Alkalinity								  [mol\,HCO$_3^-$/m$^{3}$]  & $\pVars{G}{H$_2$}{}    $      & Hydrogen gas                                [kg\,COD/m$^{3}$]         \\ 
        $\pVars{S}{SU}{}    $       & Monosaccharides                             [kg\,COD/m$^{3}$]         & $\pVars{G}{CH$_4$}{}   $      & Methane gas                                 [kg\,COD/m$^{3}$]         \\ 
        $\pVars{S}{AA}{}    $       & Amino-acids                                 [kg\,COD/m$^{3}$]         & $\pVars{G}{CO$_2$}{}   $      & Carbon dioxide gas                          [kg\,COD/m$^{3}$]         \\ 
        $\pVars{S}{FA}{}    $       & Long chain fatty acids                      [kg\,COD/m$^{3}$]         & $Q_{A}$		                & Internal recirculation flow-rate			  [m$^{3}$/d]               \\ 
        $\pVars{S}{VA}{}    $       & Total valerate                              [kg\,COD/m$^{3}$]         & $Q_{S}$		                & External recirculation flow-rate			  [m$^{3}$/d]               \\ 
        $\pVars{S}{BU}{}    $       & Total butyrate                              [kg\,COD/m$^{3}$]         & $Q_{W}$		                & Excess sludge wastage flow-rate 			  [m$^{3}$/d]               \\ 
        $\pVars{S}{PRO}{}   $       & Total propionate                            [kg\,COD/m$^{3}$]         & $Q_{G}$		                & (Normalized) biogas flow-rate 			  [m$^{3}$/d]               \\ 
        $\pVars{S}{AC}{}    $       & Total acetate                               [kg\,COD/m$^{3}$]         & $Q_{R}$		                & Reject water recycle flow-rate 			  [m$^{3}$/d]               \\ 
        $\pVars{S}{H$_2$}{} $       & Soluble hydrogen                            [kg\,COD/m$^{3}$]         & $Q_{\text{EC}}$	            & External carbon source flow-rate			  [m$^{3}$/d]               \\ 
        $\pVars{S}{CH$_4$}{}$       & Soluble methane                             [kg\,COD/m$^{3}$]         & $K_La$		                & Oxygen transfer coefficient				  [1/d]                     \\ 
        $\pVars{S}{IC}{}    $       & Inorganic carbon                            [kg\,COD/m$^{3}$]         &                               &                                                                       \\ 
        \bottomrule
    \end{tabular}}
\end{table}

\subsection{Model dynamics, measurements, and control}\label{subsubsec: BSM1-model}

The BSM2 takes the general form of a continuous-time state-space representation,
\begin{subequations}
\begin{align}
    \textstyle\frac{d}{dt}x(t) & = f(x(t), u(t), w(t) \mid \theta_x);  \label{eq: SS_Nonlinear_x}\\
          y(t) & = g(x(t), u(t) \mid \theta_y);        \label{eq: SS_Nonlinear_y}\\
          z(t) & = h(x(t), u(t) \mid \theta_z).        \label{eq: SS_Nonlinear_z}
\end{align} \label{eq: SS_Nonlinear}%
\end{subequations} 
The state equation \eqref{eq: SS_Nonlinear_x} models the evolution of the $N_x$ state variables, given their value $x(t) \in \mathbb{R}^{N_x}$, the value $u(t) \in \mathbb{R}^{N_u}$ of $N_u$ actionable input or control variables, and the value $w(t) \in \mathbb{R}^{N_w}$ of $N_w$ disturbances, at time $t \in \mathbb{R}_{\geq 0}$. 
The output equation \eqref{eq: SS_Nonlinear_y} models how the $N_y$ output variables $y(t) \in \mathbb{R}^{N_y}$ are emitted from the state $x(t)$ and actions $u(t)$. 
The performance equation \eqref{eq: SS_Nonlinear_z} models how the $N_z$ key performance indicators $z(t) \in \mathbb{R}^{N_z}$ are computed from the state $x(t)$ and actions $u(t)$. 
The vectors $\theta_x$, $\theta_y$, and $\theta_z$, collect the constant parameters in the functions $f(\cdot)$, $g(\cdot)$, and $h(\cdot)$, respectively.

\begin{figure*}[htb!] \centering
	\includegraphics[width=\textwidth]{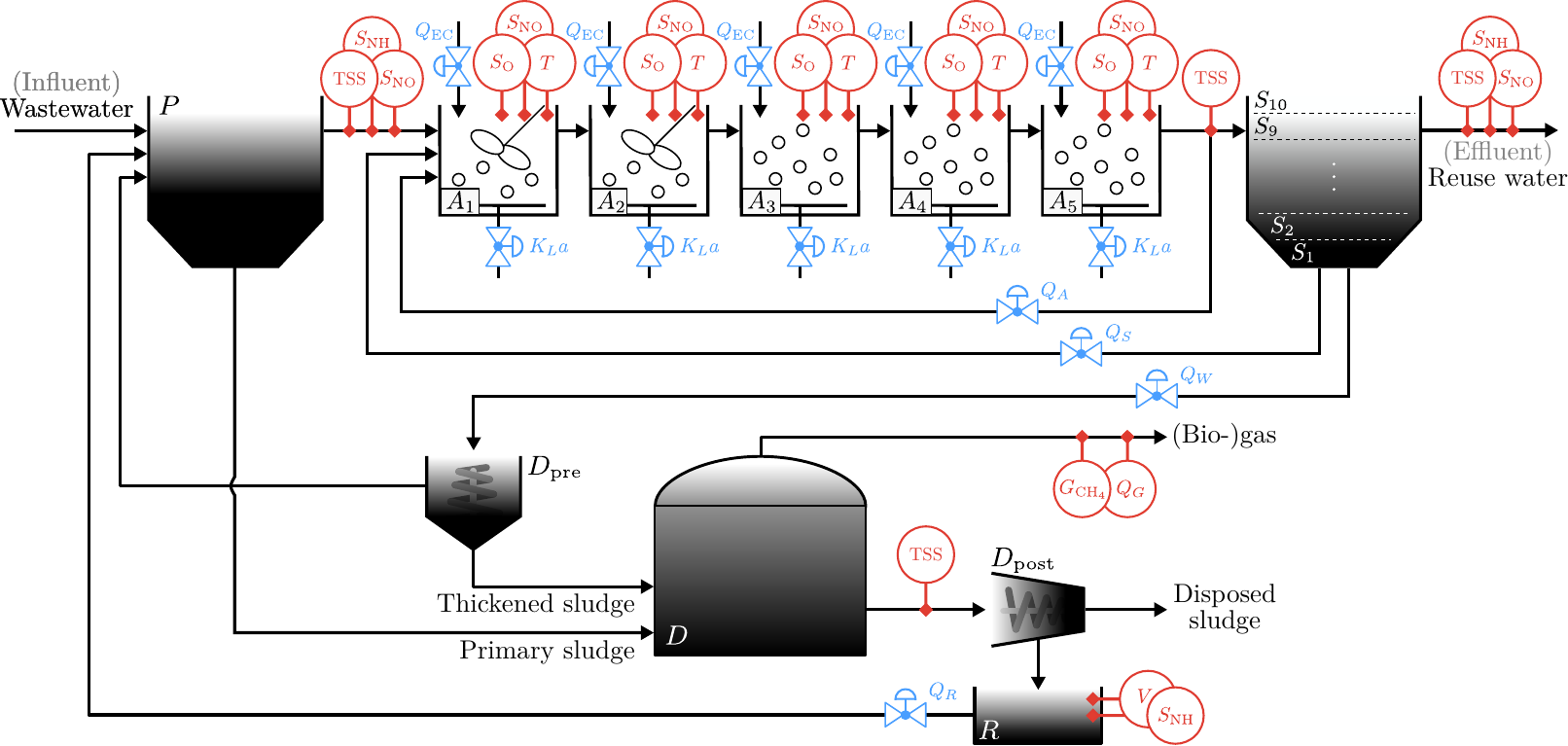}

    \caption{Water resource recovery facility: Process layout.} 
	\label{fig: BSM2}
\end{figure*}

Due to its modular layout, the plant's state is described in time by the signal $x : \mathbb{R}_{\geq 0} \to \mathbb{R}^{N_x}$,
\begin{equation}
    x = (x^P,~
         x^{A_1}, \ldots, x^{A_5},~
         x^{S_1}, \ldots, x^{S_{10}},~
         x^D,~
         x^R),
\end{equation}
defined by concatenating 
\begin{itemize}
    \item [--] The state of the primary clarifier ($P$), described by the $N_x^P = 15$ quantities
    \begin{equation*}
        x^P = (
            \pVars{Q}{   }{P},~
            \pVars{S}{I  }{P},~
            \pVars{S}{S  }{P},~
            \pVars{X}{I  }{P},~
            \pVars{X}{S  }{P},~
            \pVars{X}{BH }{P},~
            \pVars{X}{BA }{P},~
            \pVars{X}{P  }{P},~
            \pVars{S}{O  }{P},~
            \pVars{S}{NO }{P},~
            \pVars{S}{NH }{P},~
            \pVars{S}{ND }{P},~
            \pVars{X}{ND }{P},~
            \pVars{S}{ALK}{P},~
            \pVars{T}{   }{P}
        );
    \end{equation*}
    \item [--] The state of each $r$-th bioreactor ($A_r$, $r=1,\ldots,5$), described by the $N_x^{A_r} = 14$ quantities
    \begin{equation*}
        x^{A_r} = (
            \pVars{S}{I  }{A_r},~
            \pVars{S}{S  }{A_r},~
            \pVars{X}{I  }{A_r},~
            \pVars{X}{S  }{A_r},~
            \pVars{X}{BH }{A_r},~
            \pVars{X}{BA }{A_r},~
            \pVars{X}{P  }{A_r},~
            \pVars{S}{O  }{A_r},~
            \pVars{S}{NO }{A_r},~
            \pVars{S}{NH }{A_r},~
            \pVars{S}{ND }{A_r},~
            \pVars{X}{ND }{A_r},~
            \pVars{S}{ALK}{A_r},~
            \pVars{T}{   }{A_r}
        );
    \end{equation*}
    \item [--] The state of each $l$-th layer of the settler ($S_l$, $l=1,\ldots,10$), described by the $N_x^{S_l} = 9$ quantities
    \begin{equation*}
        x^{S_l} = (
            \pVars{\text{TSS}}{}{S_l},~
            \pVars{S}{I  }{S_l},~
            \pVars{S}{S  }{S_l},~
            \pVars{S}{O  }{S_l},~
            \pVars{S}{NO }{S_l},~
            \pVars{S}{NH }{S_l},~
            \pVars{S}{ND }{S_l},~
            \pVars{S}{ALK}{S_l},~
            \pVars{T}{   }{S_l}
        ); \hspace*{17em}
    \end{equation*}
    \item [--] The state of the anaerobic digester ($D$), described by the $N_x^D = 35$ quantities
    \begin{multline*}
        x^D = (
            \pVars{S}{SU}{D},~
            \pVars{S}{AA}{D},~
            \pVars{S}{FA}{D},~
            \pVars{S}{VA}{D},~
            \pVars{S}{BU}{D},~
            \pVars{S}{PRO}{D},~
            \pVars{S}{AC}{D},~
            \pVars{S}{H$_2$}{D},~
            \pVars{S}{CH$_4$}{D},~
            \pVars{S}{IC}{D},~
            \pVars{S}{IN}{D},~
            \pVars{S}{I}{D},~
            \pVars{X}{C}{D},\\
            \pVars{X}{CH}{D},~
            \pVars{X}{PR}{D},~
            \pVars{X}{LI}{D},~
            \pVars{X}{SU}{D},~
            \pVars{X}{AA}{D},~
            \pVars{X}{FA}{D},~
            \pVars{X}{C$_4$}{D},~
            \pVars{X}{PRO}{D},~
            \pVars{X}{AC}{D},~
            \pVars{X}{H$_2$}{D},~
            \pVars{X}{I}{D},~
            \pVars{S}{CAT$^+$}{D},\\
            \pVars{S}{AN$^-$}{D},~
            \pVars{S}{VA$^-$}{D},~
            \pVars{S}{BU$^-$}{D},~
            \pVars{S}{PRO$^-$}{D},~
            \pVars{S}{AC$^-$}{D},~
            \pVars{S}{HCO$_3^-$}{D},~
            \pVars{S}{NH$_3$}{D},~
            \pVars{G}{H$_2$}{D},~
            \pVars{G}{CH$_4$}{D},~
            \pVars{G}{CO$_2$}{D}
        );
    \end{multline*}
    \item [--] The state of the reject water storage tank ($R$), described by the $N_x^{R} = 15$ quantities
    \begin{equation*}
        x^R = (
            \pVars{V}{   }{R},~
            \pVars{S}{I  }{R},~
            \pVars{S}{S  }{R},~
            \pVars{X}{I  }{R},~
            \pVars{X}{S  }{R},~
            \pVars{X}{BH }{R},~
            \pVars{X}{BA }{R},~
            \pVars{X}{P  }{R},~
            \pVars{S}{O  }{R},~
            \pVars{S}{NO }{R},~
            \pVars{S}{NH }{R},~
            \pVars{S}{ND }{R},~
            \pVars{X}{ND }{R},~
            \pVars{S}{ALK}{R},~
            \pVars{T}{   }{R}
        ).
    \end{equation*}
\end{itemize}
We again refer to Table \ref{tab: WRRF_Notation} for the description of each symbol.
Note that the symbols $S_{(\cdot)}$, $X_{(\cdot)}$, and $G_{(\cdot)}$, denote the concentration of soluble, particulate, and gaseous matter, respectively.
Since modelled as ideal solid-separation processes, the thickener ($D_{\text{pre}}$) and dewatering ($D_{\text{pos}}$) units are not endowed with state variables: They behave as static transformations.
From the actuation layout (Figure \ref{fig: BSM2}), the control signal $u : \mathbb{R}_{\geq 0} \to \mathbb{R}^{N_u}$ is described by the $N_u = 14$ actionable quantities
\begin{equation}
    u = (
        Q_A,~
        Q_S,~
        Q_W,~
        Q_R,~
        \pVars{K_La}{}{A_1}, \ldots, \pVars{K_La}{}{A_5},~
        \pVars{Q}{EC}{A_1}, \ldots, \pVars{Q}{EC}{A_5}
    ).
\end{equation}
The disturbance signal $w : \mathbb{R}_{\geq 0} \to \mathbb{R}^{N_w}$ represents the influent wastewater to the primary clarifier ($P$) as described by the $N_w = 15$ non-actionable quantities
\begin{equation}
    w = (
        \pVars{Q}{\text{in}}{},~
        \pVars{S}{I  }{\text{in}},~
        \pVars{S}{S  }{\text{in}},~
        \pVars{X}{I  }{\text{in}},~
        \pVars{X}{S  }{\text{in}},~
        \pVars{X}{BH }{\text{in}},~
        \pVars{X}{BA }{\text{in}},~
        \pVars{X}{P  }{\text{in}},~
        \pVars{S}{O  }{\text{in}},~
        \pVars{S}{NO }{\text{in}},~
        \pVars{S}{NH }{\text{in}},~
        \pVars{S}{ND }{\text{in}},~
        \pVars{X}{ND }{\text{in}},~
        \pVars{S}{ALK}{\text{in}},~
        \pVars{T}{   }{\text{in}}
    ).
\end{equation}
From the instrumentation layout (Figure \ref{fig: BSM2}), the output (or measurement) signal $y : \mathbb{R}_{\geq 0} \to \mathbb{R}^{N_u}$ is described by the $N_y = 27$ measurable quantities
\begin{multline}
    y = (
        \pVars{\text{TSS}}{}{P_{\text{eff}}},~      
        \pVars{S}{NH}{P_{\text{eff}}},~  
        \pVars{S}{NO}{P_{\text{eff}}},~  
        \pVars{S}{NO}{A_1}, \ldots, \pVars{S}{NO}{A_5},~  
        \pVars{S}{NH}{A_1}, \ldots, \pVars{S}{NH}{A_5},~  
        \pVars{T}{}{A_1}, \ldots, \pVars{T}{}{A_5},~  
        \pVars{\text{TSS}}{}{A_5},~ \\
        \pVars{\text{TSS}}{}{S_{10}}, ~  
        \pVars{S}{NH}{S_{10}}, ~  
        \pVars{S}{NO}{S_{10}}, ~  
        \pVars{G}{CH$_4$}{D}, ~
        \pVars{Q}{$G$}{}, ~
        \pVars{\text{TSS}}{}{D},  ~  
        \pVars{V}{}{R},~  
        \pVars{S}{NH}{R}, 
    ),
\end{multline}
Finally, the key performance indicator signal $z : \mathbb{R}_{\geq 0} \to \mathbb{R}^{N_z}$, characterizing the tasks formulated in the main text, are described by the $N_z = 4$ quantities
\begin{equation}
    z = (\text{TSS}^{\text{eff}},~ \text{BOD}_5^{\text{eff}},~ \text{TN}^{\text{eff}},~ \text{ECI}). 
\end{equation}

The BSM2 thus defines a state-space model with $N_x = 225$ state-variables, $N_u = 14$ controllable inputs, $N_w = 15$ disturbances, $N_y = 27$ measurable quantities, and $N_z = 5$ KPIs. 
In the following, we provide the state ($f(\cdot)$) and output ($g(\cdot)$) equations associated with each individual process unit (that is, the units $P$, $A_{1 \leadsto 5}$, $S_{1 \leadsto 10}$, $D_{\text{pre}}$, $D_{\text{post}}$, $D$, and $R$). 
The values assumed for the constant parameters ($\theta_x, \theta_y$) are also reported. 
We then present the explicit expressions for the performance ($h(\cdot)$) equations, and, finally, report the nominal operating conditions used for the simulation experiments.

\subsection{Primary clarifier ($P$)} \label{subsec: Primary_Clarifier}

The primary clarifier has dynamics according to the model by Otterpohl \cite{Otterpohl1995}.
Hereafter, we use the symbol $Z$ to refer to any soluble $S_{(\cdot)}$, particulate $X_{(\cdot)}$, and gas $G_{(\cdot)}$ compound, and also to temperature $T$. 
The dynamics of the ``\textit{internal}'' flow-rate $\clX{Q^{P}}$ and each component $\clX{Z^{P}}$ in the primary clarifier are
\begin{align} \label{eq: Clarifier_Dynamics}
    \textstyle\frac{d}{dt}{Q}^{P} &= \cfrac{1}{\clP{t_m}} \left( \clW{\pVars{Q}{in}{P}} - \clX{\pVars{Q}{}{P}} \right);\\
    \textstyle\frac{d}{dt}{Z}^{P} &= \cfrac{\clW{\pVars{Q}{in}{P}}}{\clP{V^P}} \left( \clW{\pVars{Z}{}{P_{\text{in}}}} - \clX{\pVars{Z}{}{P}} \right),
\end{align}
where $\big( \clW{\pVars{Q}{in}{P}}, \clW{\pVars{Z}{}{P_{\text{in}}}} \big)$ is the influent flow-rate and concentration to the clarifier with 
\begin{equation*}
    \clW{\pVars{Q}{in}{P}} = \clW{\pVars{Q}{in}{}} + \clU{\pVars{Q}{eff}{D_{\text{pre}}}} + \clU{Q_R}; 
    \qquad
    \clW{\pVars{Z}{}{P_{\text{in}}}} = \frac{1}{\clW{\pVars{Q}{in}{P}}} \left( \clW{Q_{\text{in}}} \clW{Z^{\text{in}}} + \clU{\pVars{Q}{eff}{D_{\text{pre}}}} \clX{\pVars{Z}{eff}{D_{\text{pre}}}} + \clU{Q_R} \clX{Z^R} \right),
\end{equation*}%
given the influent stream $(\clW{Q_{\text{in}}}, \clW{Z^{\text{in}}})$ and the overflow streams $\big( \clU{\pVars{Q}{eff}{D_{\text{pre}}}}, \clX{\pVars{Z}{eff}{D_{\text{pre}}}} \big)$ and $\big( \clU{Q_R}, \clX{Z^R} \big)$ from the thickener and storage tank units, respectively (Sections \ref{subsec: Separation_Processes} and \ref{subsec: Storage_Tank}). 
To avoid confusion, we emphasize that $\clW{\pVars{Z}{}{P_{\text{in}}}}$ are the concentrations in the combined influent stream to the clarifier, whereas $\clW{Z^{\text{in}}}$ are the concentrations in the influent to the whole plant, that is, for the wastewater collected from the municipal sewage. 

The primary clarifier is modelled as a process in which solids are perfectly separated into an underflow stream $\big( \clW{\pVars{Q}{und}{P}}, \clX{\pVars{Z}{}{P_{\text{und}}}} \big)$ and an overflow stream $\big( \clW{\pVars{Q}{eff}{P}}, \clX{\pVars{Z}{}{P_{\text{eff}}}} \big)$, in which
\begin{equation*}
    \begin{aligned}
        \clW{\pVars{Q}{und}{P}} &= \clP{f_{Q_u}} \clW{\pVars{Q}{in}{P}}; \\
        \clX{\pVars{Z}{}{P_{\text{und}}}} &= \left(1 + \Big( \frac{1 - \clP{f_{Q_u}}}{\clP{f_{Q_u}}} \Big) \clP{\chi_{Z}} \cdot \eta_{\text{COD}}^P(\clX{\pVars{Q}{}{P}}) \right)  \clX{\pVars{Z}{}{P}}; \\
    \end{aligned}\ \qquad
    \begin{aligned}
        \clW{\pVars{Q}{eff}{P}} &= (1-\clP{f_{Q_u}}) \clW{\pVars{Q}{in}{P}}; \\
        \clX{\pVars{Z}{}{P_{\text{eff}}}} &= \left( 1 - \clP{\chi_{Z}} \cdot \eta_{\text{COD}}^P(\clX{\pVars{Q}{}{P}}) \right) \clX{\pVars{Z}{}{P}}. \\
    \end{aligned}
\end{equation*}%
given the normalized COD removal efficiency $\eta_{\text{COD}}^P : \mathbb{R}_{\geq 0} \to [0,1]$, computed as 
\begin{equation*}
    \eta_{\text{COD}}^P(\clX{\pVars{Q}{}{P}}) =
        \min\left\{\max\left\{0,~ \frac{\clP{f_{corr}}}{100 \clP{f_X^P}}\left( 2.88\clP{f_X^P} - 0.118 \right) \left( 1.45 + 6.15\log \frac{1440\clP{V^P}}{\clX{\pVars{Q}{}{P}}} \right)\right\},~ 1 \right\},
\end{equation*}
and the parameter $\clP{\chi_{Z}}$ which is 1 for all particulate matter and 0 otherwise (e.g., $\clP{\chi_{X_{\text{BH}}}} = 1$ but $\clP{\chi_{S_{\text{NO}}}} = 0$).

The primary clarifier is monitored through the measurable quantities $\clY{y^{P}} = (\clY{\pVars{\text{TSS}}{}{P_{\text{eff}}}}, \clY{\pVars{S}{NH}{P_{\text{eff}}}}, \clY{\pVars{S}{NO}{P_{\text{eff}}}})$, with 
    $$\clY{\pVars{\text{TSS}}{}{P_{\text{eff}}}} = 0.75\big( \clX{\pVars{X}{I}{P_{\text{eff}}}} + \clX{\pVars{X}{S}{P_{\text{eff}}}} + \clX{\pVars{X}{BH}{P_{\text{eff}}}} + \clX{\pVars{X}{BA}{P_{\text{eff}}}} + \clX{\pVars{X}{P}{P_{\text{eff}}}} \big)$$

The constant model parameters for this unit are reported in Table \ref{tab: Parameters_PrimaryClarifier}.

\begin{table}[htb!] \centering
    \caption{Primary clarifier: Model constant parameters.}
    \label{tab: Parameters_PrimaryClarifier}
    \smallskip
    {\small\begin{tabular}{@{}p{4em}p{22em}cl@{}}\toprule
                            & Description                                                   & Value & Unit  \\\midrule
        $\clP{t^m}$		    & Smoothing time constant for average flow						& 3/24  & d     \\
        $\clP{V^P}$         & Volume of the primary clarifier                               & 900   & m$^3$ \\
        $\clP{f_{Q_u}}$     & Underflow proportion of influent flow                         & 0.007 & --    \\
        $\clP{f_{corr}}$    & Correction factor removal efficiency                          & 0.65  & --    \\
        $\clP{f_X^P}$       & Average COD$_{\text{partial}}$/COD$_{\text{total}}$ ratio     & 0.85  & --    \\
        \bottomrule
    \end{tabular}}
\end{table}

\subsection{Biological reactors ($A_r$, $r = 1, \ldots, 5$)} \label{subsec: ASM1}

\newcommand{\asmIN}[2]{ \cfrac{\clW{\pVars{Q}{in}{A_r}}}{\clP{V^{A_r}}} \big( \clW{\pVars{#1}{#2}{A_{r,\text{in}}}} - \clX{\pVars{#1}{#2}{A_r}} \big) }
\newcommand{\asmX}[1]{
    \IfEqCase{#1}{%
	       {1}{ \clX{\pVars{S}{I}{A_r}} }     {2}{ \clX{\pVars{S}{S}{A_r}} }   {3}{  \clX{\pVars{X}{I}{A_r}} }      {4}{ \clX{\pVars{X}{S}{A_r}} }%
           {5}{ \clX{\pVars{X}{BH}{A_r}} }    {6}{ \clX{\pVars{X}{BA}{A_r}} }  {7}{  \clX{\pVars{X}{P}{A_r}} }      {8}{ \clX{\pVars{S}{O}{A_r}} }%
           {9}{ \clX{\pVars{S}{NO}{A_r}} }   {10}{ \clX{\pVars{S}{NH}{A_r}} }  {11}{ \clX{\pVars{S}{ND}{A_r}} }    {12}{ \clX{\pVars{X}{ND}{A_r}} }%
          {13}{ \clX{\pVars{S}{ALK}{A_r}} }  {14}{ \clX{\pVars{T}{}{A_r}} }%
	}[\PackageError{pointer}{Undefined option to pointer: #1}{}]%
}
\newcommand{\asmP}[1]{ 
    \IfEqCase{#1}{%
	     {1}{ \pVars{\mu}{H}{}(\asmX{14}) \cfrac{\asmX{2}}{\clP{\pVars{K}{S}{}} + \asmX{2}} \cfrac{\asmX{8}}{\clP{\pVars{K}{OH}{}} + \asmX{8}} \asmX{5}  }
         {2}{ \clP{\pVars{\eta}{h}{}} \pVars{\mu}{H}{}(\asmX{14}) \cfrac{\asmX{2}}{\clP{\pVars{K}{S}{}} + \asmX{2}} \cfrac{\clP{\pVars{K}{OH}{}}}{\clP{\pVars{K}{OH}{}} + \asmX{8}} \cfrac{\asmX{9}}{\clP{\pVars{K}{NO}{}} + \asmX{9}} \asmX{5} }%
		 {3}{ \pVars{\mu}{A}{}(\asmX{14}) \cfrac{\asmX{10}}{\clP{\pVars{K}{NH}{}} + \asmX{10}} \cfrac{\asmX{8}}{\clP{\pVars{K}{OA}{}} + \asmX{8}} \asmX{6} }%
         {4}{ \pVars{b}{H}{}(\asmX{14}) \asmX{5} }%
         {5}{ \pVars{b}{A}{}(\asmX{14}) \asmX{6} }%
         {6}{ \pVars{k}{A}{}(\asmX{14}) \asmX{11} \asmX{5} }%
         {7}{ \pVars{k}{h}{}(\asmX{14}) \cfrac{\asmX{4} }{\clP{\pVars{K}{X}{}} \asmX{5} + \asmX{4}} \left( \cfrac{\asmX{8}}{\clP{\pVars{K}{OH}{}} + \asmX{8}} + \clP{\pVars{\eta}{h}{}} \cfrac{\clP{\pVars{K}{OH}{}}}{\clP{\pVars{K}{OH}{}} + \asmX{8}} \cfrac{\asmX{9}}{\clP{\pVars{K}{NO}{}} + \asmX{9}}  \right) \asmX{5} }%
         {8}{ \pVars{k}{h}{}(\asmX{14}) \cfrac{\asmX{12}}{\clP{\pVars{K}{X}{}} \asmX{5} + \asmX{4}} \left( \cfrac{\asmX{8}}{\clP{\pVars{K}{OH}{}} + \asmX{8}} + \clP{\pVars{\eta}{h}{}} \cfrac{\clP{\pVars{K}{OH}{}}}{\clP{\pVars{K}{OH}{}} + \asmX{8}} \cfrac{\asmX{9}}{\clP{\pVars{K}{NO}{}} + \asmX{9}}  \right) \asmX{5} }%
         {1p2}{ \pVars{\mu}{H}{}(\asmX{14}) \cfrac{\asmX{2}}{\clP{\pVars{K}{S}{}} + \asmX{2}} \left( \cfrac{\asmX{8}}{\clP{\pVars{K}{OH}{}} + \asmX{8}} + \clP{\pVars{\eta}{h}{}}  \cfrac{\clP{\pVars{K}{OH}{}}}{\clP{\pVars{K}{OH}{}} + \asmX{8}} \cfrac{\asmX{9}}{\clP{\pVars{K}{NO}{}} + \asmX{9}} \right) \asmX{5} }%
         {KLa}{ 1.024^{\asmX{14} - 15} \clU{\pVars{K_La}{}{A_r}} \big[ \clP{\pVars{S}{O}{\text{sat}}}(\asmX{14}) - \asmX{8} \big] }%
    }[\PackageError{pointer}{Undefined option to pointer: #1}{}]%
}

The bioreactors $A_r$ ($r = 1,\ldots,5$) have dynamics according to the Activated Sluge Model no. 1 (ASM1, \cite{Henze2000}). 
In this model, each component $\clX{Z^{A_r}}$ in the $r$-th bioreactor $A_r$ has dynamics in the form 
\begin{equation} \label{eq: ASM1_Dynamics}
    \textstyle\frac{d}{dt}{Z}^{A_r} = \cfrac{\clW{Q^{A_r}_{\text{in}}}}{\clP{V^{A_r}}} \left( \clW{\pVars{Z}{}{A_{r,\text{in}}}} - \clX{Z^{A_r}} \right) + R_{Z}\big( \clX{x^{A_r}} \big),
\end{equation}
where $\big( \clW{Q^{A_r}_{\text{in}}}, \clW{\pVars{Z}{}{A_{r,\text{in}}}} \big)$ is the influent stream to $A_r$ and $R_Z\big( \clX{x^{A_r}} \big)$ denotes the net change in the state component $\clX{Z^{A_r}}$ caused by the reaction network. 
Specifically, we have that
$$
    \clW{Q^{A_r}_{\text{in}}} = 
    \begin{cases}
        \clW{\pVars{Q}{eff}{P}} + \clU{Q_A} + \clU{Q_S} + \clU{Q_{\text{EC}}^{A_1}}  & \text{ }(r = 1); \vspace*{0.5em}\\
        \clW{Q^{A_{r-1}}_{\text{in}}} + \clU{Q_{\text{EC}}^{A_r}}     & \text{ otherwise},
    \end{cases}
$$
and
$$
    \clW{\pVars{Z}{}{A_{r,\text{in}}}} = \begin{cases}
        \dfrac{1}{\clW{Q^{A_r}_{\text{in}}}} \left( \clW{\pVars{Q}{eff}{P}} \clX{\pVars{Z}{}{P_{\text{eff}}}} + \clU{Q_A} \clX{Z^{A_5}} + \clU{Q_S} \clX{Z^{S_1}} + \clU{Q_{\text{EC}}^{A_1}} \clX{Z^{\text{EC}}} \right)  & \text{ }(r = 1); \vspace*{0.5em}\\
        \dfrac{1}{\clW{Q^{A_r}_{\text{in}}}} \left( \clW{Q^{A_r}_{\text{in}}} \clX{Z^{A_{r-1}}} + \clU{Q_{\text{EC}}^{A_r}} \clP{Z^{\text{EC}}} \right)                                                          & \text{ otherwise},
    \end{cases}
$$%
with an external carbon stream of profile $\clP{Z^{\text{EC}}} = \clP{S_S^{\text{EC}}}$ if $\clX{Z^{A_r}} \equiv \clX{S_S^{A_r}}$ and $\clP{Z^{\text{EC}}} = 0$ otherwise.  
Explicitly, the dynamics in each reactor $A_r$ ($r=1,\ldots,5$) are described by the differential equations
\begin{align}
\textstyle\frac{d}{dt}\pVars{S}{I}{A_r} 
	& = \asmIN{S}{I} \\[0.5em]
\textstyle\frac{d}{dt}\pVars{S}{S}{A_r} 
	& = \asmIN{S}{S} \\
	&\quad -\cfrac{1}{\clP{\pVars{Y}{H}{}}} \asmP{1p2} \nonumber\\
	&\quad + \asmP{7} \nonumber\\[0.5em]
\textstyle\frac{d}{dt}\pVars{X}{I}{A_r} 
    & = \asmIN{X}{I} \\[0.5em]
\textstyle\frac{d}{dt}\pVars{X}{S}{A_r} 
    & = \asmIN{X}{S} \\
    &\quad - \asmP{7} \nonumber\\
    &\quad + \big[1-\clP{\pVars{f}{P}{}}\big] \asmP{4}  +\big[1-\clP{\pVars{f}{P}{}}\big] \asmP{5} \nonumber\\[0.5em]
\textstyle\frac{d}{dt}\pVars{X}{BH}{A_r} 
    & = \asmIN{X}{BH} \\ 
	&\quad + \asmP{1p2} - \asmP{4} \nonumber\\[0.5em]
\textstyle\frac{d}{dt}\pVars{X}{BA}{A_r} 
    & = \asmIN{X}{BA} \\ 
    &\quad + \asmP{3} - \asmP{5} \nonumber\\[0.5em]
\textstyle\frac{d}{dt}\pVars{X}{P}{A_r} 
    & = \asmIN{X}{P} \\
    &\quad + \clP{\pVars{f}{P}{}} \Big[ \asmP{4} + \asmP{5} \Big] \nonumber\\[0.5em]
\textstyle\frac{d}{dt}\pVars{S}{O}{A_r} 
    & = \asmIN{S}{O} + \asmP{KLa}  \\
    &\quad - \cfrac{1-\clP{\pVars{Y}{H}{}}}{\clP{\pVars{Y}{H}{}}} \asmP{1} - \cfrac{4.57-\clP{\pVars{Y}{A}{}}}{\clP{\pVars{Y}{A}{}}}\asmP{3} \nonumber\\[0.5em]
\textstyle\frac{d}{dt}\pVars{S}{NO}{A_r} 
    & = \asmIN{S}{NO}  \\
    &\quad - \cfrac{1-\clP{\pVars{Y}{H}{}}}{2.86\clP{\pVars{Y}{H}{}}} \asmP{2} + \cfrac{1}{\clP{\pVars{Y}{A}{}}} \asmP{3} \nonumber\\[0.5em]
\textstyle\frac{d}{dt}\pVars{S}{NH}{A_r} 
    & = \asmIN{S}{NH}  \\
    &\quad - \clP{\pVars{i}{XB}{}} \asmP{1p2} \nonumber\\
    &\quad - \Big(\clP{\pVars{i}{XB}{}} + \cfrac{1}{\clP{\pVars{Y}{A}{}}}\Big) \asmP{3} + \asmP{6} \nonumber\\[0.5em]
\textstyle\frac{d}{dt}\pVars{S}{ND}{A_r} 
    & = \asmIN{S}{ND}  \\
    &\quad + \asmP{8} - \asmP{6} \nonumber\\[0.5em]
\textstyle\frac{d}{dt}\pVars{X}{ND}{A_r} 
    & = \asmIN{X}{ND}  \\
    &\quad - \asmP{8} \nonumber\\
    &\quad + \Big(\clP{\pVars{i}{XB}{}} - \clP{\pVars{f}{P}{}}\clP{\pVars{i}{XP}{}} \Big) \left( \asmP{4} + \asmP{5} \right)   \nonumber\\[0.5em]
\textstyle\frac{d}{dt}\pVars{S}{ALK}{A_r} 
    & = \asmIN{S}{ALK}  \\
    &\quad - \cfrac{\clP{\pVars{i}{XB}{}}}{14} \asmP{1} + \cfrac{1}{14} \asmP{6} \nonumber\\
    &\quad + \Big(\cfrac{1-\clP{\pVars{Y}{H}{}}}{14 \times 2.86 \clP{\pVars{Y}{H}{}}} - \cfrac{\clP{\pVars{i}{XB}{}}}{14} \Big) \asmP{2} \nonumber\\ 
    &\quad - \Big(\cfrac{\clP{\pVars{i}{XB}{}}}{14}+\cfrac{1}{7\clP{\pVars{Y}{A}{}}}\Big) \asmP{3} \nonumber \\[0.5em]
\textstyle\frac{d}{dt} T^{A_r} 
    & = \asmIN{T}{} 
\end{align}

The kinetic parameters are assumed to not be affected by temperature changes inside the reactors, and thus they are set constant to their base values at 15$^{\circ}$C. 
Exceptions are the parameters $\clP{\kappa} \in \{\clP{\pVars{\mu}{H}{}}, \clP{\pVars{\mu}{A}{}}, \clP{\pVars{b}{H}{}}, \clP{\pVars{b}{A}{}}, \clP{\pVars{k}{H}{}}, \clP{\pVars{k}{A}{}}\}$, which depend on the temperature through the Arrhenius-type relationship
\begin{equation*}
    \clP{\kappa}(\clX{\pVars{T}{}{A_r}}) = \clP{\pVars{\kappa}{base}{(15^{\circ}\text{C})}} \exp\left[\log\left( \frac{\clP{\pVars{\kappa}{base}{(15^{\circ}\text{C})}}}{\clP{\pVars{\kappa}{base}{(10^{\circ}\text{C})}}} \right) \frac{\clX{\pVars{T}{}{A_r}} - 15}{5} \right],
\end{equation*}
given their base values at 15$^{\circ}$C, $\clP{\pVars{\kappa}{base}{(15^{\circ}\text{C})}}$, and at 10$^{\circ}$C, $\clP{\pVars{\kappa}{base}{(10^{\circ}\text{C})}}$ (see Table \ref{tab: Parameters_ASM1}). 
Similarly, also the oxygen saturation coefficient $\clP{S_O^{\text{sat}}}$ depends on temperature in the bioreactor according to the expression
\begin{equation*}
    \clP{S_O^{\text{sat}}}(\clX{\pVars{T}{}{A_r}}) = \left(\frac{8 \cdot 6791.5 \cdot 56.12}{10.50237016}\right)\exp \left[-66.7354 + \frac{8747.55}{\clX{\pVars{T}{}{A_r}} + 273.15} + 24.4526\log\left(\frac{\clX{\pVars{T}{}{A_r}} + 273.15}{100}\right) \right].
\end{equation*}

The bioreactors are monitored through the measurable quantities $\clY{y^{A_r}} = (\clY{\pVars{S}{O}{A_r}}, \clY{\pVars{S}{NO}{A_r}}, \clY{\pVars{T}{}{A_r}})$, for $r = 1,\ldots,4$, and $\clY{y^{A_5}} = (\clY{\pVars{S}{O}{A_5}}, \clY{\pVars{S}{NO}{A_5}}, \clY{\pVars{T}{}{A_5}}, \clY{\pVars{\text{TSS}}{}{A_5}})$, for the last reactor, with
    $$\clY{\pVars{\text{TSS}}{}{A_5}} = 0.75\big( \clX{\pVars{X}{I}{A_5}} + \clX{\pVars{X}{S}{A_5}} + \clX{\pVars{X}{BH}{A_5}} + \clX{\pVars{X}{BA}{A_5}} + \clX{\pVars{X}{P}{A_5}} \big).$$

The constant model parameters for these process units are reported in Table \ref{tab: Parameters_ASM1}.

\begin{table}[htb!] \centering
    \caption{Bioreactors (ASM1): Model constant parameters.}
    \label{tab: Parameters_ASM1}
    \smallskip
    {\small\begin{tabular}{@{}llccl@{}}\toprule
                                            & General parameter					            & \multicolumn{2}{c}{Value} 	    & Units				\\\midrule                
        $\clP{V^{A_{1\leadsto 2}}}$		    & Reactor volume (anoxic section)		        & \multicolumn{2}{c}{$1000$} 	    & m$^{3}$		    \\
        $\clP{V^{A_{3\leadsto 5}}}$	        & Reactor volume (aerobic section) 		        & \multicolumn{2}{c}{$1333$}	    & m$^{3}$		    \\
        $\clP{\pVars{S}{S}{\text{EC}}}$	    & External carbon source concentration	        & \multicolumn{2}{c}{$4\cdot10^5$} & g\,COD/m$^{3}$    \\
        $\clP{\pVars{S}{O}{\text{sat}}}$    & Oxygen saturation concentration		        & \multicolumn{2}{c}{$8$}  		    & g\,O$_2$/m$^{3}$	\\[1ex]\midrule
                                            & Stoichiometric parameter		                & \multicolumn{2}{c}{Value} 	    & Units						                                        \\\midrule                
        $\clP{\pVars{Y}{A}{}}$	            & Autotrophic yield						        & \multicolumn{2}{c}{$0.24$}        & (g\,$\pVars{X}{BA}{}$\,COD formed)/(g\,N oxidized)	            \\
        $\clP{\pVars{Y}{H}{}}$	            & Heterotrophic yield 					        & \multicolumn{2}{c}{$0.67$}        & (g\,$\pVars{X}{BH}{}$\,COD formed)/(g\,COD utilized)	            \\
        $\clP{\pVars{f}{P}{}}$	            & Fraction of biomass to particulate products 	& \multicolumn{2}{c}{$0.08$}        & (g\,$\pVars{X}{P}{}$\,COD formed)/(g\,$\pVars{X}{BH}{}$ decayed)	\\
        $\clP{\pVars{i}{XB}{}}$	            & Fraction nitrogen in biomass 				    & \multicolumn{2}{c}{$0.08$}        & g\,N/(g\,COD) in biomass                                          \\
        $\clP{\pVars{i}{XP}{}}$	            & Fraction nitrogen in particulate products 	& \multicolumn{2}{c}{$0.06$}        & g\,N/(g\,COD) in $\pVars{X}{P}{}$                                 \\[1ex]\midrule
                                            & Kinetic parameter					            & \multicolumn{2}{c}{Value} 	    & Units						                                \\\midrule                
                                            &                                               & at 10$^{\circ}$C & at 15$^{\circ}$C &                                                         \\\cmidrule{3-4}
        $\clP{\pVars{\mu}{H,base}{}}$       & Maximum heterotrophic growth rate 	        & $3.00$ & $4.00$	    & 1/d						                                \\
        $\clP{\pVars{\mu}{A,base}{}}$       & Maximum autotrophic growth rate		        & $0.30$ & $0.50$	    & 1/d						                                \\	
        $\clP{\pVars{b}{H,base}{}}$	        & Heterotrophic decay rate 			            & $0.20$ & $0.30$	    & 1/d						                                \\
        $\clP{\pVars{b}{A,base}{}}$	        & Autotrophic decay rate				        & $0.03$ & $0.05$	    & 1/d						                                \\
        $\clP{\pVars{k}{H,base}{}}$	        & Maximum specific hydrolysis rate		        & $2.50$ & $3.00$	    & g\,$\pVars{X}{S}{}$/(g\,$\pVars{X}{BH}{}$\,COD$\cdot$d)	\\
        $\clP{\pVars{k}{A,base}{}}$	        & Ammonification rate 				            & $0.04$ & $0.05$	    & m$^3$/(g\,COD\,d)			                                \\
        $\clP{\pVars{K}{S}{}}$		        & Half-saturation (heterotrophic growth) 	    &   --   & $10.0$	    & g\,COD/m$^{3}$				                            \\
        $\clP{\pVars{K}{OH}{}}$	            & Half-saturation (heterotrophic oxygen) 	    &   --   & $0.20$	    & g\,O$_2$/m$^{3}$				                            \\
        $\clP{\pVars{K}{NO}{}}$	            & Half-saturation (nitrate)				        &   --   & $0.50$	    & g\,NO$_3$-N/m$^{3}$			                            \\
        $\clP{\pVars{K}{X}{}}$	            & Half-saturation (hydrolysis)			        &   --   & $0.10$	    & g\,$\pVars{X}{S}{}$/(g\,$\pVars{X}{BH}{}$\,COD)	        \\
        $\clP{\pVars{K}{NH}{}}$	            & Half-saturation (autotrophic growth)	        &   --   & $1.00$	    & g\,NH$_4$-N/m$^{3}$			                            \\
        $\clP{\pVars{K}{OA}{}}$	            & Half-saturation (autotrophic oxygen) 	        &   --   & $0.40$	    & g\,O$_2$/m$^{3}$				                            \\		
        $\clP{\pVars{\eta}{g}{}}$	        & Anoxic growth rate correction factor 	        &   --   & $0.80$	    & --					                                    \\
        $\clP{\pVars{\eta}{h}{}}$	        & Anoxic hydrolysis rate correction factor 	    &   --   & $0.80$	    & --					                                    \\
        \bottomrule
    \end{tabular}}
\end{table}

\subsection{Secondary settler ($S_l$, $l = 1,\ldots,10$)} \label{subsec: Settler}

The layers of the secondary settler ($S_l$, $l = 1,\ldots,10$) have dynamics according to the model by Tak\'acs \cite{Takacs1991}. 
The dynamics for suspended solids in each $l$-th settler's layer, $\clX{\pVars{\text{TSS}}{}{S_l}}$, are modelled as
\begin{align*} 
\textstyle\frac{d}{dt}\pVars{\text{TSS}}{}{S_l} &= 
    \begin{cases} 
    \cfrac{\clW{\pVars{Q}{eff}{}} }{\clP{\pVars{V}{}{S_l}}} \Big( \clX{\pVars{\text{TSS}}{}{S_{l-1}}} - \clX{\pVars{\text{TSS}}{}{S_l}} \Big)   - \dfrac{1}{\clP{ \pVars{h}{}{S_l}}} J_{\text{cla}}\big(\clX{\pVars{\text{TSS}}{}{S_l}}, \clX{\pVars{\text{TSS}}{}{S_{l-1}}}\big) & (l = 10); \\
    \cfrac{\clW{\pVars{Q}{eff}{}} }{\clP{\pVars{V}{}{S_l}}} \Big( \clX{\pVars{\text{TSS}}{}{S_{l-1}}} - \clX{\pVars{\text{TSS}}{}{S_l}} \Big)   + \dfrac{1}{\clP{ \pVars{h}{}{S_l}}} \left( J_{\text{cla}}\big(\clX{\pVars{\text{TSS}}{}{S_{l+1}}}, \clX{\pVars{\text{TSS}}{}{S_l}}\big) - J_{\text{cla}}\big(\clX{\pVars{\text{TSS}}{}{S_l}}, \clX{\pVars{\text{TSS}}{}{S_{l-1}}}\big) \right) & (l = 7, \cdots, 9); \\
    \cfrac{\clW{\pVars{Q}{in}{S}} }{\clP{\pVars{V}{}{S_l}}} \Big( \clX{\pVars{\text{TSS}}{}{S_{\text{in}}}}       - \clX{\pVars{\text{TSS}}{}{S_l}} \Big) + \dfrac{1}{\clP{ \pVars{h}{}{S_l}}} \left( J_{\text{cla}}\big(\clX{\pVars{\text{TSS}}{}{S_{l+1}}}, \clX{\pVars{\text{TSS}}{}{S_l}}\big) - J_{\text{st} }\big(\clX{\pVars{\text{TSS}}{}{S_l}}, \clX{\pVars{\text{TSS}}{}{S_{l-1}}}\big) \right) & (l = 6); \\
    \cfrac{\clU{\pVars{Q}{und}{S}}}{\clP{\pVars{V}{}{S_l}}} \Big( \clX{\pVars{\text{TSS}}{}{S_{l+1}}} - \clX{\pVars{\text{TSS}}{}{S_l}} \Big)   + \dfrac{1}{\clP{ \pVars{h}{}{S_l}}} \left( J_{\text{st} }\big(\clX{\pVars{\text{TSS}}{}{S_{l+1}}}, \clX{\pVars{\text{TSS}}{}{S_l}}\big) - J_{\text{st} }\big(\clX{\pVars{\text{TSS}}{}{S_l}}, \clX{\pVars{\text{TSS}}{}{S_{l-1}}}\big) \right) & (l = 2, \cdots, 5); \\
    \cfrac{\clU{\pVars{Q}{und}{S}}}{\clP{\pVars{V}{}{S_l}}} \Big( \clX{\pVars{\text{TSS}}{}{S_{l+1}}} - \clX{\pVars{\text{TSS}}{}{S_l}} \Big)   + \dfrac{1}{\clP{ \pVars{h}{}{S_l}}} J_{\text{st}}\big(\clX{\pVars{\text{TSS}}{}{S_{l+1}}}, \clX{\pVars{\text{TSS}}{}{S_l}}\big) & (l = 1),
    \end{cases}
\end{align*}%
while the dynamics of all other components $\clX{\pVars{Z}{}{S_l}}$, including temperature, within each $l$-th layer are
\begin{equation}
\textstyle\frac{d}{dt} \pVars{Z}{}{S_l}
     = \begin{cases} 
    \cfrac{\clW{\pVars{Q}{eff}{}} }{\clP{ \pVars{V}{}{S_l}}} \Big( \clX{\pVars{Z}{}{S_{l-1}}} - \clX{\pVars{Z}{}{S_l}} \Big) & (l = 7, \cdots, 10); \\
    \cfrac{\clW{\pVars{Q}{in}{S}} }{\clP{ \pVars{V}{}{S_l}}} \Big( \clX{\pVars{Z}{}{S_{\text{in}}}}     - \clX{\pVars{Z}{}{S_l}} \Big) & (l = 6); \\
    \cfrac{\clU{\pVars{Q}{und}{S}}}{\clP{ \pVars{V}{}{S_l}}} \Big( \clX{\pVars{Z}{}{S_{l+1}}} - \clX{\pVars{Z}{}{S_l}} \Big) & (l = 1, \cdots, 5). 
    \end{cases}
\end{equation}%
The triple $(\clW{\pVars{Q}{in}{S}}, \clX{\pVars{\text{TSS}}{}{S_{\text{in}}}}, \clX{\pVars{Z}{}{S_{\text{in}}}})$ defines the influent to the settler (entering at layer $S_6$), with $\clW{\pVars{Q}{in}{S}} = \clW{\pVars{Q}{in}{A_5}} - \clU{Q_A}$, $\clX{\pVars{Z}{}{S_{\text{in}}}} = \clX{\pVars{Z}{}{A_5}}$, and $\clX{\pVars{\text{TSS}}{}{S_{\text{in}}}} = 0.75\big( \clX{\pVars{X}{I}{A_5}} + \clX{\pVars{X}{S}{A_5}} + \clX{\pVars{X}{BH}{A_5}} + \clX{\pVars{X}{BA}{A_5}} + \clX{\pVars{X}{P}{A_5}} \big)$. 
From the settler, the underflow stream is $(\clU{\pVars{Q}{und}{S}}, \clX{Z^{S_1}})$ and the overflow stream (i.e., the effluent of the plant) is $(\clW{\pVars{Q}{eff}{}}, \clX{Z^{S_{10}}})$, with $\clU{\pVars{Q}{und}{S}} = (\clU{Q_S} + \clU{Q_W})$ and $\clW{\pVars{Q}{eff}{}} = (\clU{\pVars{Q}{in}{S}} - \clU{\pVars{Q}{und}{S}})$. 
The downward and upward flux of solids are, respectively,
\begin{align} 
    J_{\text{st}}\left(\clX{\pVars{\text{TSS}}{}{S_l}}, \clX{\pVars{\text{TSS}}{}{S_{l-1}}}\right) 
        & = \min\left\{ v_s\big(\clX{\pVars{\text{TSS}}{}{S_{l-1}}}\big)\clX{\pVars{\text{TSS}}{}{S_{l-1}}}, v_s\big(\clX{\pVars{\text{TSS}}{}{S_l}}\big)\clX{\pVars{\text{TSS}}{}{S_l}} \right\}; \\
    J_{\text{cla}}\left(\clX{\pVars{\text{TSS}}{}{S_l}}, \clX{\pVars{\text{TSS}}{}{S_{l-1}}}\right) 
        & = \begin{cases} 
        J_{\text{st}}\left(\clX{\pVars{\text{TSS}}{}{S_l}}, \clX{\pVars{\text{TSS}}{}{S_{l-1}}}\right) & \text{if } \clX{\pVars{\text{TSS}}{}{S_{l-1}}} > \clP{\pVars{\text{TSS}}{}{\max}}; \\
        v_s\big(\clX{\pVars{\text{TSS}}{}{S_l}}\big)\clX{\pVars{\text{TSS}}{}{S_l}} & \text{otherwise},
        \end{cases} \label{eq: Settler_solid_flux_downward}
\end{align}%
in which $v_s(\cdot)$ is the double-exponential settling velocity function defined as
\begin{equation} 
    v_s\Big(\clX{\pVars{\text{TSS}}{}{S_l}}\Big) = \max\Big\{ 0,\, \min\Big\{ \clP{ v_0}\Big( \exp[-\clP{ r_h}(\clX{\pVars{\text{TSS}}{}{S_l}} - \clP{\pVars{f}{ns}{}} \clX{\pVars{\text{TSS}}{}{S_{\text{in}}}})] - \exp[-\clP{ r_p} (\clX{\pVars{\text{TSS}}{}{S_l}} - \clP{\pVars{f}{ns}{}} \clX{\pVars{\text{TSS}}{}{S_{\text{in}}}})] \Big),\, \clP{ v_0^{\max}} \Big\} \Big\}.
\end{equation}

The secondary settler is monitored through the measurable quantities $\clY{y^S} = (\clY{\pVars{\text{TSS}}{}{S_{10}}}, \clY{\pVars{S}{NH}{S_{10}}}, \clY{\pVars{S}{NO}{S_{10}}})$.
The constant model parameters for this unit are reported in Table \ref{tab: Parameters_Settler}.

\begin{table}[htb!] \centering
    \caption{Secondary settler: Model constant parameters.}
    \label{tab: Parameters_Settler}
    {\small\begin{tabular}{@{}llcl@{}}\toprule
                                            & Description					        & Value 	    & Units			    \\\midrule                
        $\clP{V^{S_{l}}}$		            & Volume of settler layer		        & $600$	        & m$^{3}$		    \\
        $\clP{h^{S_{l}}}$	                & Height of settler layer 		        & $0.4$         & m		            \\
        $\clP{\pVars{\text{TSS}}{}{\max}}$	& Settling threshold concentration	    & $3000$        & g\,SS/m$^{3}$     \\
        $\clP{v_0^{\max}}$	                & Maximum settling velocity			    & $250.0$	    & m/d	            \\
        $\clP{v_0}$	                        & Maximum Vesilind settling velocity	& $474.0$	    & m/d			    \\
        $\clP{\pVars{r}{h}{}}$	            & Hindered zone settling parameter	    & $0.000576$	& m$^3$/(g\,SS)	    \\
        $\clP{\pVars{r}{h}{}}$	            & Flocculant zone settling parameter	& $0.00286$	    & m$^3$/(g\,SS)	    \\
        $\clP{\pVars{f}{ns}{}}$	            & Non-settleable fraction 	            & $0.00228$	    & --			    \\		
        \bottomrule
    \end{tabular}}
\end{table}

\subsection{Thickener and Dewatering units} \label{subsec: Separation_Processes}

The thickener ($D_{\text{pre}}$) and dewatering ($D_{\text{post}}$) units are assumed ideal (static) separation processes that reduce the volume of the streams entering the digester and storage tank, respectively. 
In both units (jointly referred to as $D_{\circ}$), the influent is split into an underflow $(\clU{\pVars{Q}{und}{D_{\circ}}}, \clX{\pVars{Z}{}{D_{\circ,\text{und}}}})$\footnote{$H(\cdot)$ refers to the Heaviside step function},
\begin{equation}
    \clU{\pVars{Q}{und}{D_{\circ}}} = \left( 1 - H\big(\pVars{\eta}{COD}{D_{\circ}}(\clX{\pVars{x}{}{D_{\circ,\text{in}}}}) {-} 1 \big) \pVars{Q}{factor}{D_{\circ}}(\clX{\pVars{x}{}{D_{\circ,\text{in}}}}) \right) \clU{\pVars{Q}{in}{D_{\circ}}}, \qquad
    \clX{\pVars{Z}{}{D_{\circ,\text{und}}}} = \pVars{f}{und}{D_{\circ}}(\clX{\pVars{x}{}{D_{\circ,\text{in}}}}) \clX{\pVars{Z}{}{D_{\circ,\text{in}}}}; 
\end{equation}
and an overflow $(\clU{\pVars{Q}{eff}{D_{\circ}}}, \clX{\pVars{Z}{}{D_{\circ,\text{eff}}}})$ stream,
\begin{equation}
    \clU{\pVars{Q}{eff}{D_{\circ}}} = \clU{\pVars{Q}{in}{D_{\circ}}} - \clU{\pVars{Q}{und}{D_{\circ}}}, \qquad
    \clX{\pVars{Z}{}{D_{\circ,\text{eff}}}} = 0.002H\big(\pVars{\eta}{COD}{D_{\circ}}(\clX{\pVars{x}{}{D_{\circ,\text{in}}}}) {-} 1 \big)\pVars{f}{eff}{D_{\circ}}(\clX{\pVars{x}{}{D_{\circ,\text{in}}}}) \clX{\pVars{Z}{}{D_{\circ,\text{in}}}},
\end{equation}
with the separation factors 
\begin{subequations}
\begin{align}
    \pVars{f}{und}{D_{\circ}}(\clX{\pVars{x}{}{D_{\circ,\text{in}}}}) &= \max\big( 1,~ \pVars{\eta}{COD}{D_{\circ}}(\clX{\pVars{x}{}{D_{\circ,\text{in}}}}) \big); \\
    \pVars{f}{eff}{D_{\circ}}(\clX{\pVars{x}{}{D_{\circ,\text{in}}}}) &= \pVars{\eta}{COD}{D_{\circ}}(\clX{\pVars{x}{}{D_{\circ,\text{in}}}})/(\pVars{\eta}{COD}{D_{\circ}}(\clX{\pVars{x}{}{D_{\circ,\text{in}}}}) - 0.98),
\end{align}%
\end{subequations}
if the component $\clX{Z}$ represents a particulate, and 
    $\pVars{f}{und}{D_{\circ}}(\clX{\pVars{x}{}{D_{\circ,\text{in}}}}) = \pVars{f}{eff}{D_{\circ}}(\clX{\pVars{x}{}{D_{\circ,\text{in}}}}) = 1$ 
otherwise. 
The underflow factor is $\pVars{Q}{factor}{D_{\circ}}(\clX{\pVars{x}{}{D_{\circ,\text{in}}}}) = 1 - 0.98/\pVars{\eta}{COD}{D_{\circ}}(\clX{\pVars{x}{}{D_{\circ,\text{in}}}})$. 
Finally, the COD removal efficiency $\pVars{\eta}{COD}{D_{\circ}}(\cdot)$ is defined as
\begin{equation}
    \pVars{\eta}{COD}{D_{\circ}}(\clX{\pVars{x}{}{D_{\circ,\text{in}}}}) = \frac{\clP{f_X^{D_{\circ}}} \times 10^4}{0.75\big( \clX{\pVars{X}{I}{D_{\circ,\text{in}}}} + \clX{\pVars{X}{S}{D_{\circ,\text{in}}}} + \clX{\pVars{X}{BH}{D_{\circ,\text{in}}}} + \clX{\pVars{X}{BA}{D_{\circ,\text{in}}}} + \clX{\pVars{X}{P}{D_{\circ,\text{in}}}} \big)}.
\end{equation}

For the thickener, the influent stream\footnote{We remark that $\clX{\pVars{X}{I}{S_1}} + \clX{\pVars{X}{S}{S_1}} + \clX{\pVars{X}{BH}{S_1}} + \clX{\pVars{X}{BA}{S_1}} + \clX{\pVars{X}{P}{S_1}} = \clX{\pVars{\text{TSS}}{}{S_1}}$.} is $(\clU{\pVars{Q}{in}{D_{\text{pre}}}}, \clX{\pVars{Z}{}{D_{\text{pre,in}}}}) = (\clU{Q_W}, \clX{Z^{S_1}})$ with a removal efficiency $\clP{f_X^{D_{\text{pre}}}} = 7$. 
For the dewater, the influent stream is $(\clU{\pVars{Q}{in}{D_{\text{post}}}}, \clX{\pVars{Z}{}{D_{\text{post,in}}}}) = (\clW{\pVars{Q}{eff}{D}}, \clX{\pVars{Z}{}{D_{\text{eff}}}})$ with a removal efficiency $\clP{f_X^{D_{\text{post}}}} = 28$. 

\subsection{Biodigester: Anaerobic Digestion Model no. 1 (ADM1)} \label{subsec: ADM1}

\newcommand{\admIN}[2]{ \cfrac{\clW{\pVars{Q}{in}{D}}}{\clP{\pVars{V}{liq}{D}}} \left( \clW{\pVars{#1}{#2}{D_{\text{in}}}} - \clX{\pVars{#1}{#2}{D}} \right) }
\newcommand{\admX}[1]{
    \IfEqCase{#1}{%
	       {1}{ \clX{\pVars{S}{SU}{D}} }       {2}{ \clX{\pVars{S}{AA}{D}} }      {3}{ \clX{\pVars{S}{FA}{D}} }         {4}{ \clX{\pVars{S}{VA}{D}} }%
           {5}{ \clX{\pVars{S}{BU}{D}} }       {6}{ \clX{\pVars{S}{PRO}{D}} }     {7}{ \clX{\pVars{S}{AC}{D}} }         {8}{ \clX{\pVars{S}{H$_2$}{D}} }%
           {9}{ \clX{\pVars{S}{CH$_4$}{D}} }  {10}{ \clX{\pVars{S}{IC}{D}} }     {11}{ \clX{\pVars{S}{IN}{D}} }        {12}{ \clX{\pVars{S}{I}{D}} }%
          {13}{ \clX{\pVars{X}{C}{D}} }       {14}{ \clX{\pVars{X}{CH}{D}} }     {15}{ \clX{\pVars{X}{PR}{D}} }        {16}{ \clX{\pVars{X}{LI}{D}} }%
          {17}{ \clX{\pVars{X}{SU}{D}} }      {18}{ \clX{\pVars{X}{AA}{D}} }     {19}{ \clX{\pVars{X}{FA}{D}} }        {20}{ \clX{\pVars{X}{C$_4$}{D}} }%
          {21}{ \clX{\pVars{X}{PRO}{D}} }     {22}{ \clX{\pVars{X}{AC}{D}} }     {23}{ \clX{\pVars{X}{H$_2$}{D}} }     {24}{ \clX{\pVars{X}{I}{D}} }%
          {25}{ \clX{\pVars{S}{CAT$^+$}{D}} } {26}{ \clX{\pVars{S}{AN$^-$}{D}} } {27}{ \clX{\pVars{S}{VA$^-$}{D}} }    {28}{ \clX{\pVars{S}{BU$^-$}{D}} }%
          {29}{ \clX{\pVars{S}{PRO$^-$}{D}} } {30}{ \clX{\pVars{S}{AC$^-$}{D}} } {31}{ \clX{\pVars{S}{HCO$_3^-$}{D}} } {32}{ \clX{\pVars{S}{NH$_3$}{D}} }%
          {33}{ \clX{\pVars{G}{H$_2$}{D}} }   {34}{ \clX{\pVars{G}{CH$_4$}{D}} } {35}{ \clX{\pVars{G}{CO$_2$}{D}} }%
	}[\PackageError{pointer}{Undefined option to pointer: #1}{}]%
}
\newcommand{\admP}[1]{ 
    \IfEqCase{#1}{%
	      {1}{ \clP{\pVars{k}{dis}{}}     \admX{13} }
          {2}{ \clP{\pVars{k}{CH}{\text{hyd}}}  \admX{14} }%
		  {3}{ \clP{\pVars{k}{PR}{\text{hyd}}}  \admX{15} }%
          {4}{ \clP{\pVars{k}{LI}{\text{hyd}}}  \admX{16} }%
          {5}{ \clP{\pVars{k}{SU}   {\text{m}}}    \cfrac{ \admX{1} }{ \clP{\pVars{K}{SU}{}}    + \admX{1} } \admX{17} I_1(\clX{x^{D}}) }%
          {6}{ \clP{\pVars{k}{AA}   {\text{m}}}    \cfrac{ \admX{2} }{ \clP{\pVars{K}{AA}{}}    + \admX{2} } \admX{18} I_1(\clX{x^{D}}) }%
          {7}{ \clP{\pVars{k}{FA}   {\text{m}}}    \cfrac{ \admX{3} }{ \clP{\pVars{K}{FA}{}}    + \admX{3} } \admX{19} I_2(\clX{x^{D}}) }%
          {8}{ \clP{\pVars{k}{C$_4$}{\text{m}}} \cfrac{ \admX{4} }{ \clP{\pVars{K}{VA}{}}    + \admX{4} } \admX{20} \cfrac{ \admX{4} }{ \admX{4} + \admX{5} } I_3(\clX{x^{D}}) }%
          {9}{ \clP{\pVars{k}{C$_4$}{\text{m}}} \cfrac{ \admX{5} }{ \clP{\pVars{K}{BU}{}}    + \admX{5} } \admX{20} \cfrac{ \admX{5} }{ \admX{4} + \admX{5} } I_3(\clX{x^{D}}) }%
         {10}{ \clP{\pVars{k}{PR}   {\text{m}}}    \cfrac{ \admX{6} }{ \clP{\pVars{K}{PRO}{}}   + \admX{6} } \admX{20} I_4(\clX{x^{D}}) }%
         {11}{ \clP{\pVars{k}{AC}   {\text{m}}}    \cfrac{ \admX{7} }{ \clP{\pVars{K}{AC}{}}    + \admX{7} } \admX{21} I_5(\clX{x^{D}}) }%
         {12}{ \clP{\pVars{k}{H$_2$}{\text{m}}} \cfrac{ \admX{8} }{ \clP{\pVars{K}{H$_2$}{}} + \admX{8} } \admX{23} I_6(\clX{x^{D}}) }%
         {13}{ \clP{\pVars{k}{SU}   {\text{dec}}}    \admX{17} }%
         {14}{ \clP{\pVars{k}{AA}   {\text{dec}}}    \admX{18} }%
         {15}{ \clP{\pVars{k}{FA}   {\text{dec}}}    \admX{19} }%
         {16}{ \clP{\pVars{k}{C$_4$}{\text{dec}}} \admX{20} }%
         {17}{ \clP{\pVars{k}{PRO}  {\text{dec}}}   \admX{21} }%
         {18}{ \clP{\pVars{k}{AC}   {\text{dec}}}    \admX{22} }%
         {19}{ \clP{\pVars{k}{H$_2$}{\text{dec}}} \admX{23} }%
         {H2}{ \clP{K_La^{D}}\big( \admX{8} - \clP{\pVars{K}{H$_2$}{H}}\clP{R}\clP{\pVars{T}{ad}{}} \admX{33}  \big) }%
        {CH4}{ \clP{K_La^{D}}\big( \admX{9} - \clP{\pVars{K}{CH$_4$}{H}}\clP{R}\clP{\pVars{T}{ad}{}} \admX{34} \big) }%
        {CO2}{ \clP{K_La^{D}}\big( \admX{10} - \admX{31} - \clP{\pVars{K}{CO$_2$}{H}}\clP{R}\clP{\pVars{T}{ad}{}} \admX{35} \big) }%
    }[\PackageError{pointer}{Undefined option to pointer: #1}{}]%
}

The anaerobic digester is subjected to influents summarized by $(\clW{\pVars{Q}{in}{D}}, \clW{\pVars{Z}{}{D_{\text{in}}}})$ as 
\begin{equation} \label{eq: ADM_Influent}
    \clW{\pVars{Q}{in}{D}} = \clW{\pVars{Q}{und}{P}} + \clU{\pVars{Q}{und}{D_{\text{pre}}}}, 
    \qquad 
    \clW{\pVars{\widetilde Z}{}{D_{\text{in}}}} = \cfrac{1}{\clW{\pVars{Q}{in}{D}}} \left( \clW{\pVars{Q}{und}{P}} \clX{\pVars{Z}{}{P_{\text{und}}}} + \clU{\pVars{Q}{und}{D_{\text{pre}}}} \clX{\pVars{Z}{}{D_{\text{pre,und}}}} \right),
\end{equation}
where $(\clW{\pVars{Q}{und}{P}}, \clX{\pVars{Z}{}{P_{\text{und}}}})$ is the underflow from the primary clarifier (Section \ref{subsec: Primary_Clarifier}) and $(\clU{\pVars{Q}{und}{D_{\text{pre}}}}, \clX{\pVars{Z}{}{D_{\text{pre,und}}}})$ is the underflow from the thickener (Section \ref{subsec: Separation_Processes}) on the wastage sludge from the secondary settler. 
This influent profile (in terms of ASM1 process variables) is then converted into the influent $\clW{\pVars{Z}{}{D_{\text{in}}}} = \mathtt{ASM2ADM}(\clW{\pVars{\widetilde Z}{}{D_{\text{in}}}})$ for the ADM1 model.
The dynamics of each component $\clX{\pVars{Z}{}{D}}$ within the anaerobic biodigester are thus modelled as
\begin{flalign}
\textstyle\frac{d}{dt}\pVars{S}{SU}{D}
    & = \admIN{S}{SU} \\ 
    & + \admP{2} + (1{-}\clP{\pVars{f}{FA,LI}{}}) \admP{4} - \admP{5} \nonumber\\
\textstyle\frac{d}{dt}\pVars{S}{AA}{D}
    & = \admIN{S}{AA} \\ 
    & + \admP{3} - \admP{6} \nonumber\\
\textstyle\frac{d}{dt}\pVars{S}{FA}{D} 
    & = \admIN{S}{FA} \\
    & + \clP{\pVars{f}{FA,LI}{}} \admP{4}
    - \admP{7} \nonumber\\ 
\textstyle\frac{d}{dt}\pVars{S}{VA}{D}
    & = \admIN{S}{VA} \\
    & + (1{-}\clP{\pVars{Y}{AA}{}})\clP{\pVars{f}{FA,LI}{}} \admP{6} 
      - \admP{8} \nonumber\\  
\textstyle\frac{d}{dt}\pVars{S}{BU}{D}
    & = \admIN{S}{BU} \\ 
    & + (1{-}\clP{\pVars{Y}{SU}{}})\clP{\pVars{f}{BU,SU}{}} \admP{5} 
      + (1{-}\clP{\pVars{Y}{AA}{}})\clP{\pVars{f}{BU,AA}{}} \admP{6} \nonumber\\
    & - \admP{9} \nonumber\\  
\textstyle\frac{d}{dt}\pVars{S}{PRO}{D}
    & = \admIN{S}{PRO} \\ 
    & + (1{-}\clP{\pVars{Y}{SU}{}})\clP{\pVars{f}{PRO,SU}{}} \admP{5} 
      + (1{-}\clP{\pVars{Y}{AA}{}})\clP{\pVars{f}{PRO,AA}{}} \admP{6} \nonumber\\
    & + 0.54(1-\clP{\pVars{Y}{C$_4$}{}}) \admP{8}
      - \admP{10} \nonumber\\  
\textstyle\frac{d}{dt}\pVars{S}{AC}{D}
    & = \admIN{S}{AC} \\ 
    & + (1{-}\clP{\pVars{Y}{SU}{}})\clP{\pVars{f}{AC,SU}{}} \admP{5} 
      + (1{-}\clP{\pVars{Y}{AA}{}})\clP{\pVars{f}{AC,AA}{}} \admP{6} \nonumber\\
    & + 0.7 (1{-}\clP{\pVars{Y}{FA}{}}) \admP{7}
      + 0.31(1{-}\clP{\pVars{Y}{C$_4$}{}}) \admP{8} \nonumber\\
    & + 0.8 (1{-}\clP{\pVars{Y}{C$_4$}{}}) \admP{9} 
      + 0.57(1{-}\clP{\pVars{Y}{PRO}{}}) \admP{10} \nonumber\\
    & - \admP{11} \nonumber\\  
\textstyle\frac{d}{dt}\pVars{S}{H$_2$}{D}
    & = \admIN{S}{H$_2$} \\ 
    & + (1{-}\clP{\pVars{Y}{SU}{}})\clP{\pVars{f}{H$_2$,SU}{}} \admP{5} 
      + (1{-}\clP{\pVars{Y}{AA}{}})\clP{\pVars{f}{H$_2$,AA}{}} \admP{6} \nonumber\\
    & + 0.3 (1{-}\clP{\pVars{Y}{FA}{}})  \admP{7}
      + 0.15(1{-}\clP{\pVars{Y}{C$_4$}{}}) \admP{8} \nonumber\\
    & + 0.2 (1{-}\clP{\pVars{Y}{C$_4$}{}}) \admP{9} 
      + 0.43(1{-}\clP{\pVars{Y}{PRO}{}}) \admP{10} \nonumber\\
    & - \admP{12} - \admP{H2} \nonumber\\  
\textstyle\frac{d}{dt}\pVars{S}{CH$_4$}{D}
    & = \admIN{S}{CH$_4$} \\ 
    & (1{-}\clP{\pVars{Y}{AC}{}}) \admP{11} + (1{-}\clP{\pVars{Y}{H$_2$}{}}) \admP{12} \nonumber\\ 
    & - \admP{CH4} \nonumber\\ 
\textstyle\frac{d}{dt}\pVars{S}{IC}{D}
    & = \admIN{S}{IC} \\ 
    & - (\clP{\pVars{f}{S$_I$,C}{}\pVars{C}{S$_I$}{}} + \clP{\pVars{f}{CH,C}{}\pVars{C}{CH}{}} + \clP{\pVars{f}{PRO,C}{}\pVars{C}{PRO}{}} + \clP{\pVars{f}{LI,C}{}\pVars{C}{LI}{}} + \clP{\pVars{f}{X$_I$,C}{}\pVars{C}{X$_I$}{}} - \clP{\pVars{C}{X$_C$}{}}) \admP{1} \nonumber\\
    & - (\clP{\pVars{C}{SU}{}} - \clP{\pVars{C}{CH}{}} ) \admP{2} + (\clP{\pVars{C}{AA}{}} - \clP{\pVars{C}{PRO}{}}) \admP{3} \nonumber\\
    & - ((1{-}\clP{\pVars{f}{FA,LI}{}})\clP{\pVars{C}{SU}{}} + \clP{\pVars{f}{FA,LI}{}\pVars{C}{FA}{}} - \clP{\pVars{C}{LI}{}}) \admP{4} \nonumber\\
    & - ((1{-}\clP{\pVars{Y}{SU}{}})(\clP{\pVars{f}{BU,SU}{} \pVars{C}{BU}{}} + \clP{\pVars{f}{PRO,SU}{} \pVars{C}{PRO}{}} + \clP{\pVars{f}{AC,SU}{} \pVars{C}{AC}{}}) + \clP{\pVars{Y}{SU}{} \pVars{C}{BAC}{}} - \clP{\pVars{C}{SU}{}}) \admP{5} \nonumber\\
    & - \Big( (1{-}\clP{\pVars{Y}{AA}{}})(\clP{\pVars{f}{VA,AA}{} \pVars{C}{VA}{}} {+} \clP{\pVars{f}{BU,AA}{} \pVars{C}{BU}{}} {+} \clP{\pVars{f}{PRO,AA}{} \pVars{C}{PRO}{}} {+} \clP{\pVars{f}{AC,AA}{} \pVars{C}{AC}{}}) \nonumber\\[-1ex]
    &       \hspace*{23em}+ \clP{\pVars{Y}{AA}{} \pVars{C}{BAC}{}} - \clP{\pVars{C}{AA}{}} \Big) \admP{6} \nonumber\\
    & - (0.7 (1{-}\clP{\pVars{Y}{FA}{}})    \clP{\pVars{C}{AC}{}}  + \clP{\pVars{Y}{FA}{} \pVars{C}{BAC}{}} - \clP{\pVars{C}{FA}{}}) \admP{7} \nonumber\\
    & - (0.54(1{-}\clP{\pVars{Y}{C$_4$}{}}) \clP{\pVars{C}{PRO}{}} + 0.31(1-\clP{\pVars{Y}{C$_4$}{}}) \clP{\pVars{C}{AC}{}} + \clP{\pVars{Y}{C$_4$}{} \pVars{C}{BAC}{}} {-} \clP{\pVars{C}{VA}{}}) \admP{8} \nonumber\\
    & - (0.8 (1{-}\clP{\pVars{Y}{C$_4$}{}}) \clP{\pVars{C}{AC}{}}  + \clP{\pVars{Y}{C$_4$}{} \pVars{C}{BAC}{}} - \clP{\pVars{C}{BU}{}}) \admP{9} \nonumber\\
    & - (0.57 (1{-}\clP{\pVars{Y}{PRO}{}})  \clP{\pVars{C}{AC}{}}  + \clP{\pVars{Y}{PRO}{} \pVars{C}{BAC}{}} - \clP{\pVars{C}{PRO}{}}) \admP{10} \nonumber\\
    & - ((1{-}\clP{\pVars{Y}{AC}{}}) \clP{\pVars{C}{CH$_4$}{}} + \clP{\pVars{Y}{AC}{} \pVars{C}{BAC}{}} - \clP{\pVars{C}{AC}{}}) \admP{11} \nonumber\\
    & - ((1{-}\clP{\pVars{Y}{H$_2$}{}}) \clP{\pVars{C}{CH$_4$}{}} + \clP{\pVars{Y}{AC}{} \pVars{C}{BAC}{}}) \admP{12} \nonumber\\
    & - (-\clP{\pVars{C}{BAC}{}} + \clP{\pVars{C}{SU}{}}) \big( \admP{13} + \admP{14} + \admP{15} + \admP{16} + \admP{17} \nonumber\\[-0.5ex]
    &       \qquad+ \admP{18} + \admP{19} \big) \nonumber\\
    & - \admP{CO2} \nonumber\\
\textstyle\frac{d}{dt}\pVars{S}{IN}{D}
    & = \admIN{S}{IN} \\ 
    & + (\clP{N_{X_C}}{-}\clP{\pVars{f}{X$_I$,C}{} N_{X_I}}{-}\clP{\pVars{f}{S$_I$,C}{} N_{X_I}}{-}\clP{\pVars{f}{PRO,C}{} \pVars{N}{AA}{}}) \admP{1} \nonumber\\
    & - \clP{\pVars{Y}{SU}{}} \clP{\pVars{N}{BAC}{}}  \admP{5} 
      + \clP{\pVars{N}{AA}{}}{-}\clP{\pVars{Y}{AA}{}} \clP{\pVars{B}{BAC}{}} \admP{6} \nonumber\\
    & - \clP{\pVars{Y}{FA}{}} \clP{\pVars{N}{BAC}{}}  \admP{7}
      - \clP{\pVars{Y}{C$_4$}{}} \clP{\pVars{N}{BAC}{}}  \admP{8} \nonumber\\
    & - \clP{\pVars{Y}{C$_4$}{}} \clP{\pVars{N}{BAC}{}}  \admP{9} 
      - \clP{\pVars{Y}{PRO}{}} \clP{\pVars{N}{BAC}{}} \admP{10} \nonumber\\
    & - \clP{\pVars{Y}{AC}{}} \clP{\pVars{N}{BAC}{}}  \admP{11} \nonumber 
      - \clP{\pVars{Y}{H$_2$}{}} \clP{\pVars{N}{BAC}{}}  \admP{12} \nonumber\\ 
    & + (\clP{\pVars{N}{BAC}{}}{-}\clP{N_{X_C}})\big[ \admP{13}+\admP{14}+\admP{15}+\admP{16}+\admP{17} \nonumber\\
    & \qquad +\admP{18}+\admP{19} \big] \nonumber\\
\textstyle\frac{d}{dt}\pVars{S}{I}{D}
    & = \admIN{S}{I} \quad + \clP{\pVars{f}{S$_I$,C}{}} \admP{1} \\
\textstyle\frac{d}{dt}\pVars{X}{C}{D}
    & = \admIN{X}{C} \quad - \admP{1} \\
    & + \admP{13} + \admP{14} + \admP{15} + \admP{16} + \admP{17} + \admP{18} + \admP{19} \nonumber\\
\textstyle\frac{d}{dt}\pVars{X}{CH}{D}
    & = \admIN{X}{CH} \quad + \clP{\pVars{f}{CH,C}{}} \admP{1} - \admP{2} \\
\textstyle\frac{d}{dt}\pVars{X}{PR}{D}
    & = \admIN{X}{PR} \quad + \clP{\pVars{f}{PRO,C}{}} \admP{1} - \admP{3} \\
\textstyle\frac{d}{dt}\pVars{X}{LI}{D}
    & = \admIN{X}{LI} \quad + \clP{\pVars{f}{LI,C}{}} \admP{1} - \admP{4} \\
\textstyle\frac{d}{dt}\pVars{X}{SU}{D}
    & = \admIN{X}{SU} \quad + \clP{\pVars{Y}{SU}{}} \admP{5} - \admP{13} \\
\textstyle\frac{d}{dt}\pVars{X}{AA}{D}
    & = \admIN{X}{AA} \quad + \clP{\pVars{Y}{AA}{}} \admP{6} - \admP{14} \\
\textstyle\frac{d}{dt}\pVars{X}{FA}{D}
    & = \admIN{X}{FA} \quad + \clP{\pVars{Y}{FA}{}} \admP{7} - \admP{15} \\
\textstyle\frac{d}{dt}\pVars{X}{C$_4$}{D}
    & = \admIN{X}{C$_4$} \\
    & + \clP{\pVars{Y}{C$_4$}{}} \admP{8} + \clP{\pVars{Y}{C$_4$}{}} \admP{9} - \admP{16} \nonumber\\
\textstyle\frac{d}{dt}\pVars{X}{PRO}{D}
    & = \admIN{X}{PRO} \quad + \clP{\pVars{Y}{PRO}{}} \admP{10} - \admP{17} \\
\textstyle\frac{d}{dt}\pVars{X}{AC}{D}
    & = \admIN{X}{AC} \quad + \clP{\pVars{Y}{AC}{}} \admP{11} - \admP{18} \\
\textstyle\frac{d}{dt}\pVars{X}{H$_2$}{D}
    & = \admIN{X}{H$_2$} \quad + \clP{\pVars{Y}{H$_2$}{}} \admP{12} - \admP{19} \\
\textstyle\frac{d}{dt}\pVars{X}{I}{D}
    & = \admIN{X}{I} \quad + \clP{\pVars{f}{X$_I$,C}{}} \admP{1} \\
\textstyle\frac{d}{dt}\pVars{S}{CAT$^+$}{D}
    & = \admIN{S}{CAT$^+$} \\
\textstyle\frac{d}{dt}\pVars{S}{AN$^-$}{D}
    & = \admIN{S}{AN$^-$} \\
\textstyle\frac{d}{dt}\pVars{S}{VA$^-$}{D}
    & = \admIN{S}{VA$^-$} \quad - \clP{\pVars{k}{VA}{A/B}}\left((\clP{\pVars{K}{VA}{A}} + \Phi_{\text{ion}}(\clX{x^{D}}))\admX{27} - \clP{\pVars{K}{VA}{A}}\admX{4} \right) \\
\textstyle\frac{d}{dt}\pVars{S}{BU$^-$}{D}
    & = \admIN{S}{BU$^-$} \quad - \clP{\pVars{k}{BU}{A/B}}\left((\clP{\pVars{K}{BU}{A}} + \Phi_{\text{ion}}(\clX{x^{D}}))\admX{28} - \clP{\pVars{K}{BU}{A}}\admX{5} \right) \\
\textstyle\frac{d}{dt}\pVars{S}{PRO$^-$}{D}
    & = \admIN{S}{PRO$^-$} \quad - \clP{\pVars{k}{PRO}{A/B}}\left((\clP{\pVars{K}{PRO}{A}} + \Phi_{\text{ion}}(\clX{x^{D}}))\admX{29} - \clP{\pVars{K}{PRO}{A}}\admX{6} \right) \\
\textstyle\frac{d}{dt}\pVars{S}{AC$^-$}{D}
    & = \admIN{S}{AC$^-$} \quad - \clP{\pVars{k}{AC}{A/B}}\left((\clP{\pVars{K}{AC}{A}} + \Phi_{\text{ion}}(\clX{x^{D}}))\admX{30} - \clP{\pVars{K}{AC}{A}}\admX{7} \right) \\
\textstyle\frac{d}{dt}\pVars{S}{HCO$_3^-$}{D}
    & = \admIN{S}{HCO$_3^-$} \quad - \clP{\pVars{k}{CO$_2$}{A/B}}\left((\clP{\pVars{K}{CO$_2$}{A}} + \Phi_{\text{ion}}(\clX{x^{D}}))\admX{31} - \clP{\pVars{K}{CO$_2$}{A}}\admX{10} \right) \\
\textstyle\frac{d}{dt}\pVars{S}{NH$_3$}{D}
    & = \admIN{S}{NH$_3$} \quad - \clP{\pVars{k}{IN}{A/B}}\left((\clP{\pVars{K}{IN}{A}} + \Phi_{\text{ion}}(\clX{x^{D}}))\admX{32} - \clP{\pVars{K}{IN}{A}}\admX{11} \right) \\
\textstyle\frac{d}{dt}\pVars{G}{H$_2$}{D}
    & = \cfrac{ 1 }{ \clP{V^{D}_{\text{gas}}} } \left[ \clP{V^{D}_{\text{liq}}} \admP{H2} - Q_G(\clX{x^{D}}) \clX{\pVars{G}{H$_2$}{D}} \right]   \\
\textstyle\frac{d}{dt}\pVars{G}{CH$_4$}{D}
    & = \cfrac{ 1 }{ \clP{V^{D}_{\text{gas}}} } \left[ \clP{V^{D}_{\text{liq}}} \admP{CH4} - Q_G(\clX{x^{D}}) \clX{\pVars{G}{CH$_4$}{D}} \right]   \\
\textstyle\frac{d}{dt}\pVars{G}{CO$_2$}{D}
    & = \cfrac{ 1 }{ \clP{V^{D}_{\text{gas}}} } \left[ \clP{V^{D}_{\text{liq}}} \admP{CO2} - Q_G(\clX{x^{D}}) \clX{\pVars{G}{CO$_2$}{D}}  \right] 
\end{flalign}
The physico-chemical parameters are assumed independent of the temperature in the digester: 
The exceptions are the parameters $\clP{K_{(\cdot)}} \in \{ \clP{\pVars{K}{W}{}}, \clP{\pVars{K}{CO$_2$}{}}, \clP{\pVars{K}{IN}{}}, \clP{\pVars{K}{CO$_2$}{H}}, \clP{\pVars{K}{CH$_4$}{H}}, \clP{\pVars{K}{H$_2$}{H}}, \clP{\pVars{P}{H$_2$O}{}} \}$, which are corrected as 
\begin{equation*}
    \clP{K_{(\cdot)}} = \clP{\pVars{K}{$(\cdot)$,base}{}} \exp\left[ \frac{\clP{\pVars{\Delta H}{$(\cdot)$,base}{}}}{100\clP{R}} (\clP{\pVars{T}{base}{-1}} - \clP{\pVars{T}{ad}{-1}}) \right].
\end{equation*}
The differential equations above also depend on the inhibition rates $I : (\clX{x^{D}(t)}) \to (I_{1,\ldots,6}(t))$, defined as 
\begin{flalign} \label{eq: Inhibition_Rates}
    I_1(\clX{x^{D}}) & = \cfrac{\admX{11}}{\clP{\pVars{K}{S,IN}{}} + \admX{11}}~\cfrac{(\clP{\pVars{K}{pH}{}})^{n_A}}{\Phi_{\text{ion}}(\clX{x^{D}})^{\clP{n_{AA}}} + (\clP{\pVars{K}{pH}{}})^{n_A}} \\ 
    I_2(\clX{x^{D}}) & = \cfrac{\clP{\pVars{K}{H$_2$,FA}{\text{inh}}}}{\clP{\pVars{K}{H$_2$,FA}{\text{inh}}} + \admX{8}}~\cfrac{\admX{11}}{\clP{\pVars{K}{S,IN}{}} + \admX{11}}~\cfrac{(\clP{\pVars{K}{pH}{}})^{n_A}}{\Phi_{\text{ion}}(\clX{x^{D}})^{\clP{n_{AA}}} + (\clP{\pVars{K}{pH}{}})^{n_A}} \\
    I_3(\clX{x^{D}}) & = \cfrac{\clP{\pVars{K}{H$_2$,C$_4$}{\text{inh}}}}{\clP{\pVars{K}{H$_2$,C$_4$}{\text{inh}}} + \admX{8}}~\cfrac{\admX{11}}{\clP{\pVars{K}{S,IN}{}} + \admX{11}}~\cfrac{(\clP{\pVars{K}{pH}{}})^{n_A}}{\Phi_{\text{ion}}(\clX{x^{D}})^{\clP{n_{AA}}} + (\clP{\pVars{K}{pH}{}})^{n_A}} \\
    I_4(\clX{x^{D}}) & = \cfrac{\clP{\pVars{K}{H$_2$,PRO}{\text{inh}}}}{\clP{\pVars{K}{H$_2$,PRO}{\text{inh}}} + \admX{8}}~\cfrac{\admX{11}}{\clP{\pVars{K}{S,IN}{}} + \admX{11}}~\cfrac{(\clP{\pVars{K}{pH}{}})^{n_A}}{\Phi_{\text{ion}}(\clX{x^{D}})^{\clP{n_{AA}}} + (\clP{\pVars{K}{pH}{}})^{n_A}} \\
    I_5(\clX{x^{D}}) & = \cfrac{\clP{\pVars{K}{NH$_3$}{\text{inh}}}}{\clP{\pVars{K}{NH$_3$}{\text{inh}}} + \admX{32}}~\cfrac{\admX{11}}{\clP{\pVars{K}{S,IN}{}} + \admX{11}}~\cfrac{(\clP{\pVars{K}{pH}{}})^{n_{AC}}}{\Phi_{\text{ion}}(\clX{x^{D}})^{\clP{n_{AC}}} + (\clP{\pVars{K}{pH}{}})^{n_{AC}}} \\
    I_6(\clX{x^{D}}) & = \cfrac{\admX{11}}{\clP{\pVars{K}{S,IN}{}} + \admX{11}}~\cfrac{(\clP{\pVars{K}{pH}{}})^{n_{H_2}}}{\Phi_{\text{ion}}(\clX{x^{D}})^{\clP{n_{H_2}}} + (\clP{\pVars{K}{pH}{}})^{n_{H_2}}},
\end{flalign}
with $\clP{\pVars{K}{pH}{}} = 10^{-0.5(\clP{\pVars{\text{pH}}{UL,H$_2$}{}} + \clP{\pVars{\text{pH}}{LL,H$_2$}{}})}$, $\clP{n_{AA}} = 3(\clP{\pVars{\text{pH}}{UL,AA}{}}-\clP{\pVars{\text{pH}}{LL,AA}{}})^{-1}$, $\clP{n_{AC}} = 3(\clP{\pVars{\text{pH}}{UL,AC}{}}-\clP{\pVars{\text{pH}}{LL,AC}{}})^{-1}$, and $\clP{n_{H_2}} = 3(\clP{\pVars{\text{pH}}{UL,H$_2$}{}}-\clP{\pVars{\text{pH}}{LL,H$_2$}{}})^{-1}$. 
The ionization rate $\Phi_{\text{ion}} : \mathbb{R}^{35}_{\geq 0} \to \mathbb{R}$ is given by 
\begin{equation} \label{eq: Ionization_Rates}
    \Phi_{\text{ion}}(\clX{x^D}) = \frac{\sqrt{\clX{\phi}^2 + 4\clP{K_{w}}} - \clX{\phi}}{2}, \quad 
        \clX{\phi} = \admX{11}{+}\admX{25}{-}\admX{26}{-}\frac{\admX{27}}{208}{-}\frac{\admX{28}}{160}{-}\frac{\admX{29}}{112}{-}\frac{\admX{30}}{64}{-}\admX{31}{-}\admX{32}.
\end{equation} 
Finally, the effluent gas flow-rate $Q_G(\cdot)$ is computed from the concentrations inside the bioreactor, 
\begin{equation} \label{eq: Gas_Outflow}
    Q_G(\clX{x^D}) = \max\left\{ 0,\ \clP{k_P}\left[ \clP{R}~\clP{\pVars{T}{ad}{}}\left( \cfrac{\admX{33}}{16} + \cfrac{\admX{34}}{64} + \admX{35} \right) + \clP{\pVars{P}{H$_2$O}{}} - \clP{P_{\text{atm}}} \right] \right\}.
\end{equation} 

The effluent of the anaerobic digester (in terms of ADM1 process variables) is expressed (in terms of ASM1 process variables) as the stream $(\clW{\pVars{Q}{eff}{D}}, \clX{\pVars{Z}{}{D_{\text{eff}}}})$ with $\clW{\pVars{Q}{eff}{D}} = \clW{\pVars{Q}{in}{D}}$ and $\clX{\pVars{Z}{}{D_{\text{eff}}}} = \mathtt{ADM2ASM}(\clX{\pVars{Z}{}{D}})$.
The digester is monitored through the measurable quantities $\clY{y^D} = (\clY{\pVars{\text{TSS}}{}{D_{\text{eff}}}}, \clY{\pVars{Q}{G}{}}, \clY{\pVars{G}{CH$_4$}{D}})$, with 
    $$\clY{\pVars{\text{TSS}}{}{D_{\text{eff}}}} = 0.75\big( \clX{\pVars{X}{I}{D_{\text{eff}}}} + \clX{\pVars{X}{S}{D_{\text{eff}}}} + \clX{\pVars{X}{BH}{D_{\text{eff}}}} + \clX{\pVars{X}{BA}{D_{\text{eff}}}} + \clX{\pVars{X}{P}{D_{\text{eff}}}} \big).$$
The constant model parameters for this unit are reported in Table \ref{tab: Parameters_Digester_A}.

\begin{table}[htb!] \centering
    \caption{Anaerobic Digester (ADM1): Model constant parameters.}
    \label{tab: Parameters_Digester_A}
    \smallskip
    {\small\begin{tabular}{@{}p{4em}p{26em}p{7em}p{8em}@{}}\toprule
                                            & General parameter					                & Value 	    & Units				\\\midrule                
        $\clP{\pVars{V}{liq}{D}}$		    & Volume of liquid phase (digester)		            & $3400$ 	    & m$^{3}$		    \\
        $\clP{\pVars{V}{gas}{D}}$	        & Volume of gaseous phase (digester) 		        & $300$	        & m$^{3}$		    \\
        $\clP{R}$	                        & Universal gas constant                            & $0.083145$    & bar/(M$\cdot$K)    \\
        $\clP{\pVars{P}{atm}{}}$	        & Atmospheric pressure                              & $1.013$       & bar    \\
        $\clP{\pVars{T}{base}{}}$	        & Reference temperature for the Van't Hoff equation & $298.15$      & K    \\
        $\clP{\pVars{T}{ad}{}}$	            & Operational temperature of the digester	        & $308.15$      & K    \\[1ex]\midrule
                                            & Stoichiometric parameter		                & Value 	    & Units						                                        \\\midrule                
        $\clP{\pVars{f}{H$_2$,SU}{}}$       & Hydrogen from sugars                          & 0.1900        & -- \\    
        $\clP{\pVars{f}{BU,SU}{}}$          & Butyrate from sugars                          & 0.1300        & -- \\    
        $\clP{\pVars{f}{PRO,SU}{}}$         & Propionate from sugars                        & 0.2700        & -- \\    
        $\clP{\pVars{f}{AC,SU}{}}$          & Acetate from sugars                           & 0.4100        & -- \\    
        $\clP{\pVars{f}{VA,AA}{}}$          & Valerate from amino acids                     & 0.2300        & -- \\    
        $\clP{\pVars{f}{H$_2$,AA}{}}$       & Hydrogen from amino acids                     & 0.0600        & -- \\    
        $\clP{\pVars{f}{BU,AA}{}}$          & Butyrate from amino acids                     & 0.2600        & -- \\    
        $\clP{\pVars{f}{PRO,AA}{}}$         & Propionate from amino acids                   & 0.0500        & -- \\    
        $\clP{\pVars{f}{AC,AA}{}}$          & Acetate from amino acids                      & 0.4000        & -- \\    
        $\clP{\pVars{f}{S$_I$,C}{}}$        & Soluble inerts from composites                & 0.1000        & -- \\
        $\clP{\pVars{f}{X$_I$,C}{}}$        & Particulate inerts from composites            & 0.2000        & -- \\
        $\clP{\pVars{f}{CH,C}{}}$           & Carbohydrates from composites                 & 0.2000        & -- \\
        $\clP{\pVars{f}{PR,C}{}}$           & Proteins from composites                      & 0.2000        & -- \\
        $\clP{\pVars{f}{LI,C}{}}$           & Lipids from composites                        & 0.3000        & -- \\
        $\clP{\pVars{f}{FA,LI}{}}$          & Fatty acids from lipids                       & 0.9500        & -- \\    
        $\clP{\pVars{N} {I}     {}}$        & Nitrogen content of inerts                    & 0.0600/14.0   & kmol\,N/(kg\,COD) \\ 
        $\clP{\pVars{N} {BAC}   {}}$        & Nitrogen content of bacteria                  & 0.0800/14.0   & kmol\,N/(kg\,COD) \\ 
        $\clP{\pVars{N} {X$_C$} {}}$        & Nitrogen content of composites                & 0.0376/14.0   & kmol\,N/(kg\,COD) \\ 
        $\clP{\pVars{N} {AA}    {}}$        & Nitrogen content of amino acids and proteins  & 0.0070        & kmol\,N/(kg\,COD) \\ 
        $\clP{\pVars{C} {SU}    {}}$        & Carbon content of monosaccharides             & 0.0313        & kmol\,C/(kg\,COD) \\ 
        \bottomrule
        & & & \hfill (Continued)
    \end{tabular}}
\end{table}
\addtocounter{table}{-1}
\begin{table}[htb!] \centering
    \caption{Anaerobic Digester (ADM1): Model constant parameters (continued).}
    \label{tab: Parameters_Digester_B}
    \smallskip
    {\small\begin{tabular}{@{}p{5em}p{26em}p{7em}p{8em}@{}}\toprule
                                                    & Stoichiometric parameter		                                     & Value 	            & Units						                                        \\\midrule                
        $\clP{\pVars{C} {AA}    {}}$        & Carbon content of amino acids                 & 0.0300        & kmol\,C/(kg\,COD) \\ 
        $\clP{\pVars{C} {FA}    {}}$        & Carbon content of long chain fatty acids      & 0.0217        & kmol\,C/(kg\,COD) \\ 
        $\clP{\pVars{C} {VA}    {}}$        & Carbon content of total valerate              & 0.0240        & kmol\,C/(kg\,COD) \\ 
        $\clP{\pVars{C} {BU}    {}}$        & Carbon content of total butyrate              & 0.0250        & kmol\,C/(kg\,COD) \\ 
        $\clP{\pVars{C} {PRO}   {}}$        & Carbon content of total propinate             & 0.0268        & kmol\,C/(kg\,COD) \\ 
        $\clP{\pVars{C} {AC}    {}}$        & Carbon content of total acetate               & 0.0313        & kmol\,C/(kg\,COD) \\ 
        $\clP{\pVars{C} {CH$_4$}{}}$        & Carbon content of methane                     & 0.0156        & kmol\,C/(kg\,COD) \\ 
        $\clP{\pVars{C} {X$_C$} {}}$        & Carbon content of composites                  & 0.02786       & kmol\,C/(kg\,COD) \\ 
        $\clP{\pVars{C} {S$_I$} {}}$                & Carbon content of soluble inerts              & 0.0300        & kmol\,C/(kg\,COD) \\ 
        $\clP{\pVars{C} {CH}    {}}$                & Carbon content of carbohydrates               & 0.0313        & kmol\,C/(kg\,COD) \\ 
        $\clP{\pVars{C} {PR}    {}}$                & Carbon content of proteins                    & 0.0300        & kmol\,C/(kg\,COD) \\ 
        $\clP{\pVars{C} {LI}    {}}$                & Carbon content of lipids                      & 0.0220        & kmol\,C/(kg\,COD) \\ 
        $\clP{\pVars{C} {X$_I$} {}}$                & Carbon content of particulate inerts                               & 0.0300               & kmol\,C/(kg\,COD) \\ 
        $\clP{\pVars{C} {BAC}   {}}$                & Carbon content of bacteria                                         & 0.0313               & kmol\,C/(kg\,COD) \\ 
        $\clP{\pVars{Y} {SU}    {}}$                & Yield of monosaccharides                                           & 0.1000               & kg\,COD/(kg\,COD) \\ 
        $\clP{\pVars{Y} {AA}    {}}$                & Yield of amino acids                                               & 0.0800               & kg\,COD/(kg\,COD) \\ 
        $\clP{\pVars{Y} {FA}    {}}$                & Yield of long chain fatty acids                                    & 0.0600               & kg\,COD/(kg\,COD) \\ 
        $\clP{\pVars{Y} {PRO}   {}}$                & Yield of propinate                                                 & 0.0400               & kg\,COD/(kg\,COD) \\ 
        $\clP{\pVars{Y} {C$_4$} {}}$                & Yield of valerate and butyrate degraders                           & 0.0600               & kg\,COD/(kg\,COD) \\ 
        $\clP{\pVars{Y} {AC}    {}}$                & Yield of total acetate                                             & 0.0500               & kg\,COD/(kg\,COD) \\ 
        $\clP{\pVars{Y} {H$_2$} {}}$                & Yield of hydrogen                                                  & 0.0600               & kg\,COD/(kg\,COD) \\[1ex]\midrule
                                                    & Kinetic parameter					                                 & Value 	            & Units						                                \\\midrule
        $\clP{\pVars{k}{dis}{}}$                    & Disintegration rate                                                &  0.50                &       1/d \\ 
        $\clP{\pVars{k}{CH}{hyd}}$                  & First-order hydrolysis rate (carbohydrates)                        & 10.0                 & 1/d \\ 
        $\clP{\pVars{k}{PR}{hyd}}$                  & First-order hydrolysis rate (proteins)                             & 10.0                 & 1/d \\ 
        $\clP{\pVars{k}{LI}{hyd}}$                  & First-order hydrolysis rate (lipids)                               & 10.0                 & 1/d \\ 
        $\clP{\pVars{k}{SU}{\text{dec}}}$           & First-order decay rate (monosaccharides)                           &  0.02                & 1/d \\ 
        $\clP{\pVars{k}{AA}{\text{dec}}}$           & First-order decay rate (amino acids)                               &  0.02                & 1/d \\ 
        $\clP{\pVars{k}{FA}{\text{dec}}}$           & First-order decay rate (long chain fatty acids)                    &  0.02                & 1/d \\ 
        $\clP{\pVars{k}{C$_4$}{\text{dec}}}$        & First-order decay rate (valerate and butyrate degraders)           &  0.02                & 1/d \\ 
        $\clP{\pVars{k}{PRO}{\text{dec}}}$          & First-order decay rate (total propinate)                           &  0.02                & 1/d \\ 
        $\clP{\pVars{k}{AC}{\text{dec}}}$           & First-order decay rate (total acetate)                             &  0.02                & 1/d \\ 
        $\clP{\pVars{k}{H$_2$}{\text{dec}}}$        & First-order decay rate (hydrogen)                                  &  0.02                & 1/d \\ 
        $\clP{\pVars{K}{SU}{}}$                     & Half-saturation (monosaccharides)                                  &  0.50                & kg\,COD/m$^{3}$ \\ 
        $\clP{\pVars{K}{AA}{}}$                     & Half-saturation (amino acids)                                      &  0.30                & kg\,COD/m$^{3}$ \\ 
        $\clP{\pVars{K}{FA}{}}$                     & Half-saturation (long chain fatty acids)                           &  0.40                & kg\,COD/m$^{3}$ \\ 
        $\clP{\pVars{K}{C$_4$}{}}$                  & Half-saturation (valerate and butyrate degraders)                  &  0.20                & kg\,COD/m$^{3}$ \\ 
        $\clP{\pVars{K}{PRO}{}}$                    & Half-saturation (total propinate)                                  &  0.10                & kg\,COD/m$^{3}$ \\ 
        $\clP{\pVars{K}{AC}{}}$                     & Half-saturation (total acetate)                                    &  0.15                & kg\,COD/m$^{3}$ \\ 
        $\clP{\pVars{K}{H$_2$}{}}$                  & Half-saturation (hydrogen)                                         & 7$\cdot 10^{-6}$     & kg\,COD/m$^{3}$ \\ 
        $\clP{\pVars{K}{S,IN}{}}$                   & Half-saturation (inorganic nitrogen)                               &  1$\cdot 10^{-4}$    & kmol\,N/m$^3$ \\ 
        $\clP{\pVars{K}{H$_2$,FA}{\text{inh}}}$     & 50\% inhibitory concentration (H$_2$ to fatty acids)               & 5$\cdot 10^{-6}$     & kg\,COD/m$^3$ \\ 
        $\clP{\pVars{K}{H$_2$,C$_4$}{\text{inh}}}$  & 50\% inhibitory concentration (H$_2$ to C$_4$)                     & 1$\cdot 10^{-5}$     & kg\,COD/m$^3$ \\ 
        $\clP{\pVars{K}{H$_2$,PRO}{\text{inh}}}$    & 50\% inhibitory concentration (H$_2$ to propinate)                 & 3.5$\cdot 10^{-5}$   & kg\,COD/m$^3$ \\ 
        $\clP{\pVars{K}{NH$_3$}{\text{inh}}}$       & 50\% inhibitory concentration (ammonia)                            &  0.0018              & kmol\,NH$_4$-N/m$^3$ \\ 
        $\clP{\pVars{k}{SU}{m}}$                    & Monod maximum uptake rate (monosaccharides)                        & 30.0                 & 1/d \\ 
        $\clP{\pVars{k}{AA}{m}}$                    & Monod maximum uptake rate (amino acids)                            & 50.0                 & 1/d \\ 
        \bottomrule
        & & & \hfill (Continued)
    \end{tabular}}
\end{table}
\addtocounter{table}{-1}
\begin{table}[htb!] \centering
    \caption{Anaerobic Digester (ADM1): Model constant parameters (continued).}
    \label{tab: Parameters_Digester_C}
    \smallskip
    {\small\begin{tabular}{@{}p{5em}p{28em}p{6em}p{7em}@{}}\toprule
                                                    & Stoichiometric parameter		                                     & Value 	            & Units						                                        \\\midrule                
        $\clP{\pVars{k}{FA}{m}}$                    & Monod maximum uptake rate (long chain fatty acids)                 &  6.00                & 1/d \\ 
        $\clP{\pVars{k}{C$_4$}{m}}$                 & Monod maximum uptake rate (C$_4$)                                  & 20.0                 & 1/d \\         $\clP{\pVars{k}{PRO}{m}}$                   & Monod maximum uptake rate (total propinate)                        & 13.0                 & 1/d \\ 
        $\clP{\pVars{k}{AC}{m}}$                    & Monod maximum uptake rate (total acetate)                          &  8.00                & 1/d \\ 
        $\clP{\pVars{k}{H$_2$}{m}}$                 & Monod maximum uptake rate (hydrogen)                               & 35.0                 & 1/d \\         $\clP{\pVars{\text{pH}}{UL,AC}{}}$          & Upper limit of pH for 50\% inhibition (total acetate)              &  7.00                & -- \\ 
        $\clP{\pVars{\text{pH}}{LL,AC}{}}$          & Lower limit of pH for 50\% inhibition (total acetate)              &  6.00                & -- \\ 
        $\clP{\pVars{\text{pH}}{UL,AA}{}}$          & Upper limit of pH for 50\% inhibition (amino acids)                &  5.50                & -- \\         $\clP{\pVars{\text{pH}}{LL,AA}{}}$          & Lower limit of pH for 50\% inhibition (amino acids)                &  4.00                & -- \\ 
        $\clP{\pVars{\text{pH}}{UL,H$_2$}{}}$       & Upper limit of pH for 50\% inhibition (hydrogen)                   &  6.00                & -- \\ 
        $\clP{\pVars{\text{pH}}{LL,H$_2$}{}}$       & Lower limit of pH for 50\% inhibition (hydrogen)                   &  5.00                & -- \\[1ex]\midrule
                                                    & Physico-chemical parameter					                     & Value 	            & Units	\\\midrule
        $\clP{\pVars{K}{W,base}{A}}$                & Acid-base equilibrium constant (hydroxide)                         & $10^{-14.0}$         & M \\
        $\clP{\pVars{K}{CO$_2$,base}{A}}$           & Acid-base equilibrium constant (bicarbonate)                       & $10^{-6.35}$         & M \\
        $\clP{\pVars{K}{IN,base}{A}}$               & Acid-base equilibrium constant (inorganic nitrogen)                & $10^{-9.25}$         & M \\
        $\clP{\pVars{K}{VA}{A}}$                    & Acid-base equilibrium constant (total valerate)                    & $10^{-4.86}$         & M \\
        $\clP{\pVars{K}{BU}{A}}$                    & Acid-base equilibrium constant (total butyrate)                    & $10^{-4.82}$         & M \\
        $\clP{\pVars{K}{PRO}{A}}$                   & Acid-base equilibrium constant (total propinate)                   & $10^{-4.88}$         & M \\
        $\clP{\pVars{K}{AC}{A}}$                    & Acid-base equilibrium constant (total acetate)                     & $10^{-4.76}$         & M \\
        $\clP{\pVars{K}{VA}{A/B}}$                  & Acid-base kinetic parameter (total valerate)                       & 10$^{10}$            & 1/(M$\cdot$d) \\
        $\clP{\pVars{K}{BU}{A/B}}$                  & Acid-base kinetic parameter (total butyrate)                       & 10$^{10}$            & 1/(M$\cdot$d) \\
        $\clP{\pVars{K}{PRO}{A/B}}$                 & Acid-base kinetic parameter (total propinate)                      & 10$^{10}$            & 1/(M$\cdot$d) \\
        $\clP{\pVars{K}{AC}{A/B}}$                  & Acid-base kinetic parameter (total acetate)                        & 10$^{10}$            & 1/(M$\cdot$d) \\
        $\clP{\pVars{K}{CO$_2$}{A/B}}$              & Acid-base kinetic parameter (carbon dioxide)                       & 10$^{10}$            & 1/(M$\cdot$d) \\
        $\clP{\pVars{K}{IN}{A/B}}$                  & Acid-base kinetic parameter (inorganic nitrogen)                   & 10$^{10}$            & 1/(M$\cdot$d) \\
        $\clP{\pVars{K}{CO$_2$,base}{H}}$           & Henry's law coefficient (carbon dioxide)                           & 0.035                & M$_{\text{liq}}$/bar \\
        $\clP{\pVars{K}{CH$_4$,base}{H}}$           & Henry's law coefficient (methane)                                  & 0.0014               & M$_{\text{liq}}$/bar \\
        $\clP{\pVars{K}{H$_2$,base}{H}}$            & Henry's law coefficient (hydrogen)                                 & 7.8$\cdot 10^{-4}$   & M$_{\text{liq}}$/bar \\
        $\clP{\pVars{P}{H$_2$O,base}{}}$            & Water vapour saturation pressure coefficient                       & 0.0313               & bar \\
        $\clP{\pVars{\Delta H}{W}{A}}$              & Standard enthalpy change for reaction (hydroxide)                  & 55900                & -- \\
        $\clP{\pVars{\Delta H}{CO$_2$}{A}}$         & Standard enthalpy change for reaction (bicarbonate)                & 7646                 & -- \\
        $\clP{\pVars{\Delta H}{IN}{A}}$             & Standard enthalpy change for reaction (inorganic nitrogen)         & 51965                & -- \\
        $\clP{\pVars{\Delta H}{CO$_2$}{H}}$         & Standard enthalpy change for reaction (carbon dioxide)             & -4180                & -- \\
        $\clP{\pVars{\Delta H}{CH$_4$}{H}}$         & Standard enthalpy change for reaction (methane)                    & -14240               & -- \\
        $\clP{\pVars{\Delta H}{H$_2$}{H}}$          & Standard enthalpy change for reaction (hydrogen)                   & -19410               & -- \\
        $\clP{\pVars{\Delta H}{H$_2$O}{}}$          & Standard enthalpy change for reaction (water vapour)               & 5290$\cdot$\clP{R}   & -- \\
        $\clP{\pVars{K_La}{}{D}}$                   & Oxygen transfer coefficient (digester)                             & 200.0                & 1/d \\
        $\clP{k_P}$                                 & Pipe resistance coefficient                                        & 5$\cdot 10^{4}$      & m$^3$/(bar$\cdot$d) \\
        \bottomrule
    \end{tabular}}
\end{table}

\subsubsection*{The \texttt{ASM2ADM} and \texttt{ADM2ASM} interfaces}

The dynamical model of the anaerobic digester (ADM1) is integrated with the dynamical model of the remaining units through the \texttt{ASM2ADM} and \texttt{ADM2ASM} interfaces: 
These are routines that convert between equivalent state-components, mostly by distributing mass according to the target process representation. 
For the explicit equations, we refer to (Nopens \textit{et~al~}, 2009) \cite{Nopens2009} and the official implementation of the BSM2 \cite{WWTModels}.

\subsection{Reject water storage tank} \label{subsec: Storage_Tank}

The dynamics of the volume $\clX{V^{R}}$ and each component $\clX{Z^{R}}$ in the reject water storage tank unit are 
\begin{align}
    \textstyle\frac{d}{dt}{V}^{R} &= \clW{\pVars{Q}{eff}{D_{\text{post}}}} - \clU{Q_{R}}; \\
    \textstyle\frac{d}{dt}{Z}^{R} &= \cfrac{1}{\clX{V^{R}}} \left[ \clW{\pVars{Q}{eff}{D_{\text{post}}}}\clX{\pVars{Z}{}{D_{\text{post,eff}}}} - \clU{Q_{R}}\clX{Z^{R}} \right].
\end{align}
This unit is monitored through the measurable quantities $\clY{y^R} = (\clX{V^R}, \clX{\pVars{S}{NH}{R}})$

\subsection{Key performance indicators (KPIs)} \label{subsec: KPIs}

The key performance indicators (KPIs) are computed from the process state as 
\begin{flalign}
\clY{\text{TSS}^{\text{eff}}} &= \clX{\text{TSS}^{S_{10}}}; \\
\clY{\text{BOD}_5^{\text{eff}}} &= 0.25(\pVars{S}{S}{S_{10}} + \pVars{X}{S}{S_{10}} + 0.8(\pVars{X}{BH}{S_{10}} + \pVars{X}{BA}{S_{10}}));\\
\clY{\text{TN}^{\text{eff}}} &= \pVars{S}{NO}{S_{10}} + \pVars{S}{NH}{S_{10}} + \pVars{S}{ND}{S_{10}} + \pVars{X}{ND}{S_{10}} + \clP{\pVars{i}{XB}{}}(\pVars{X}{BH}{S_{10}} + \pVars{X}{BA}{S_{10}}) + \clP{\pVars{i}{XP}{}}(\pVars{X}{I}{S_{10}} + \pVars{X}{P}{S_{10}}); \\
\clY{\text{ECI}} &= \text{AE} + \text{PE} + \text{ME} - \clP{\eta_{E}} \text{MP} + \max(0, \text{HE} - \clP{\eta_{H}}\text{MP}), 
\end{flalign}
with methane conversion efficiencies $\clP{\eta_E} = 6$ kWh/kg\,CH$_4$ and $\clP{\eta_H} = 7$ kWh/kg\,CH$_4$. 
The particulate matter in the settler layer $S_{10}$ is computed as $\clX{\pVars{X}{$(\cdot)$}{S_{10}}} = (\text{TSS}^{S_{10}}/\text{TSS}^{A_{5}})\clX{\pVars{X}{$(\cdot)$}{A_{5}}}$.
The aeration (AE), pumping (PE), mixing (ME), and heating (HE) energy, and the methane production (MP) are computed as
\begin{flalign}
\clY{\text{AE}} &= \frac{\clP{\pVars{S}{O}{\text{sat}}}}{1.8\cdot1000} \sum_{r=1}^5 \clP{V^{A_r}}\clU{\pVars{K_La}{}{A_r}}; \\
\clY{\text{PE}} &= \frac{1}{1000}\left( 4\clU{Q_A} + 8\clU{Q_R} + 50\clU{Q_W} + 75\clW{\pVars{Q}{und}{P}} + 60\clU{\pVars{Q}{und}{D_{\text{pre}}}} + 4\clW{\pVars{Q}{eff}{D_{\text{post}}}} + 4\clU{Q_R} \right); \\
\clY{\text{ME}} &= \frac{120}{1000}\left( \clP{\pVars{V}{liq}{D}} + \sum_{r=1}^5 \big(1-H(\clU{\pVars{K_La}{}{A_r}} {-} 20) \big)\pVars{V}{}{A_r} \right); \\
\clY{\text{HE}} &= \frac{24 \cdot 4186}{86.4 \cdot 1000}  \big((\clP{\pVars{T}{ad}{}}-273.15) - \clW{\pVars{T}{}{D_{\text{in}}}}\big)\clW{\pVars{Q}{in}{D}}; \\
\clY{\text{MP}} &= \frac{16\clP{\pVars{P}{atm}{}}}{\clP{R}\clP{\pVars{T}{ad}{}}} \frac{\clX{\pVars{G}{CH$_4$}{D}}/64}{\clX{\pVars{G}{H$_2$}{D}}/16 + \clX{\pVars{G}{CH$_4$}{D}}/64 + \clX{\pVars{G}{CO$_2$}{D}} + \clP{\pVars{P}{H$_2$O}{}}/(\clP{R}\clP{\pVars{T}{ad}{}})} Q_G(\clX{x^D}). 
\end{flalign}

The parameters used for computing the KPIs are as reported in Tables \ref{tab: Parameters_PrimaryClarifier}--\ref{tab: Parameters_Digester_A}.

\subsection{Nominal operating conditions}

In Tables \ref{tab: Operating_Point_A}-\ref{tab: Operating_Point_B}, we present the reference nominal operating point $p^{\text{ref}} = (\clX{\bar x^{\text{ref}}}, \clU{\bar u^{\text{ref}}}, \clW{\bar w^{\text{ref}}})$ used in this work both for the tuning of the Output MPC (Section 3, main text) and for the initial conditions of the process simulation (Section 4, main text). 
This fixed point refers to a (pseudo)steady-state reached when the process is operated in open-loop with the controllable inputs and disturbances kept constant and equal to the reported values \cite{Gernaey2014}. 
The open-loop controllable inputs are defined based on expert knowledge and conventional practices. 
The constant disturbances correspond to a constant influent scenario obtained by the flow-weighted averaging of the first-year of wastewater data (see Figure 3, main text). 

\subsection{Smoothification of discontinuities}

The numerics of the Output MPC require the functions $f(\cdot)$, $g(\cdot)$, and $h(\cdot)$, to be differentiable with respect to the state, controls, and disturbances. 
As such, this work considers a representation of the BSM2 model obtained by replacing discontinuities in the state- and output-equations with smooth approximations: 
The $\min(\cdot)$ and $\max(\cdot)$ functions between two terms are replaced by log-sum-exp functions. 
The Heaviside step function $H(x)$ is replaced by a logistic function approximation.
Finally, the conditional functions of the form
$$
    l(\cdot) = 
    \begin{cases}
        l_{\text{true}}(\cdot)  & \text{if } s(\cdot) \geq 0 \\
        l_{\text{false}}(\cdot) & \text{otherwise} \\
    \end{cases}
$$
are written equivalently as $l(\cdot) = H(s(\cdot)) l_{\text{true}}(\cdot) + (1{-}H(s(\cdot)))l_{\text{false}}(\cdot)$, then smoothified by having the step functions replaced with their logistic function approximation.

\clearpage
\begin{landscape}
    \begin{table}[htb!] \centering
        \caption{Benchmark Model No. 2: Nominal operating point $(\clX{\bar x^{\text{ref}}}, \clU{\bar u^{\text{ref}}}, \clW{\bar w^{\text{ref}}})$ for all process units, except the anaerobic digester, in scientific notation.}
        \label{tab: Operating_Point_A}
        \smallskip
        {\small\begin{tabular}{@{}llcccccccccc@{}}\toprule
            Variable                    & Units                &                     &                 in  &                  $P$ &               $A_1$ &                $A_2$ &               $A_3$ &               $A_4$ &               $A_5$ &                $R$  &                        \\\midrule 
            $\pVars{V}{}{}$             & m$^3$                &                     &                     &                      &                     &                      &                     &                     &                     & $8.00\text{e}{+}01$ &                        \\
            $\pVars{Q}{}{}$             & m$^3$/d              &                     & $2.06\text{e}{+}04$ &  $2.09\text{e}{+}04$ &                     &                      &                     &                     &                     &                     &                        \\ 
            $\pVars{S}{I}{}$            & g\,COD/m$^3$         &                     & $2.56\text{e}{+}01$ &  $2.72\text{e}{+}01$ & $2.72\text{e}{+}01$ &  $2.72\text{e}{+}01$ & $2.72\text{e}{+}01$ & $2.72\text{e}{+}01$ & $2.72\text{e}{+}01$ & $1.40\text{e}{+}02$ &                        \\ 
            $\pVars{S}{S}{}$            & g\,COD/m$^3$         &                     & $5.81\text{e}{+}01$ &  $5.74\text{e}{+}01$ & $2.38\text{e}{+}00$ &  $1.33\text{e}{+}00$ & $9.71\text{e}{-}01$ & $7.86\text{e}{-}01$ & $6.72\text{e}{-}01$ & $2.60\text{e}{+}02$ &                        \\ 
            $\pVars{X}{I}{}$            & g\,COD/m$^3$         &                     & $9.27\text{e}{+}01$ &  $9.21\text{e}{+}01$ & $1.47\text{e}{+}03$ &  $1.47\text{e}{+}03$ & $1.47\text{e}{+}03$ & $1.47\text{e}{+}03$ & $1.47\text{e}{+}03$ & $3.63\text{e}{+}02$ &                        \\ 
            $\pVars{X}{S}{}$            & g\,COD/m$^3$         &                     & $3.64\text{e}{+}02$ &  $3.59\text{e}{+}02$ & $5.83\text{e}{+}01$ &  $5.35\text{e}{+}01$ & $4.10\text{e}{+}01$ & $3.27\text{e}{+}01$ & $2.77\text{e}{+}01$ & $5.71\text{e}{+}01$ &                        \\ 
            $\pVars{X}{BH}{}$           & g\,COD/m$^3$         &                     & $5.01\text{e}{+}01$ &  $5.11\text{e}{+}01$ & $1.95\text{e}{+}03$ &  $1.95\text{e}{+}03$ & $1.95\text{e}{+}03$ & $1.95\text{e}{+}03$ & $1.95\text{e}{+}03$ & $0.00\text{e}{+}00$ &                        \\ 
            $\pVars{X}{BA}{}$           & g\,COD/m$^3$         &                     & $0.00\text{e}{+}00$ &  $7.11\text{e}{-}02$ & $1.24\text{e}{+}02$ &  $1.24\text{e}{+}02$ & $1.25\text{e}{+}02$ & $1.25\text{e}{+}02$ & $1.25\text{e}{+}02$ & $0.00\text{e}{+}00$ &                        \\ 
            $\pVars{X}{P}{}$            & g\,COD/m$^3$         &                     & $0.00\text{e}{+}00$ &  $4.75\text{e}{-}01$ & $8.34\text{e}{+}02$ &  $8.35\text{e}{+}02$ & $8.36\text{e}{+}02$ & $8.38\text{e}{+}02$ & $8.39\text{e}{+}02$ & $1.37\text{e}{+}01$ &                        \\ 
            $\pVars{S}{O}{}$            & g\,O$_2$/m$^3$       &                     & $0.00\text{e}{+}00$ &  $3.37\text{e}{-}02$ & $2.61\text{e}{-}02$ &  $3.89\text{e}{-}04$ & $9.97\text{e}{-}01$ & $2.88\text{e}{+}00$ & $2.58\text{e}{+}00$ & $0.00\text{e}{+}00$ &                        \\ 
            $\pVars{S}{NO}{}$           & g\,N/m$^3$           &                     & $0.00\text{e}{+}00$ &  $1.13\text{e}{-}01$ & $4.80\text{e}{+}00$ &  $3.23\text{e}{+}00$ & $6.32\text{e}{+}00$ & $8.11\text{e}{+}00$ & $8.73\text{e}{+}00$ & $0.00\text{e}{+}00$ &                        \\ 
            $\pVars{S}{NH}{}$           & g\,N/m$^3$           &                     & $2.26\text{e}{+}01$ &  $2.35\text{e}{+}01$ & $4.77\text{e}{+}00$ &  $5.08\text{e}{+}00$ & $1.92\text{e}{+}00$ & $4.15\text{e}{-}01$ & $1.30\text{e}{-}01$ & $1.56\text{e}{+}03$ &                        \\ 
            $\pVars{S}{ND}{}$           & g\,N/m$^3$           &                     & $4.91\text{e}{+}00$ &  $5.58\text{e}{+}00$ & $1.01\text{e}{+}00$ &  $7.49\text{e}{-}01$ & $6.86\text{e}{-}01$ & $6.23\text{e}{-}01$ & $5.66\text{e}{-}01$ & $4.78\text{e}{-}01$ &                        \\ 
            $\pVars{X}{ND}{}$           & g\,N/m$^3$           &                     & $1.48\text{e}{+}01$ &  $1.59\text{e}{+}01$ & $3.47\text{e}{+}00$ &  $3.33\text{e}{+}00$ & $2.73\text{e}{+}00$ & $2.32\text{e}{+}00$ & $2.06\text{e}{+}00$ & $2.20\text{e}{+}00$ &                        \\ 
            $\pVars{S}{ALK}{}$          & mol\,HCO$_3^-$/m$^3$ &                     & $7.00\text{e}{+}00$ &  $6.96\text{e}{+}00$ & $5.29\text{e}{+}00$ &  $5.42\text{e}{+}00$ & $4.98\text{e}{+}00$ & $4.74\text{e}{+}00$ & $4.68\text{e}{+}00$ & $1.06\text{e}{+}02$ &                        \\ 
            $\pVars{T}{}{}$             & $^{\circ}$C          &                     & $1.31\text{e}{+}01$ &  $1.48\text{e}{+}01$ & $1.48\text{e}{+}01$ &  $1.48\text{e}{+}01$ & $1.48\text{e}{+}01$ & $1.48\text{e}{+}01$ & $1.48\text{e}{+}01$ & $1.48\text{e}{+}01$ &                        \\[1.5ex] 
            $Q_A$                       & m$^3$/d              & $6.19\text{e}{+}04$ &                     &                      &                     &                      &                     &                     &                     &                     &                        \\ 
            $Q_S$                       & m$^3$/d              & $2.06\text{e}{+}04$ &                     &                      &                     &                      &                     &                     &                     &                     &                        \\ 
            $Q_W$                       & m$^3$/d              & $3.00\text{e}{+}02$ &                     &                      &                     &                      &                     &                     &                     &                     &                        \\ 
            $Q_R$                       & m$^3$/d              & $1.00\text{e}{+}02$ &                     &                      &                     &                      &                     &                     &                     &                     &                        \\ 
            $\pVars{K_La}{}{}$          & 1/d                  &                     &                     &                      & $0.00\text{e}{+}00$ &  $0.00\text{e}{+}00$ & $1.20\text{e}{+}02$ & $1.20\text{e}{+}02$ & $6.00\text{e}{+}01$ &                     &                        \\
            $\pVars{Q}{EC}{}$           & m$^3$/d              &                     &                     &                      & $0.00\text{e}{+}00$ &  $0.00\text{e}{+}00$ & $0.00\text{e}{+}00$ & $0.00\text{e}{+}00$ & $0.00\text{e}{+}00$ &                     &                        \\[2ex] 
            \toprule
            Variable                    & Units                &                 $S_1$ &                $S_2$ &               $S_3$ &               $S_4$ &               $S_5$ &               $S_6$ &               $S_7$ &               $S_8$ &               $S_9$ &            $S_{10}$   \\\midrule 
            $\pVars{\text{TSS}}{}{}$    & g\,SS/m$^3$          &   $6.54\text{e}{+}03$ &  $1.40\text{e}{+}03$ & $3.88\text{e}{+}02$ & $3.97\text{e}{+}02$ & $3.89\text{e}{+}02$ & $3.95\text{e}{+}02$ & $7.75\text{e}{+}01$ & $3.30\text{e}{+}01$ & $2.00\text{e}{+}01$ & $1.36\text{e}{+}01$   \\
            $\pVars{S}{I}{}$            & g\,COD/m$^3$         &   $2.72\text{e}{+}01$ &  $2.72\text{e}{+}01$ & $2.72\text{e}{+}01$ & $2.72\text{e}{+}01$ & $2.72\text{e}{+}01$ & $2.72\text{e}{+}01$ & $2.72\text{e}{+}01$ & $2.72\text{e}{+}01$ & $2.72\text{e}{+}01$ & $2.72\text{e}{+}01$   \\ 
            $\pVars{S}{S}{}$            & g\,COD/m$^3$         &   $6.72\text{e}{-}01$ &  $6.72\text{e}{-}01$ & $6.72\text{e}{-}01$ & $6.72\text{e}{-}01$ & $6.72\text{e}{-}01$ & $6.72\text{e}{-}01$ & $6.72\text{e}{-}01$ & $6.72\text{e}{-}01$ & $6.72\text{e}{-}01$ & $6.72\text{e}{-}01$   \\ 
            $\pVars{S}{O}{}$            & g\,O$_2$/m$^3$       &   $2.58\text{e}{+}00$ &  $2.58\text{e}{+}00$ & $2.58\text{e}{+}00$ & $2.58\text{e}{+}00$ & $2.58\text{e}{+}00$ & $2.58\text{e}{+}00$ & $2.58\text{e}{+}00$ & $2.58\text{e}{+}00$ & $2.58\text{e}{+}00$ & $2.58\text{e}{+}00$   \\ 
            $\pVars{S}{NO}{}$           & g\,N/m$^3$           &   $8.73\text{e}{+}00$ &  $8.73\text{e}{+}00$ & $8.73\text{e}{+}00$ & $8.73\text{e}{+}00$ & $8.73\text{e}{+}00$ & $8.73\text{e}{+}00$ & $8.73\text{e}{+}00$ & $8.73\text{e}{+}00$ & $8.73\text{e}{+}00$ & $8.73\text{e}{+}00$   \\ 
            $\pVars{S}{NH}{}$           & g\,N/m$^3$           &   $1.30\text{e}{-}01$ &  $1.30\text{e}{-}01$ & $1.30\text{e}{-}01$ & $1.30\text{e}{-}01$ & $1.30\text{e}{-}01$ & $1.30\text{e}{-}01$ & $1.30\text{e}{-}01$ & $1.30\text{e}{-}01$ & $1.30\text{e}{-}01$ & $1.30\text{e}{-}01$   \\ 
            $\pVars{S}{ND}{}$           & g\,N/m$^3$           &   $5.66\text{e}{-}01$ &  $5.66\text{e}{-}01$ & $5.66\text{e}{-}01$ & $5.66\text{e}{-}01$ & $5.66\text{e}{-}01$ & $5.66\text{e}{-}01$ & $5.66\text{e}{-}01$ & $5.66\text{e}{-}01$ & $5.66\text{e}{-}01$ & $5.66\text{e}{-}01$   \\ 
            $\pVars{S}{ALK}{}$          & mol\,HCO$_3^-$/m$^3$ &   $4.68\text{e}{+}00$ &  $4.68\text{e}{+}00$ & $4.68\text{e}{+}00$ & $4.68\text{e}{+}00$ & $4.68\text{e}{+}00$ & $4.68\text{e}{+}00$ & $4.68\text{e}{+}00$ & $4.68\text{e}{+}00$ & $4.68\text{e}{+}00$ & $4.68\text{e}{+}00$   \\ 
            $\pVars{T}{}{}$             & $^{\circ}$C          &   $1.48\text{e}{+}01$ &  $1.48\text{e}{+}01$ & $1.48\text{e}{+}01$ & $1.48\text{e}{+}01$ & $1.48\text{e}{+}01$ & $1.48\text{e}{+}01$ & $1.48\text{e}{+}01$ & $1.48\text{e}{+}01$ & $1.48\text{e}{+}01$ & $1.48\text{e}{+}01$   \\ 
            \bottomrule
        \end{tabular}}
    \end{table}	
\end{landscape}    

\clearpage
\begin{landscape}
    \begin{table}[htb!] \centering
        \caption{Benchmark Model No. 2: Nominal operating point $(\clX{\bar x^{\text{ref}}}, \clU{\bar u^{\text{ref}}}, \clW{\bar w^{\text{ref}}})$ for the anaerobic digester, in scientific notation.}
        \label{tab: Operating_Point_B}
        \smallskip
        {\small\begin{tabular}{@{}llc@{}}\toprule
            Variable                 & Units           &         $D$       \\\midrule
            $\pVars{S}{SU}{}$        & kg\,COD/m$^{3}$ & $1.21\text{e}{-}02$ \\
            $\pVars{S}{AA}{}$        & kg\,COD/m$^{3}$ & $5.43\text{e}{-}03$ \\
            $\pVars{S}{FA}{}$        & kg\,COD/m$^{3}$ & $1.04\text{e}{-}01$ \\
            $\pVars{S}{VA}{}$        & kg\,COD/m$^{3}$ & $1.20\text{e}{-}02$ \\
            $\pVars{S}{BU}{}$        & kg\,COD/m$^{3}$ & $1.37\text{e}{-}02$ \\
            $\pVars{S}{PRO}{}$       & kg\,COD/m$^{3}$ & $1.71\text{e}{-}02$ \\
            $\pVars{S}{AC}{}$        & kg\,COD/m$^{3}$ & $8.12\text{e}{-}02$ \\
            $\pVars{S}{H$_2$}{}$     & kg\,COD/m$^{3}$ & $2.45\text{e}{-}07$ \\
            $\pVars{S}{CH$_4$}{}$    & kg\,COD/m$^{3}$ & $5.52\text{e}{-}02$ \\
            $\pVars{S}{IC}{}$        & kg\,COD/m$^{3}$ & $9.14\text{e}{-}02$ \\
            $\pVars{S}{IN}{}$        & kg\,COD/m$^{3}$ & $9.05\text{e}{-}02$ \\
            $\pVars{S}{I}{}$         & kg\,COD/m$^{3}$ & $1.13\text{e}{-}01$ \\
            $\pVars{X}{C}{}$         & kg\,COD/m$^{3}$ & $1.07\text{e}{-}01$ \\
            $\pVars{X}{CH}{}$        & kg\,COD/m$^{3}$ & $2.03\text{e}{-}02$ \\
            $\pVars{X}{PR}{}$        & kg\,COD/m$^{3}$ & $8.05\text{e}{-}02$ \\
            $\pVars{X}{LI}{}$        & kg\,COD/m$^{3}$ & $4.34\text{e}{-}02$ \\
            $\pVars{X}{SU}{}$        & kg\,COD/m$^{3}$ & $3.16\text{e}{-}01$ \\
            $\pVars{X}{AA}{}$        & kg\,COD/m$^{3}$ & $9.08\text{e}{-}01$ \\
            $\pVars{X}{FA}{}$        & kg\,COD/m$^{3}$ & $3.43\text{e}{-}01$ \\
            $\pVars{X}{C$_4$}{}$     & kg\,COD/m$^{3}$ & $3.28\text{e}{-}01$ \\
            $\pVars{X}{PRO}{}$       & kg\,COD/m$^{3}$ & $9.97\text{e}{-}02$ \\
            $\pVars{X}{AC}{}$        & kg\,COD/m$^{3}$ & $6.71\text{e}{-}01$ \\
            $\pVars{X}{H$_2$}{}$     & kg\,COD/m$^{3}$ & $2.83\text{e}{-}01$ \\
            $\pVars{X}{I}{}$         & kg\,COD/m$^{3}$ & $1.64\text{e}{+}01$ \\
            $\pVars{S}{CAT$^+$}{}$   & kmol/m$^{3}$    & $0.00\text{e}{-}00$ \\
            $\pVars{S}{AN$^-$}{}$    & kmol/m$^{3}$    & $5.30\text{e}{-}03$ \\
            $\pVars{S}{VA$^-$}{}$    & kg\,COD/m$^{3}$ & $1.20\text{e}{-}02$ \\
            $\pVars{S}{BU$^-$}{}$    & kg\,COD/m$^{3}$ & $1.36\text{e}{-}02$ \\
            $\pVars{S}{PRO$^-$}{}$   & kg\,COD/m$^{3}$ & $1.71\text{e}{-}02$ \\
            $\pVars{S}{AC$^-$}{}$    & kg\,COD/m$^{3}$ & $8.09\text{e}{-}02$ \\
            $\pVars{S}{HCO$_3^-$}{}$ & kmol/m$^{3}$    & $8.19\text{e}{-}02$ \\
            $\pVars{S}{NH$_3$}{}$    & kmol/m$^{3}$    & $1.72\text{e}{-}03$ \\
            $\pVars{G}{H$_2$}{}$     & kg\,COD/m$^{3}$ & $1.08\text{e}{-}05$ \\
            $\pVars{G}{CH$_4$}{}$    & kg\,COD/m$^{3}$ & $1.65\text{e}{+}00$ \\ 
            $\pVars{G}{CO$_2$}{}$    & kg\,COD/m$^{3}$ & $1.35\text{e}{-}02$ \\
            \bottomrule
        \end{tabular}}
    \end{table}	
\end{landscape} 

\clearpage
\section{Output-MPC: Overview, tuning parameters, and implementation details} \label{sec: OutMPC}

\begin{figure}[htb!] \centering   
   	\includegraphics[width=0.47\columnwidth]{OutMPC_Diagram}  
    \hfill
   	\includegraphics[width=0.47\columnwidth]{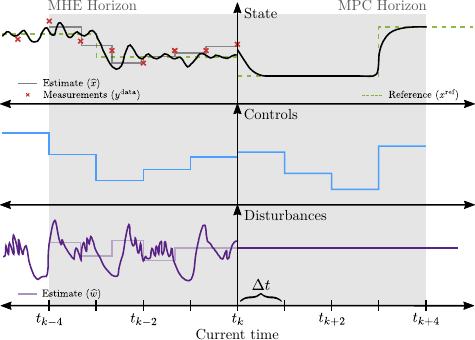}  
	
	\caption{Main components of the Output MPC (left) and an illustration of the \textit{discretize-then-optimize} strategy (right).} 
    \label{fig: OutputMPC_Scheme} 
\end{figure}

The output-feedback model predictive control (Output MPC, Figure \ref{fig: OutputMPC_Scheme}) proposed in this work (Section 3, main text) is a realisation of the model-based control framework presented in (Neto\textit{~et~al.}, 2024) \cite{NMC_BSM1_Control}. 
We refer to this preliminary work for the full details on designing an Output MPC to operate a general system for which a state-space model is available. 
Our specific design is tailored to achieve an efficient and robust operation of biological wastewater treatment plants to achieve energy-autonomous water resource recovery.

The Output MPC is a digital control solution which operates the process in cycles of duration $\Delta t > 0$, the \textit{control period}, each starting at time $t_k = k\Delta t$ ($k = 1,2,\ldots$). 
To operate the WWTP, we consider a control period of $\Delta t = 1/96~[\text{days}] = 15~[\text{min}]$. 
The \textit{discretize-and-optimize} approach \cite{Betts2010} to implement an Output MPC consists of first discretizing the dynamical model of the process in time, assuming that its inputs are held constant in between control cycles (see Figure \ref{fig: OutputMPC_Scheme}), then reformulating its associated optimization problems as discrete-time optimal control problems. 
Because finite-horizon, these optimization problems can be recast as conventional nonlinear programs and thus solved by specialized software \cite{Boyd2004, Rawlings2020}.

In the following, we provide further details the main components of the Output MPC: 
The operating point optimizer (OPO), the model predictive controller (MPC), the moving horizon estimator (MHE). 
We present the optimization problem associated with each component, explicitly define its tuning parameters, and discuss implementation details related to their digitalization and numerical solution. 

\subsection{The operating point optimizer (OPO)} \label{subsec: OPO}

For a cycle starting at $t_k$, the OPO computes the pair $(x_{t_k}^{\text{ref}}, u_{t_k}^{\text{ref}}) \in \mathbb{R}^{N_x} \times \mathbb{R}^{N_u}$ solving
\begin{subequations}
\begin{align} 
    \underset{x_{t_k}, u_{t_k}}{\text{minimize}}
        & \quad \left\| W_z^{\text{ref}}(h(x_{t_k}, u_{t_k}) {-} \bar{z}^{\text{ref}}_{t_k}) \right\|^2_2 + \left\| W_u^{\text{ref}} (u_{t_k} {-} \bar{u}^{\text{ref}}_{t_k}) \right\|^2_2 \label{eq: OPO_a}\\
    \text{subject to} 
        & \quad 0 = f(x_{t_k}, u_{t_k}, \bar{w}^{\text{ref}}_{t_k}),	                       \label{eq: OPO_b}\\
        & \quad x_{t_k} \in \mathcal{X}^{\text{ref}}, \\ 
        & \quad u_{t_k} \in \mathcal{U}^{\text{ref}}, \\ 
        & \quad h(x_{t_k}, u_{t_k}) \in \mathcal{Z}^{\text{ref}}. \label{eq: OPO_c}
\end{align} \label{eq: OPO}%
\end{subequations}
The reference point $(\bar{z}^{\text{ref}}_{t_k}, \bar{u}^{\text{ref}}_{t_k}, \bar{w}^{\text{ref}}_{t_k})$ is chosen so that $\bar{z}^{\text{ref}}_{t_k}$ is the desired KPI values at $t_k$ (see Table 3, main text), with $(\bar{u}^{\text{ref}}_{t_k}, \bar{w}^{\text{ref}}_{t_k})$ set constant to the nominal operating values reported in Table \ref{tab: Operating_Point_A}. 
Explicitly, 
\begin{align*}
    \bar{z}^{\text{ref}}_{t_k} &= \begin{cases}
        \big[ 10 ~~ 4 ~~  7.5 \big]^{\tran} & \text{if } t_k \in [0, 31) \cup [181, 212) \cup [243, 273) \cup [304, 334) \\
        \big[ 10 ~~ 4 ~~ 22.5 \big]^{\tran} & \text{if } t_k \in  [31, 59) \cup [151, 181) \cup [212, 243) \cup [273, 304) \cup [334, 365) \\
        \big[ 10 ~~ 4 ~~ 37.5 \big]^{\tran} & \text{if } t_k \in [59, 151)  
    \end{cases} \\ 
    \bar{u}^{\text{ref}}_{t_k} &= \big[ 61944 ~~ 20648 ~~ 300 ~~ 100 ~~ 0 ~~ 0 ~~ 120 ~~ 120 ~~ 60 ~~ 0 ~~ 0 ~~ 0 ~~ 0 ~~ 0 \big]^{\tran} \\
    \bar{w}^{\text{ref}}_{t_k} &= \big[ 20648 ~~ 25.685 ~~ 58.176 ~~ 92.78 ~~ 364.79 ~~ 50.126 ~~ 0 ~~ 0 ~~ 0 ~~ 0 ~~ 22.603 ~~ 4.9144 ~~ 14.889 ~~ 7 ~~ 13.11 \big]^{\tran}
\end{align*}
The weighting matrices $\{ W_z^{\text{ref}}, W_u^{\text{ref}} \}$ are defined as the positive semidefinite block-diagonal matrices
\begin{equation*}
    W_z^{\text{ref}} = \begin{bmatrix}
        1 &   &   &   \\ 
          & 1 &   &   \\
          &   & 1 &   \\
          &   &   & 0
    \end{bmatrix}
    \quad \text{ and } \quad
    W_u^{\text{ref}} = 0.6\begin{bmatrix}
        10^{-4} &         &         &         &            &                 \\
                & 10^{-3} &         &         &            &                 \\
                &         & 10^{-3} &         &            &                 \\
                &         &         & 10^{-3} &            &                 \\
                &         &         &         &            &                 \\
                &         &         &         & 10^{-2}I_5 &                 \\
                &         &         &         &            & \frac{1}{3}I_5  
        \end{bmatrix}.
\end{equation*}
The weighting for the inputs ($W_u^{\text{ref}}$) is chosen to ``normalize'' each variable (given their range of admissible values), then scaled by a factor of $0.6$ to prioritize matching the KPI references. 
The state dynamics $f(\cdot)$ are taken directly from the equations presented in Section \ref{sec: BSM2_SS}-- because the OPO is a static optimization problem, the model does not need to be discretized in time. 
The constraint sets $\{ \mathcal{X}^{\text{ref}}, \mathcal{U}^{\text{ref}}, \mathcal{Z}^{\text{ref}} \}$ are defined as 
\begin{align*}
    \mathcal{X}^{\text{ref}} &= \{ x \in \mathbb{R}^{N_x} : 0 \leq x, ~~ 0 \leq x_{211} \leq 320 \}; \\
    \mathcal{U}^{\text{ref}} &= \left\{ u \in \mathbb{R}^{N_u} : 
        \begin{bmatrix}
            0\\
            0\\
            0\\
            0\\
            \mathbf{0}_{5}\\
            \mathbf{0}_{5}
        \end{bmatrix} 
        \preceq u \preceq 
        \begin{bmatrix}
            92230 \hfill \\
            36892 \hfill \\
            1844 \hfill \\
            500 \hfill \\
            320 \cdot \mathbf{1}_{5} \hfill\\
            5 \cdot \mathbf{1}_{5} \hfill
        \end{bmatrix} 
    \right\};\\
    \mathcal{Z}^{\text{ref}} &= \{ z \in \mathbb{R}^{N_z} : z_4 \leq 0 \}.
\end{align*}
The state constraints enforce that, at the operating point, all the state-variables must be positive and that the volume of the reject water storage tank ($x_{211} \equiv V^{R}$) has a limit of $320$ m$^3$.
The control actions are restricted to satisfy the actuation limits imposed by the equipment, with the explicit values provided by the BSM2.
The KPI constraints enforce that the energy cost index ($\text{ECI} \equiv z_4$) of the process must be nonpositive, thus enforcing energy-autonomous operating conditions.
The constraints in Eqs. \eqref{eq: OPO_a}--\eqref{eq: OPO_c} can thus be directly replaced by the inequalities above.

Under this setup, Problem \eqref{eq: OPO} is a nonconvex quadratic program with $N_x + N_u = 239$ decision variables, $N_x = 225$ equality constraints, and $(N_x + 1) + 2N_u + 1 = 255$ inequality constraints. 
We obtain a (locally) optimal solution using an interior-point solver \cite{IPOPT2006}, whose convergence depends on the initial guess provided to the optimizer: 
At each $t_k$, the initial guess is the previous operating point $p^{\text{ref}}_{t_{k-1}} = ({x}^{\text{ref}}_{t_{k-1}}, {u}^{\text{ref}}_{t_{k-1}}, \bar{w}^{\text{ref}}_{t_{k-1}})$, with $p^{\text{ref}}_{0} = (\bar{x}^{\text{ref}}, \bar{u}^{\text{ref}}, \bar{w}^{\text{ref}})$ taken from Tables \ref{tab: Operating_Point_A}--\ref{tab: Operating_Point_B}, as this yields consistent and satisfactory results.

\clearpage
\subsection{The model predictive controller (MPC)} \label{subsec: MPC}

For a cycle starting at $t_k$, the MPC is tasked with computing the optimal control and state-response signals $(u^{\star}, x^{\star}) \in \{ [t_{k},t_{k}{+}H_c] \to \mathbb{R}^{N_u} \} \times \{ [t_{k},t_{k}{+}H_c] \to \mathbb{R}^{N_x} \}$ solving
\begin{subequations}
\begin{align}
    \underset{u(\cdot),~ x(\cdot)}{\text{minimize}}
        & \quad \int_{t_{k}}^{t_k + H_c}{ \left( \| W_x \big( x(t) - x^{\text{ref}}_{t_k} \big) \|^2_2 + \| W_u \big( u(t) - u^{\text{ref}}_{t_k} \big) \|^2_2 \right) dt} + \| W_f \big( x(t_k{+}H_c) - x^{\text{ref}}_{t_k} \big) \|^2_2	\label{eq: MPC_cost}\\
    \underset{t \in [t_k,t_k+H_c]}{\text{subject to}}
        & \quad \textstyle\frac{d}{dt} x(t) = A_{t_k} (x(t) - x_{t_k}^{\text{ref}}) + B_{t_k} (u(t) - u_{t_k}^{\text{ref}}) + G_{t_k} (\widehat{w}(t_k) - \bar{w}_{t_k}^{\text{ref}}),			\label{eq: MPC_dynamics}\\[-1.5ex]
        & \quad x(t) \in \mathcal{X}^c, \\ 
        & \quad u(t) \in \mathcal{U}^c,	\\
        & \quad x(t_k) = \widehat{x}(t_k),
\end{align} \label{eq: MPC}%
\end{subequations}
with the state-space matrices $A_{t_k} \in \mathbb{R}^{N_x \times N_x}$, $B_{t_k} \in \mathbb{R}^{N_x \times N_u}$, and $G_{t_k} \in \mathbb{R}^{N_x \times N_w}$ obtained as
\begin{equation*}
    A_{t_k} {=} (\partial f / \partial x)\vert_{p_{t_k}}, ~ B_{t_k} {=} (\partial f / \partial u)\vert_{p_{t_k}}, ~ G_{t_k} {=} (\partial f / \partial w)\vert_{p_{t_k}},
\end{equation*} 
given the reference and linearization point $p_{t_k} = ({x}^{\text{ref}}_{t_k}, {u}^{\text{ref}}_{t_k}, \bar{w}^{\text{ref}}_{t_k})$ obtained by the OPO.
The problem, as posed, is a continuous-time dynamic optimization and thus it is not amenable to a digital implementation. 
Following the \textit{discretize-then-optimize} approach \cite{Betts2010}, we proceed by first discretizing the action and state signals in time: 
Considering the \textit{control period} $\Delta t$, we assume a zero-order hold of the inputs ($u(t) = u(n \Delta t)$, for all $t \in [n\Delta t, (n{+}1)\Delta t)$) and then define the discrete-time signals $x[n] = x(n\Delta t)$ and $u[n] = u(n\Delta t)$, for the time-indices $n \in [k, k{+}1, \ldots, k{+}N_c]$ with $N_c = \lfloor H_c/\Delta t \rfloor$. 
The zero-order hold assumption allows us to analytically obtain a discrete-time equivalent of the affine dynamics in Eq. \eqref{eq: MPC_dynamics},
\begin{align*}
    \delta x[n{+}1] &= \delta x[n] + \int_{0}^{\Delta t} \big( A_{t_k} \delta x(t) + B_{t_k} \delta u(t) + G_{t_k} \delta \widehat{w}(t_k) \big)dt \\
                    &= A_{d,t_k} \delta x[n] + B_{d,t_k} \delta u[n] + G_{d,t_k} \delta \widehat{w}[k],
\end{align*}
in which we define the deviation variables $\delta x(t) = (x(t) - x_{t_k}^{\text{ref}})$, $\delta u(t) = (u(t) - u_{t_k}^{\text{ref}})$, and $\delta \widehat{w}(t_k) = (\widehat{w}(t_k) - w_{t_k}^{\text{ref}})$, for ease of notation, and the discrete-time state-space matrices are obtained through the matrix exponential
\begin{equation*}
    \left[ \begin{array}{ccc}
   A_{d,t_k} & B_{d,t_k} & G_{d,t_k} \\
             & I_{N_u}   &           \\
             &           & I_{N_w}   \\
\end{array}\right]
 = 
 \exp\left(\left[ \begin{array}{ccc}
    A_{t_k} & B_{t_k}  & G_{t_k}  \\
            & 0I_{N_u} &          \\
            &          & 0I_{N_w} \\
 \end{array}\right] \Delta t\right).
\end{equation*}
Given this discretization, the integral in Eq. \eqref{eq: MPC_cost} can be replaced by a sum.
The Problem \eqref{eq: MPC} thus becomes a discrete-time optimal control problem,
\begin{subequations}
\begin{align}
    \underset{\delta u[\cdot],~ \delta x[\cdot]}{\text{minimize}}
        & \quad \sum_{n = k}^{k + N_c - 1}{ \left( \| W_x \delta x[n] \|^2_2 + \| W_u \delta u[n] \|^2_2 \right) dt} + \| W_f \delta x[k{+}N_c] \|^2_2	\label{eq: MPC_discr_cost}\\
    \text{subject to}
        & \quad \delta x[n{+}1] = A_{d,t_k} \delta x[n] + B_{d,t_k} \delta u[n] + G_{d,t_k} \delta w[k],  & n = k, \ldots, k{+}N_c{-}1	\label{eq: MPC_discr_dynamics}\\
        & \quad \delta x[n] + x^{\text{ref}}_{t_k} \in \mathcal{X}^{c} ,                             & n = k, \ldots, k{+}N_c~~~~ \label{eq: MPC_discr_const_a}\\
        & \quad \delta u[n] + u^{\text{ref}}_{t_k} \in \mathcal{U}^{c} ,	                          & n = k, \ldots, k{+}N_c{-}1  \label{eq: MPC_discr_const_b}\\
        & \quad \delta x[k] = \widehat{x}(t_k) - x^{\text{ref}}_{t_k},
\end{align} \label{eq: MPC_discr}%
\end{subequations}
which is now a finite-dimensional optimization problem amenable to numerical solution.
We remark that, in this optimization, the vectors $\{ \widehat{x}(t_k), \widehat{w}(t_k) \}$ are data taken from the solution of the MHE (Section \ref{subsec: MHE}).
From the solution of Problem \eqref{eq: MPC_discr}, only the control action $u^{\star}[k] = \delta u^{\star}[k] + u^{\text{ref}}_{t_k}$ is deployed to the physical plant, and then it is repeated for $\Delta t = 1/96$ [days] until the next MPC cycle starting at $t_{k+1} = (k+1)\Delta t$.

In our design, the control horizon corresponds to $N_c = 24$ time-steps (or, equivalently, $H_c = 1/3$ days of operation).
The MPC is tuned with the weighting matrices $\{ W_x, W_u \}$ defined as the block-diagonal matrices
\begin{align*}
    W_x &= \begin{bmatrix} 
        0.01 &     &     &        &        &      &   &   &   &       &     &      &     &       \\
             & 0.1 &     &        &        &      &   &   &   &       &     &      &     &       \\
             &     & 0.1 &        &        &      &   &   &   &       &     &      &     &       \\
             &     &     & 0.1I_5 &        &      &   &   &   &       &     &      &     &       \\
             &     &     &        & 0.1I_5 &      &   &   &   &       &     &      &     &       \\
             &     &     &        &        & 0.01 &   &   &   &       &     &      &     &       \\
             &     &     &        &        &      & 1 &   &   &       &     &      &     &       \\
             &     &     &        &        &      &   & 2 &   &       &     &      &     &       \\
             &     &     &        &        &      &   &   & 2 &       &     &      &     &       \\
             &     &     &        &        &      &   &   &   & 0.001 &     &      &     &       \\
             &     &     &        &        &      &   &   &   &       & 0.3 &      &     &       \\
             &     &     &        &        &      &   &   &   &       &     & 0.03 &     &       \\
             &     &     &        &        &      &   &   &   &       &     &      & 0.3 &       \\
             &     &     &        &        &      &   &   &   &       &     &      &     & 0.001
    \end{bmatrix}C_{t_k}; \\
    W_u &= 3\begin{bmatrix}
        10^{-4} &         &         &         &            &                 \\
                & 10^{-3} &         &         &            &                 \\
                &         & 10^{-3} &         &            &                 \\
                &         &         & 10^{-3} &            &                 \\
                &         &         &         &            &                 \\
                &         &         &         & 10^{-2}I_5 &                 \\
                &         &         &         &            & \frac{1}{3}I_5  
    \end{bmatrix}.
\end{align*}
Similarly to the OPO, the weights for the inputs ($W_u$) are chosen to ``normalize'' each variable (given their range of admissible values), then scaled by a factor of $3$ to avoid the controller to deviate significantly from the operating conditions $u^{\text{ref}}_{t_k}$. 
The weighting matrix penalizing the state deviations is of the form $W_x = W_y^c C_{t_k}$, with the Jacobian $C_{t_k} = (\partial g / \partial x)\vert_{p_{t_k}}$: 
As such, the weighting factors in $W_y^c$ can be understood as penalizing the $N_y = 27$ measurable outputs directly, thus greatly simplifying the tuning of the MPC. 
The matrix $W_y^c$ is chosen to prioritize matching the references for the measurable quantities $\{ \text{TSS}^{S_{10}}, \pVars{S}{NH}{S_{10}}, \pVars{S}{NO}{S_{10}}, Q_G \}$.
Finally, the terminal weighting matrix $W_f$ is chosen as the solution to the discrete-time algebraic Riccati equation,
\begin{equation*}
    W_f^{\tran}W_f = A_{t_k}^{\tran} W_f^{\tran}W_f A_{t_k} - (B_{t_k}^{\tran} W_f^{\tran}W_f A_{t_k})^{\tran}(W_u^{\tran} W_u + B_{t_k}^{\tran} W_f^{\tran}W_f B_{t_k})^{-1}(B_{t_k}^{\tran} W_f^{\tran}W_f A_{t_k}) + W_x^{\tran} W_x,
\end{equation*}
as this choice can ensure exponential stability for the resulting closed-loop system \cite{Rawlings2020}.

The constraint sets $\{ \mathcal{X}^{c}, \mathcal{U}^{c} \}$ are defined as 
\begin{align*}
    \mathcal{X}^{c} = \{ x \in \mathbb{R}^{N_x} : 0 \leq x_{211} \leq 320 \}
    \quad \text{and} \quad
    \mathcal{U}^{c} = \left\{ u \in \mathbb{R}^{N_u} : 
        \begin{bmatrix}
            0\\
            0\\
            0\\
            0\\
            \mathbf{0}_{5}\\
            \mathbf{0}_{5}
        \end{bmatrix} 
        \preceq u \preceq 
        \begin{bmatrix}
            92230 \hfill \\
            36892 \hfill \\
            1844 \hfill \\
            500 \hfill \\
            320 \cdot \mathbf{1}_{5} \hfill\\
            5 \cdot \mathbf{1}_{5} \hfill
        \end{bmatrix} 
    \right\}.
\end{align*}
These constraints enforce the controller to avoid emptying or overflowing the reject water storage tank ($\text{ECI} \equiv z_4 \in [0, 320]$ m$^3$) and to satisfy the actuation limits imposed by the equipment.
The constraints in Eqs. \eqref{eq: MPC_discr_const_a}--\eqref{eq: MPC_discr_const_b} can thus be directly replaced by the inequalities above.

Under this setup, Problem \eqref{eq: MPC_discr} is a convex quadratic program with $(N_c)N_x + (N_c{-}1)N_u = 5722$ decision variables, $(N_c{-}1)N_x = 5175$ equality constraints, and $1 + 2N_u = 29$ inequality constraints. 
Despite being a large-scale problem, this program has a special structure which can be exploited by sparsity-exploiting numerical methods, and thus it can still be solved efficiently \cite{Wang2009}.
Moreover, the problem is convex and thus it can solved globally irrespective of the initial guess provided to the optimizer (provided that it is feasible).

\clearpage
\subsection{The moving horizon estimator (MHE)}  \label{subsec: MHE}

For a cycle starting at $t_k$, the MHE is tasked with computing the optimal disturbance and state estimates $(\widehat{w}, \widehat{x}) \in \{ [t_{k}{-}H_e, t_{k}] \to \mathbb{R}^{N_w} \} \times \{ [t_{k}{-}H_e,t_{k}] \to \mathbb{R}^{N_x} \}$ solving
\begin{subequations}
\begin{align}
    \underset{w(\cdot),~ x(\cdot)}{\text{minimize}}
        & \quad \| W_i \big( x(t_k{-}H_e) - \widehat{x}^{\text{ref}}_{t_k} \big) \|^2_2 + \int_{t_{k}{-}H_e}^{t_k}{ \left( \| W_y \big( y(t) - y^{\text{data}}(t) \big) \|^2_2 + \| W_w \big( w(t) - \bar{w}^{\text{ref}}_{t_k} \big) \|^2_2 \right) dt}	\label{eq: MHE_cost}\\
    \underset{t \in [t_k{-}H_e,t_k]}{\text{subject to}}
        & \quad \textstyle\frac{d}{dt} x(t) = A_{t_k} (x(t) - x_{t_k}^{\text{ref}}) + B_{t_k} (u^{\star}(t) - u_{t_k}^{\text{ref}}) + G_{t_k} (w(t_k) - \bar{w}_{t_k}^{\text{ref}}),			\label{eq: MHE_dynamics}\\[-1.5ex]
        & \quad                        y(t) = g(x^{\text{ref}}_{t_k}) + C_{t_k} (x(t) {-} x_{t_k}^{\text{ref}}),			                                                                        \label{eq: MHE_measurements}\\
        & \quad x(t) \in \mathcal{X}^e, \\ 
        & \quad w(t) \in \mathcal{W}^e,
\end{align} \label{eq: MHE}%
\end{subequations}
with state-space matrices $A_{t_k} \in \mathbb{R}^{N_x \times N_x}$, $B_{t_k} \in \mathbb{R}^{N_x \times N_u}$, $G_{t_k} \in \mathbb{R}^{N_x \times N_w}$, and $C_{t_k} \in \mathbb{R}^{N_y \times N_x}$, obtained as
\begin{equation*}
    A_{t_k} {=} (\partial f / \partial x)\vert_{p_{t_k}}, ~ B_{t_k} {=} (\partial f / \partial u)\vert_{p_{t_k}}, ~ G_{t_k} {=} (\partial f / \partial w)\vert_{p_{t_k}}, ~ C_{t_k} {=} (\partial g / \partial x)\vert_{p_{t_k}},
\end{equation*}
given the reference and linearization point $p_{t_k} = ({x}^{\text{ref}}_{t_k}, {u}^{\text{ref}}_{t_k}, \bar{w}^{\text{ref}}_{t_k})$ obtained by the OPO.
Similar to the MPC, this is a continuous-time dynamic optimization and thus the dynamics in Eq. \eqref{eq: MHE_dynamics} must be discretized in time to allow for a digital implementation. 
Again, considering the \textit{control period} $\Delta t$, we assume a zero-order hold of the inputs ($w(t) = w(n \Delta t)$, for all $t \in [n\Delta t, (n{+}1)\Delta t)$) and then define the discrete-time signals $x[n] = x(n\Delta t)$ and $w[n] = u(n\Delta t)$, for the time-indices $n \in [k{-}N_e, \ldots, k{-}1, k]$ with $N_e = \lfloor H_e/\Delta t \rfloor$. 
The measurements are represented in discrete-time as $y[n] = y(n\Delta t)$.
The discrete-time equivalent of the affine dynamics in Eq. \eqref{eq: MHE_dynamics} and measurement in Eq. \eqref{eq: MHE_measurements} are
\begin{align*}
    \delta x[n{+}1] &= A_{d,t_k} \delta x[n] + B_{d,t_k} \delta u^{\star}[n] + G_{d,t_k} \delta w[k], \\
    \delta     y[n] &= C_{t_k} \delta x[n],
\end{align*}
in which we define the deviation variables $\delta x(t) = (x(t) - x_{t_k}^{\text{ref}})$, $\delta u^{\star}(t) = (u^{\star}(t) - u_{t_k}^{\text{ref}})$, $\delta w(t_k) = (w(t_k) - w_{t_k}^{\text{ref}})$, and $\delta y(t) = y(t) - g(x^{\text{ref}}_{t_k})$, for ease of notation. 
The discrete-time state-space matrices are obtained as in Section \ref{subsec: MPC}. 
Defining $\delta y^{\text{data}}(t) = y^{\text{data}}(t) - g(x^{\text{ref}}_{t_k})$ and $\delta \widehat{x}_{t_k}^{\text{ref}} = \widehat{x}_{t_k}^{\text{ref}} - x_{t_k}^{\text{ref}}$, and replacing the integral in Eq. \eqref{eq: MHE_cost} with a sum, the Problem \eqref{eq: MHE} becomes the discrete-time optimal estimation problem
\begin{subequations}
\begin{align}
    \underset{\delta w[\cdot],~ \delta x[\cdot]}{\text{minimize}}
        & \quad \| W_i (\delta x[k{-}N_e] - \delta \widehat{x}_{t_k}^{\text{ref}}) \|^2_2 + \sum_{n = k{-}N_e{+}1}^{k}{ \left( \| W_y (C_{t_k} \delta x[n] - \delta y^{\text{data}}[n]) \|^2_2 + \| W_w \delta w[n] \|^2_2 \right) dt}	\label{eq: MPC_discr_cost}\\
    \text{subject to}
        & \quad \delta x[n{+}1] = A_{d,t_k} \delta x[n] + B_{d,t_k} \delta u^{\star}[n] + G_{d,t_k} \delta w[k],  \hspace*{04.0em} n = k{-}N_e, \ldots, k{-}1  \label{eq: MHE_discr_dynamics}\\
        & \quad \delta x[n] + x^{\text{ref}}_{t_k} \in \mathcal{X}^{e} ,                                          \hspace*{17.6em} n = k{-}N_e, \ldots, k      \label{eq: MHE_discr_const_a}\\
        & \quad \delta w[n] + \bar{w}^{\text{ref}}_{t_k} \in \mathcal{W}^{e} ,	                                  \hspace*{17.1em} n = k{-}N_e, \ldots, k{-}1  \label{eq: MHE_discr_const_b}
\end{align} \label{eq: MHE_discr}%
\end{subequations}
which is now a finite-dimensional program amenable to numerical solution.
In Problem \eqref{eq: MHE_discr}, the vectors $\{ y^{\text{data}}[k{-}N_e], \ldots, y^{\text{data}}[k] \}$ are historical measurement data collected from the process instrumentation, whereas $\{ u^{\star}[k{-}N_e], \ldots, u^{\star}[k{-}1] \}$ are data taken from the solutions of the MPC that were deployed to the system. 
The initial state reference is defined recursively as $\widehat{x}_{t_k}^{\text{ref}} = \widehat{x}(t_{k+1})$ with $\widehat{x}(\cdot)$ being the estimates obtained from the previous MHE cycle. 
Note that the disturbance $\delta w[k]$ is not constrained by Eqs. \eqref{eq: MHE_discr_dynamics}--\eqref{eq: MHE_discr_const_b}, and thus $\delta \widehat{w}[k] = 0$ is always the optimal solution;
with a slight abuse of notation, we assign $\delta \widehat{w}[k] \coloneqq \delta \widehat{w}[k-1]$ prior to providing the MHE estimates to the Problem \eqref{eq: MPC_discr} solved by the MPC.

In our design, the MHE is configured to only estimate the values of the $N_{\widetilde{w}} = 8$ disturbance variables $\widetilde{w} = (w_{2}, \ldots, w_{6}, w_{11}, \ldots, w_{13})$:
The quantities $\{ w_1, w_{15} \} \equiv \{ \pVars{Q}{}{\text{in}}, \pVars{T}{}{\text{in}} \}$ are assumed to be measured in real-time and $\{ w_{7}, w_{8}, w_{9}, w_{10}, w_{14} \} \equiv \{ \pVars{X}{BA}{\text{in}}, \pVars{X}{P}{\text{in}}, \pVars{S}{O}{\text{in}}, \pVars{S}{NO}{\text{in}}, \pVars{S}{ALK}{\text{in}}  \}$ are known to be constant.
The estimation horizon is set to $N_e = 24$ time-steps (or, equivalently, $H_e = 1/3$ days of historical data).
Our estimator is then tuned with the weighting matrices $\{ W_y, W_w\}$ defined from the inverses of the block-diagonal matrices
\begin{align*}
    W_y^{-1} &= \begin{bmatrix} 
        1 &     &     &        &        &        &   &      &     &     &   &      &   &      &      \\
          & 0.1 &     &        &        &        &   &      &     &     &   &      &   &      &      \\
          &     & 0.1 &        &        &        &   &      &     &     &   &      &   &      &      \\
          &     &     & 0.1I_5 &        &        &   &      &     &     &   &      &   &      &      \\
          &     &     &        & 0.1I_5 &        &   &      &     &     &   &      &   &      &      \\
          &     &     &        &        & 0.1I_5 &   &      &     &     &   &      &   &      &      \\
          &     &     &        &        &        & 3 &      &     &     &   &      &   &      &      \\
          &     &     &        &        &        &   &  0.3 &     &     &   &      &   &      &      \\
          &     &     &        &        &        &   &      & 0.1 &     &   &      &   &      &      \\
          &     &     &        &        &        &   &      &     & 0.1 &   &      &   &      &      \\
          &     &     &        &        &        &   &      &     &     & 3 &      &   &      &      \\
          &     &     &        &        &        &   &      &     &     &   & 0.01 &   &      &      \\
          &     &     &        &        &        &   &      &     &     &   &      & 3 &      &      \\
          &     &     &        &        &        &   &      &     &     &   &      &   & 0.01 &      \\
          &     &     &        &        &        &   &      &     &     &   &      &   &      & 0.01 
    \end{bmatrix}; \\W_w^{-1} &= \begin{bmatrix}
          5 &    &    &     &    &   &   &   &   \\
            & 28 &    &     &    &   &   &   &   \\
            &    & 33 &     &    &   &   &   &   \\
            &    &    & 129 &    &   &   &   &   \\
            &    &    &     & 18 &   &   &   &   \\
            &    &    &     &    & 9 &   &   &   \\
            &    &    &     &    &   & 2 &   &   \\
            &    &    &     &    &   &   & 6 &   \\
            &    &    &     &    &   &   &   & 3
    \end{bmatrix}.
\end{align*}
The above tuning is motivated from the \textit{maximum a posteriori} interpretation of the MHE:
In this case, the matrices $W_y^{-1}$ and $W_w^{-1}$ correspond to the covariance matrices of the unmeasured disturbances $\widetilde{w}(\cdot)$ and of the measurement noise assumed to be corrupting the data $y^{\text{data}}(\cdot)$.
The former is inspired by the sensor noise magnitudes recommended by the benchmark \cite{Gernaey2014}-- which are also used in our simulations to corrupt the measurement data--, while the latter is obtained by computing the sample covariance of the first year of wastewater data (Section 2, main text) and then extracting its diagonal components.
Finally, the initial weighting matrix $W_i$ is chosen as the solution of the discrete-time algebraic Riccati equation,
\begin{equation*}
    W_i W_i^{\tran} = A_{t_k} W_i W_i^{\tran} A_{t_k}^{\tran} - (C_{t_k} W_i W_i^{\tran} A_{t_k}^{\tran})^{\tran}(W_w W_w^{\tran} + C_{t_k} W_i W_i^{\tran} C_{t_k}^{\tran})^{-1}(C_{t_k} W_i W_i^{\tran} A_{t_k}^{\tran}) + W_y W_y^{\tran}.
\end{equation*}

The constraint sets $\{ \mathcal{X}^{e}, \mathcal{\widetilde W}^{e} \}$ are defined as 
\begin{align*}
    \mathcal{X}^{e} = \mathbb{R}^{N_x}
    \quad \text{and} \quad
    \mathcal{\widetilde W}^{e} = \mathbb{R}^{N_{\widetilde{w}}}_{\geq 0} = \{ \widetilde{w} \in \mathbb{R}^{N_{\widetilde{w}}} : 0 \preceq \widetilde{w} \}.
\end{align*}
These constraints enforce the MHE to estimate the disturbances to be positive (as they correspond to (bio)chemical concentrations).
The state estimates are constrained only by the dynamics Eq. \eqref{eq: MPC_discr_dynamics}. 
The constraints in Eqs. \eqref{eq: MHE_discr_const_a}--\eqref{eq: MHE_discr_const_b} can thus be directly replaced by the inequalities above.

Under this setup, Problem \eqref{eq: MHE_discr} is a convex quadratic program with $(N_e)N_x + (N_e{-}1)N_{\widetilde{w}} = 5560$ decision variables, $(N_e{-}1)N_x = 5175$ equality constraints, and $N_{\widetilde{w}} = 8$ inequality constraints. 
Although not identical, this optimization enjoys the same structure of the Problem \eqref{eq: MPC_discr}, and thus can be solved efficiently by the same sparsity-exploiting methods \cite{Wang2009}.
Moreover, the problem is also convex and thus it can solved globally irrespective of the initial guess provided to the optimizer (provided that it is feasible).

\clearpage
\bibliographystyle{cas-model2-names}    
